\documentclass[sn-aps]{sn-jnl}

\jyear{2022}%

\theoremstyle{thmstyleone}%
\theoremstyle{thmstyletwo}%

\theoremstyle{thmstylethree}%
\usepackage[shortlabels]{enumitem}

\usepackage{url}
\usepackage{xcolor}
\usepackage{color,soul}
\usepackage{graphicx}

\usepackage{rotating}

\usepackage{multirow}

\usepackage{subcaption}
\usepackage[export]{adjustbox}

\usepackage{amsmath}

\def\imagetop#1{\vtop{\null\hbox{#1}}}

\begin{document}

\title[Transonic buffet and separation bubble modes]{On the co-existence of transonic buffet and separation-bubble modes for the OALT25 laminar-flow wing section} 


\author*[1]{\fnm{Markus} \sur{Zauner}}\email{m.zauner@soton.ac.uk}

\author[1]{\fnm{Pradeep} \sur{Moise}}\email{pradeep890@gmail.com}



\author[1]{\fnm{Neil D.} \sur{Sandham}}\email{n.sandham@soton.ac.uk}

\affil*[1]{\orgdiv{Aerodynamics and Flight Mechanics Group}, \orgname{ University of Southampton}, \orgaddress{\street{University Rd, Highfield}, \city{Southampton}, \postcode{SO17 1BJ}, \state{Hampshire}, \country{UK}}}





\abstract{Transonic buffet is an unsteady flow phenomenon that limits the safe flight envelope of modern aircraft. Scale-resolving simulations with span-periodic boundary conditions are capable of providing new insights into its flow physics. The present contribution shows the co-existence of multiple modes of flow unsteadiness over an unswept laminar-flow wing section, appearing in the following order of increasing frequency: (a) a low-frequency transonic buffet mode, (b) an intermediate-frequency separation bubble mode, and (c) high-frequency wake modes associated with vortex shedding. Simulations are run over a range of Reynolds and Mach numbers to connect the lower frequency modes from moderate to high Reynolds numbers and from pre-buffet to established buffet conditions. The intermediate frequency mode is found to be more sensitive to Reynolds-number effects compared to those of Mach number, which is the opposite trend to that observed for transonic buffet. Spectral proper orthogonal decomposition is used to extract the spatial structure of the modes. The buffet mode involves coherent oscillations of the suction-side shock structure, consistent with previous studies including global mode analysis. The laminar separation-bubble mode at intermediate frequency is fundamentally different, with a phase relationship between separation and reattachment that does not correspond to a simple `breathing' mode and is not at the same Strouhal number observed for shock-induced separation bubbles. Instead, a Strouhal number based on separation bubble length and reverse flow magnitude is found to be independent of Reynolds number within the range of cases studied.}


\maketitle

\section{Introduction}
\label{sec:intro}

Transonic buffet refers to a self-sustained passive aerodynamic instability at high speeds that typically occurs near stall conditions and leads to strong and detrimental fluctuations of aerodynamic forces over wings and control surfaces or engine components of modern aircraft. As a consequence, these load oscillations can induce violent structural vibrations, known as buffeting, limiting the safe flight envelope of civil aircraft and performance capabilities of military platforms \citep{Jones1973, John1974}. 
Even though buffet-related phenomena have been known for a long time \citep{Duncan1934}, we still lack a complete understanding of the underlying mechanism \citep{Mabey1968, Lee2001, Giannelis2017}. While transonic buffet is typically accompanied by a single oscillating shock wave, the latter's role (\textit{i.e.} cause or consequence) with respect to intermittent flow-separation effects still remains an open question \cite{Paladini2019}. 
While two-dimensional buffet phenomena can co-exist for swept wings \citep{Paladini2019a} as well as half-wing-body configurations \citep{Timme2016a, Masini2020}, three-dimensional phenomena like buffet cells \cite{Iovnovich2012}, which have been shown to be essentially stall cells \cite{Plante2019}, are also likely to be present. 
In the scope of the present study, we focus our review on buffet over unswept wing sections, known as `2D buffet'. 
The main purpose of the present numerical study of ONERA's OALT25 profile is to disentangle the multiple modes present near stall of free-transitional sections. To do this, we first need to review a broad range of literature.

We observe two main philosophies in the modern literature, describing buffet either as an acoustic feedback mechanism or as a global instability.
In a first attempt to explain the buffet mechanism, Erickson \textit{et al.} \cite{Erickson1947} proposed a model of transonic buffet based on acoustic feedback, where acoustic waves originating from the trailing edge propagate upstream and interact with the shock wave. After modifications by \cite{Tijdeman1977}, Lee \cite{LEE1990} supplemented this model with downstream-convecting disturbances, previously observed by \cite{Roos1980}. Despite large popularity and further modifications (\textit{e.g.}, \cite{Hartmann2013,Stanewsky1990}), such acoustic feedback models lack general validity and have been shown inaccurate in several numerical as well as experimental studies \citep{Garnier2010, Fukushima2017, Sugioka2018, Zauner2019c, Moise2022}.

Crouch and co-workers \cite{Crouch2007, Crouch2009} presented a more  rigorous way to analyse buffet onset by solving an eigenvalue problem of the Reynolds-Averaged Navier-Stokes equations linearised around a two-dimensional baseflow. This allowed them to identify a global mode becoming unstable at onset conditions and frequencies typical of buffet. Even though these modes suggest strong fluctuations of the shock wave location, they also highlight strong coupling between shock dynamics and boundary-layer separation phenomena. In addition to direct global stability analysis, \cite{Sartor2015} solved also its adjoint problem, which suggests regions optimal to influence associated global modes. 
While \cite{Sartor2015} highlights the strong sensitivity of transonic buffet to boundary-layer characteristics, the shock wave itself does not appear in adjoint modes, which raises questions about the role of shock waves in transonic buffet.

The studies discussed so far have considered only fully-turbulent or tripped boundary layers upstream of the main shock wave. Buffet phenomena over laminar-flow wings with delayed natural transition of the laminar boundary layer have recently attracted increasing attention to support the design more efficient next-generation aircraft. Industry is particularly interested in answering the question of how buffet characteristics (such as onset conditions, frequency, and amplitude) of laminar-flow wings are different from conventional supercritical wings. While transition modelling for Reynolds-Averaged Navier Stokes simulations can be problematic, a recent surge in computational resources in combination with powerful massively-parallelised CFD codes enables scale-resolving simulations of practical test cases relevant for buffet research.
Dandois \textit{et al.} \cite{Dandois2018} carried out the first wall-resolved Large-Eddy-Simulation of a laminar-flow wing at buffet conditions considering the OALT25 profile.
They observed rather localised oscillations of the shock foot ($6\%$ of the chord length) at a relatively high Strouhal number (based on chord length and free-stream velocity) of $St \approx 1.2$, compared to typical large-scale motion of the entire shock wave ($20\%$ of the chord length) at $St\approx 0.07$ applying boundary-layer tripping \citep{Brion2017}. The authors suggested two fundamentally different mechanisms leading to periodic lift and shock oscillations. 
On the one hand, the instability observed for tripped boundary layers resembles the global mode described in \cite{Crouch2007}. On the other hand, the more localised oscillations for free-transitional boundary layers appeared more reminiscent of separation-bubble breathing phenomena, which are typically observed for shock waves impinging on a laminar boundary layer over a flat plate \citep{Piponniau2009}. 
Based on whether the boundary layer upstream of the shock wave is fully turbulent or laminar, \cite{Dandois2018} referred to the former as `turbulent buffet' and the latter as `laminar buffet', acknowledging the similarities in terms of aerodynamic consequences (\textit{i.e.} shock oscillations and periodic load fluctuations), but differences in terms of spatio-temporal scales (\textit{i.e.} low- versus intermediate-frequency oscillations) and origin (\textit{i.e.} instability of the entire flow field versus a localised laminar separation bubble). 
Conducting experimental studies of the same test case, \cite{Brion2019} could essentially confirm the numerical results of \cite{Dandois2016}, but observed, in addition to the sharp spectral peak at $St=1.1$, a weak bump in spectra of pressure probes around $St=0.05-0.06$. 

In the scope of the European TFAST project, several numerical (\textit{e.g.} \citep{Grossi2014, Szubert2016, Sznajder2016a, Memmolo2018}) as well as experimental studies (\textit{e.g.} \citep{Davidson2016a, Placek2016b, Placek2016}) were carried out for Dassault Aviation's V2C laminar-flow profile at $Re \approx 2,\!600,\!000$. 
In terms of the spatial organisation of flow oscillations, buffet compared well between the V2C and OALT25 test cases subjected to tripped boundary layers, even though the dominant frequency of the former case ($St\approx0.1$) is significantly higher compared to the latter ($St\approx0.06$). 
Interestingly, tripping boundary layers on the V2C airfoil at various locations did not show significant effect on buffet frequencies and amplitudes \citep{Placek2016, Davidson2016a}.
Using LES and RANS methods, \cite{Moise2022, Moise2023} carried out an extensive parametric study for the V2C airfoil at Reynolds numbers in the range of $Re = 500,\!000-3,\!000,\!000$.
While multiple shock waves appear at free-transitional conditions, \cite{Moise2023} showed that boundary-layer tripping leads to the appearance of a single shock wave.
For free-transitional cases, \cite{Moise2022} observed no significant Reynolds-number sensitivity in the features of the buffet instability at $St \approx 0.1$.
An increase in Reynolds numbers, on the other hand, was found to reduce the number of shock waves suggesting that further increase in Reynolds numbers would lead to single back and forth oscillating shock wave as seen as in experiments.
By means of Spectral Proper Orthogonal Decomposition (SPOD) and global linear stability analysis of results from  Reynolds-Averaged Navier-Stokes equations it was shown that the underlying buffet instability is essentially the same for tripped and free-transitional boundary layers. Furthermore, the shape of buffet modes for the V2C profile resemble those of OAT15A \cite{Sartor2015} and NACA0012 \cite{Crouch2007} profiles.

For free-transitional conditions, \cite{Moise2022} also reported SPOD modes at $St\approx1.5$, which are mainly dominant in the wake region, hence labelled as wake modes. Similar flow structures were also extracted by \cite{Zauner2019c} applying dynamic mode decomposition (DMD) to Direct Numerical Simulation (DNS) data at $Re = 500,\!000$. Various other studies (\textit{e.g.}, \cite{Grossi2014a, Memmolo2018}) have reported similar structures around $St \approx 1$ for higher Reynolds numbers. 

Spectra of experimental data in \citep{Placek2016b} suggest a weak secondary bump at Strouhal numbers substantially larger compared to the one linked to buffet ($St\approx0.4$). 
Also \cite{Zauner2019c} reported a weak broadband bump in DNS spectra and extracted DMD modes in the frequency range of $St=0.4-0.8$, which are composed of acoustic waves and fluctuations in boundary and shear layers at wave lengths comparable to the size of the laminar separation bubble, suggesting a connection to the separation-bubble instability reported in \cite{Dandois2018} for the OALT15. Even though the physical relevance of these weak DMD modes has not been confirmed so far, \cite{Boerner2021} reported similar intermediate-frequency phenomena at $St \approx 0.42$ for experimental studies of a transonic low pressure turbine.


The objectives of this study are to understand and characterise the nature of `laminar buffet' reported in \cite{Dandois2018} and how it links to the rest of the literature on transonic buffet by performing scale-resolving simulations of free-transitional flows over the OALT25 airfoil, where variations of $Re$ lead to different onset-conditions.

Before outlining the structure of the present paper, we want to introduce some definitions used within the scope of this work.
Using scale-resolving simulations for our analysis, we consider Reynolds numbers of the order of $Re \sim O(10^6)$ as \textit{high} compared to $Re \sim O(10^5)$ associated with \textit{moderate} Reynolds numbers. Below and beyond these bands, Reynolds numbers are labelled as \textit{low} and \textit{flight} Reynolds numbers, respectively.


As we encounter in literature a range of relevant flow phenomena at different time scales, we classify them into \textit{low-} ($St\sim O(10^{-2})$), \textit{intermediate-} ($St\sim O(10^{-1})$), \textit{high-frequency} phenomena ($St\sim O(10^{0})$). Phenomena at frequencies beyond $St\sim O(10^{1})$ are associated with linear boundary-layer instabilities and small-scale turbulence \citep{Zauner2019c, Zauner2017a}.
For present test cases, we observe transonic buffet in the low-frequency range, where the spatial organisation of oscillations of the flow field agrees well with the globally unstable `turbulent buffet' mode described by \cite{Crouch2007}. As we will see later, these phenomena do not show significant Reynolds-number sensitivity for the investigated test cases.
{Intermediate-frequency phenomena}, are confirmed to be associated with separation bubble unsteadiness, which was labelled as `laminar buffet' in previous studies \cite{Dandois2018}.
In the present study, we will for the most part refrain from expressions like `turbulent' and `laminar' buffet to avoid confusion, since both modes can co-exist for the considered test cases with natural transition. 
{High-frequency phenomena} appear to be mostly dominant within the airfoil wake and are associated with wake modes \citep{Moise2022}.

Now that we have classified transonic buffet as a low-frequency instability, we characterise flow conditions according to its dominance. While the flow field at \textit{pre-buffet} conditions is dominated by unsteadiness at Strouhal numbers $St>0.1$, we associate the first appearance of a clear peak at low frequencies with \textit{incipient buffet}. An increase of relevant flow parameters (\textit{e.g.}, Mach number or angle of attack) would lead to \textit{developed buffet} (or \textit{deterrent buffet} from the pilot's perspective), where low-frequency phenomena clearly dominate the global flow dynamics around the wing section. Further increasing the aerodynamic load will lead to a decay of low-frequency oscillations, \textit{i.e.}, \textit{buffet offset}, and eventually, instabilities at higher frequencies will again dominate the flow field, which will be referred to as \textit{post-buffet}. We should note that post-buffet does not necessarily coincide with stall. 


After describing our numerical approach in section \ref{sec:approach}, we will validate our simulations of ONERA's OALT25 airfoil against those of \cite{Dandois2018} and experiments of \cite{Brion2017} at $Re=3,\!000,\!000$ in the first part of section \ref{sec:Validation}. Then we will study Reynolds number effects in the second part of section \ref{sec:Validation}. 
Having confirmed all relevant phenomena at reduced Reynolds numbers of $Re=500,\!000$, we will study Mach-number effects at moderate $Re$ in section \ref{sec:Results}, which also allows us to compare the present simulations with previous work on the V2C profile. Before concluding our paper, we will suggest a scaling approach for the intermediate frequency (`laminar buffet') mode and further scrutinise the underlying mechanism in section \ref{sec:discussion}.

\section{Methodology}
\label{sec:approach}

\subsection{Flow solver}

For the present work, numerical simulations are performed using the in-house code, SBLI \citep{Yao2009}, which has been well-validated and used for studies on shock-wave/boundary-layer interaction \citep{Touber2009, Sansica2015}, subsonic wing sections \citep{Jones2008, DeTullio2018}, as well as transonic buffet \citep{Zauner2019a, Zauner2020, Moise2023}. This code has shown good performance for massively-parallelised simulations using structured multi-block grids on several high-performance computer architectures.
The compressible three-dimensional Navier-Stokes equations are normalised by the airfoil chord $c$, the freestream density, streamwise velocity, and temperature and solved in a non-dimensional form using fourth-order finite difference schemes (central at interior and the Carpenter scheme \citep{Carpenter1999} at boundaries) for spatial discretisation. A low-storage third-order Runge-Kutta scheme is used for temporal discretisation. A total variation diminishing scheme is employed to capture features of shock waves \citep{Sansica2015}.
Zonal characteristic boundary conditions \cite{Sandberg2006_char} are enforced at the outlet, whereas integral characteristic boundary conditions \cite{Sandhu1994} are applied at the remaining outer boundaries.

We will use for the present contribution a Spectral-Error based Implicit Large-Eddy Simulation (SE-ILES) scheme. 
For this method, spectral error indicators are computed locally every $N_E$ time steps according to \cite{Jacobs2017} and are used to control a sixth-order filter \citep{Visbal2002} in order to remove scales, which are not sufficiently resolved by the grid. Details of this approach can be found in \cite{Zauner2019b}, where this method has been validated against DNS for a transonic buffet test case and successfully applied in multiple studies \citep{Zauner2020, Moise2022, Moise2023}.

\subsection{Grid generation}
\label{subSec:grid}
The grids used in the present study have been designed and generated using an open-source parametric grid-generation tool for high-fidelity grids developed at the University of Southampton \citep{polygridwizz, Zauner2018b}. High-order polynomials are used to define the geometry of grid lines and corresponding grid-point distributions. Such body-fitted grids are not necessarily limited to a single geometry, as grid lines are designed with respect to their position (coordinates) and local curvature with respect to grid points along the airfoil contour. Therefore, it is possible to use grids with very similar characteristics and resolution for different airfoil geometries.
Here, grids previously used for simulations of Dassault Aviation's V2C airfoil and already assessed in terms of spatial resolution \citep{Zauner2018b, Zauner2018c}, were adapted to the OALT25 geometry.

\begin{figure}
\vspace{0.25cm}
\begin{tabular}[t]{@{}l@{}l}
a) & b) \\
\imagetop{\includegraphics[width=0.5\columnwidth]{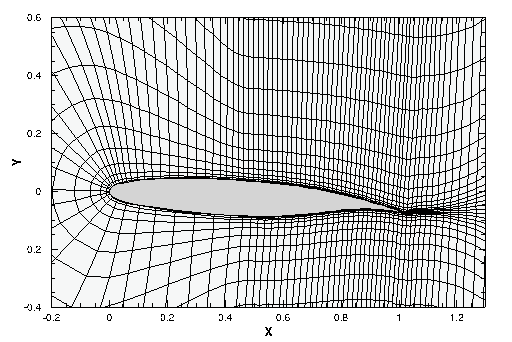}} &  \imagetop{\includegraphics[width=0.5\columnwidth]{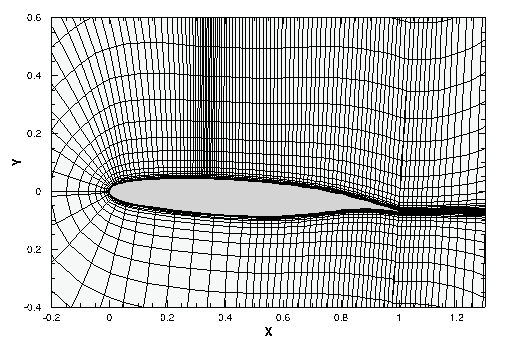}} 
\end{tabular}
  \caption{
  Approximately every $15^{th}$ grid point is shown for (a) the standard grid targeting $Re=500,\!000$ and (b) a refined grid targeting $Re=3,\!000,\!000$.
  }
\label{fig:Grids}
\end{figure}

In the present work we will use two different CH-type grids for the OALT25 profile, referred to as `standard' and `refined' grids. The topology and resolution of the standard grid is very similar to grids used in \cite{Zauner2020, Moise2022} for the V2C airfoil and have been shown adequate to capture the main phenomena of interest. For both simulations, the C-block of the grid has a radius of $7.5c$, while the two H-blocks capture $5.5c$ of the wake behind the airfoil. The spanwise extent of the domain is limited to $0.05c$.
The refined grid contains additional grid points within the region close to the airfoil surface and is only used for simulations at high Reynolds numbers. Close-ups are shown for grids in figure \ref{fig:Grids}, plotting only every $15^{th}$ grid point. 

The standard grid in figure \ref{fig:Grids}(a), consists of approximately 90 million points (one block of $1495\times 550\times 50$ points and two blocks of $799\times 564\times 50$), where the blunt trailing edge of $0.5\%c$ contains 30 grid points. The grid clustering is relatively denser close to the aerofoil surface, in its wake, and in the region where the shock wave is expected. The wall-normal and wall-tangential spacings at the airfoil surface vary between $1\times10^{-4}$ to $1.7\times10^{-4}$ and $4\times10^{-4}$ to $2\times10^{-3}$, respectively. 

The refined grid is used for a validation case at increased Reynolds numbers of $Re=3,\!000,\!000$ only and contains approximately 200 million points (one block of $1935\times 550\times 100$ points and two blocks of $799\times 564\times 100$). For comparison, the grid of a validated simulation by \cite{Dandois2018}, targeting the same Reynolds number, contained about 400 million grid points, for a two times wider domain with a larger wall-normal extent of $100c$.
Around $x\approx0.3$, it is notable that the grid is relatively denser in the wall-tangential direction. This would allow a future use of the same grid for studies with boundary-layer tripping applied, as done in \cite{Moise2023}.

\subsection{Fourier spectral analysis}

We apply Fourier transformation techniques in order to analyse one-dimensional signals such as aerodynamic histories.
A typical simulation run time covers about $160$ convective time units. Neglecting an initial transient of approximately $25$ time units, results in a total signal length of $140$ time units. 
Assuming a signal is split into $50\%$ overlapping bins containing about $35$ time units, each bin contains a sufficient number of cycles at intermediate frequencies ($>10$ cycles for $St=0.3$). However, with the same bin size low-frequency cycles are poorly captured.  To ovecome this limitation, we gradually decrease the number of bins (leading to an increase in bin size) as frequency is reduced. The resulting `composed' spectra retain the low-frequency content, while reducing the noise in the high-frequency content.

\begin{figure}[hbt!]
\centering
\includegraphics[width=.42\textwidth]{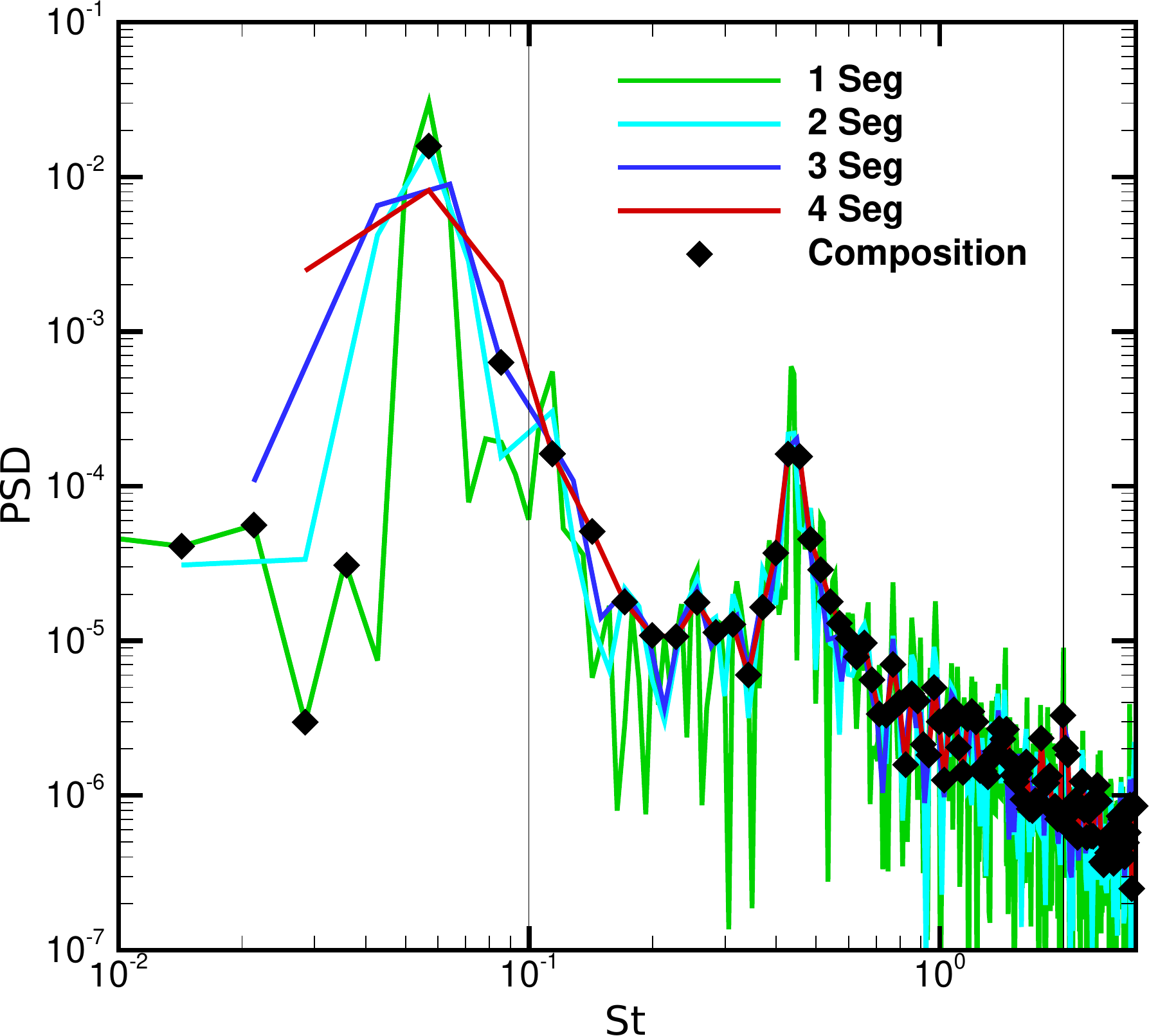}
\caption{Welch spectra of lift coefficient at $M=0.71$. 
The red curve denotes the spectrum using bins of about $35$ convective time units. Dark blue, light blue, and green curves denote Welch spectra with bins containing $46$, $70$, and $140$ time units, respectively. Symbols mark Fourier modes, which are selected for a composed spectrum. 
}
\label{fig:FFTexample}
\end{figure}

To illustrate the robustness of this strategy, an example of such a composed Fourier spectrum is shown in figure \ref{fig:FFTexample} considering a representative $C_L$ signal at $M=0.71$. 
The red dotted curve denotes the Welch spectrum using $35$ time units per bin. For Strouhal numbers $St>0.1$, modes are selected only from the red spectrum. The low-frequency part of the spectrum ($St<0.1$) is expanded by modes from spectra using larger bins containing $47$ (dark blue curve), $70$ (light blue curve), and $140$ time units (green curve). 
This leads to the final composed spectrum, marked by black symbols. 


\subsection{Spectral proper orthogonal decomposition}
Coherent features in the flow field were extracted using spectral proper orthogonal decomposition \citep{Lumley1970, Glauser1987, Towne2018}. This decomposition finds an orthonormal basis that ideally represents a given ensemble of realisations of a stochastic process. It can be shown that the ideal basis which is spatio-temporally coherent consists of the eigenfunctions, $\boldsymbol{\psi}$, of the cross-spectral density tensor, $\boldsymbol{S}$ \cite{Towne2018}, which satisfy
\begin{equation}
\int_\Omega \boldsymbol{S}(\boldsymbol{x}, \boldsymbol{x}', St)\boldsymbol{W}(\boldsymbol{x}')\boldsymbol{\psi}_i(\boldsymbol{x}',St) d\boldsymbol{x}' = \lambda_i(St) \boldsymbol{\psi}_i(\boldsymbol{x},St).
    \label{eqnSPOD}
\end{equation}
Here, $\boldsymbol{x}, \boldsymbol{x}' \in \Omega$ are any two points in the domain, $\Omega$, $St$ is the frequency/Strouhal number, $\boldsymbol{W}$ is a weight function related to the relevant inner product
, $\lambda$ is the eigenvalue and the subscript $i$ denotes the $i$-th eigenvalue/eigenfunction. The index $i$ is such that $\lambda_i$ are sorted in descending order, implying that $i=1$, represents the most-energetic mode. Thus, the eigenfunction $\boldsymbol{\psi}_i(\boldsymbol{x},St_0)$ represents the spatial structure of an SPOD mode that oscillates in time with a frequency, $St_0$. The spatio-temporal variation of this mode is given by
\begin{equation}
\boldsymbol{\zeta_i}(\boldsymbol{x},t) = \Re(\boldsymbol{\psi}_i \exp(\mathrm{i}2\pi St_0 \: t)),
\end{equation}
where $\Re(\cdot)$ denotes real part and $t$ refers to time.

The numerical code provided in \cite{SCHMIDT201998} was used for performing SPOD. It adopts the Welch approach for Fourier transforms and is a streaming algorithm, which computes only the first few of the most-energetic modes. Here, only the first two dominant modes are examined (\textit{i.e.}, $i = 1$ and 2). The domain chosen is the $z = 0$ plane. Snapshots at different time instants are constructed by arranging the density, velocity components and pressure on this plane into a column vector. The snapshots were sampled at time intervals of 0.064 (sampling frequency $F_s \approx 15$). To compute the Fourier transform using the Welch approach, snapshots are divided into blocks of $T_B \approx 40$. Note that the lowest frequency associated with coherent oscillations occurs for $St > 0.05$, implying that at least two oscillation cycles are captured in a block. The weighting function, $\boldsymbol{W}$ is chosen based on the approximate volume associated with each grid point. To compare the spatial structures of SPOD modes obtained for different cases, the phase within an oscillation cycle, $\phi = 2\pi St_0 \: t$, of the spatio-temporal SPOD mode, 

\begin{equation}
\boldsymbol{\zeta}(\boldsymbol{x},\phi) = \Re(\boldsymbol{\psi}(\boldsymbol{x},St_0) \exp(\mathrm{i}\phi)),
\end{equation}
must be chosen appropriately. Here, we choose $\phi = 0$ as occurring when  $\Im(\boldsymbol{\psi}(x_\mathrm{TE},St_0) \exp(\mathrm{i}\phi)) = 0$, where $\Im(\cdot)$ corresponds to the imaginary part and $\boldsymbol{x}_\mathrm{TE}$ represents the point on the upper corner of the trailing edge. Thus, contours shown in figures~\ref{fig:SPODModesReeffect} and \ref{fig:SPODModesRe500k} are at $\phi = 0$. Movies showing the spatial structure's variation with phase are provided in the supplementary material. Further details can be found in \cite{Moise2022} and \cite{Moise2023}.  

\subsection{Test cases}

For all present simulations, we consider ONERA's OALT25 laminar-flow wing geometry fixed at an angle of attack of $\alpha=4^{\circ}$.
The fluid is considered to be air, which can be modelled at present conditions as a perfect gas with a specific heat ratio of $\kappa=1.4$, satisfying Fourier's law of heat conduction with a Prandtl number of $Pr=0.72$. It is also assumed to be Newtonian and satisfying Sutherland's law, with the Sutherland coefficient as $T_s=110.4$ at a reference temperature $T_r=268.67~K$.

Starting from $M=0.735$ and $Re=500,\!000$, either Mach or Reynolds numbers are varied in order to study their effect on low- and intermediate-frequency oscillations separately. An overview of the test matrix and information about dominant spectral peaks is provided in table \ref{tab:test2}. 

\begin{table}
	\caption{Test matrix.\label{tab:test2}}
	\begin{center}
		\begin{tabular}{rc|cc|cc|c}
  		    $Re$ & $M$ & \multicolumn{2}{c|}{Low-frequency} & \multicolumn{2}{c|}{Intermediate-frequency} & Grid\\
		    & & $St$ & PSD$(C_L')$ & $St$ & PSD$(C_L')$ & \\
		    \hline
            $3,\!000,\!000$     & 0.735 & 0.082 & 0.00325 & 0.924 & 0.00327 & refined\\
            $3,\!000,\!000$     & 0.735 & 0.091 & 0.00780 & 1.039 & 0.00349 & standard\\
            $2,\!000,\!000$     & 0.735 & 0.080 & 0.01097 & 0.875 & 0.00196 & standard\\
            $1,\!000,\!000$     & 0.735 & 0.081 & 0.05455 & 0.535 & 0.00062 & standard\\
		    $500,\!000$         & 0.735 & 0.082 & 0.15385 & 0.393 & 0.00030 & standard\\
		    $500,\!000$         & 0.710 & 0.060 & 0.01580 & 0.427 & 0.00016 & standard\\
		    $500,\!000$         & 0.700 & 0.035 & 0.00068 & 0.417 & 0.00012 & standard\\
		    $500,\!000$         & 0.690 & 0.039 & 0.00019 & 0.469 & 0.00002 & standard\\
		    $500,\!000$         & 0.680 & 0.060 & 0.00004 & 0.595 & 0.00002 & standard\\
		    $500,\!000$         & 0.670 & 0.035 & 0.00002 & 0.831 & - & standard\\
		    $500,\!000$         & 0.750 & 0.104 & 0.06275 & 0.469 & 0.00015 & standard\\
		    $500,\!000$         & 0.800 & 0.208 & 0.00138 & 0.547 & 0.00004 & standard\\
		    \hline
		\end{tabular}
	\end{center}
\end{table}


\section{Reynolds-number effect for $M=0.735$}
\label{sec:Validation}




In this section, we consider a case in which intermediate-frequency oscillations are well established at wind-tunnel conditions with $M=0.735$ and $Re=3,\!000,\!000$ and progressively reduce the Reynolds number down to $Re=500,\!000$. Fourier spectra and SPOD are used to identify and characterise the principal modes considered in this study and their sensitivity to $Re$. We can then justify the use of more cost-effective simulations at moderate $Re$ in subsequent sections. Since the $Re=3,\!000,\!000$ million case is the same as that previously studied by \cite{Dandois2018} we start by making a cross validation between the results from two separate codes.


\paragraph{Cross-validation at $\boldsymbol{Re=3,\!000,\!000}$}

Studying the OALT25 profile, \cite{Dandois2018} observed oscillations which were fundamentally different from those corresponding to cases with tripped boundary layers. Test cases subjected to fully-turbulent separation bubbles and shock/boundary-layer interactions typically show large-scale shock motion and lift fluctuations at low Strouhal numbers around $St<0.1$. For free-transitional test cases, however, shock oscillations were more localised and mainly limited to the shock foot. These lift oscillations occurred at significantly higher frequencies corresponding to $St \approx 1.1$. These observations were confirmed by wind-tunnel experiments at similar free-transitional conditions of $M=0.735$ and $Re=3,\!000,\!000$ \citep{Brion2019}, where an additional peak was observed in spectra of lift-fluctuations at significantly lower frequencies around $St\approx0.06$, which is in the range of typical buffet frequencies observed for the OALT25 profile. 

\begin{figure}
\vspace{0.25cm}
\begin{tabular}[t]{@{}l@{}l}
a) & b) \\
\imagetop{\includegraphics[width=0.5\columnwidth]{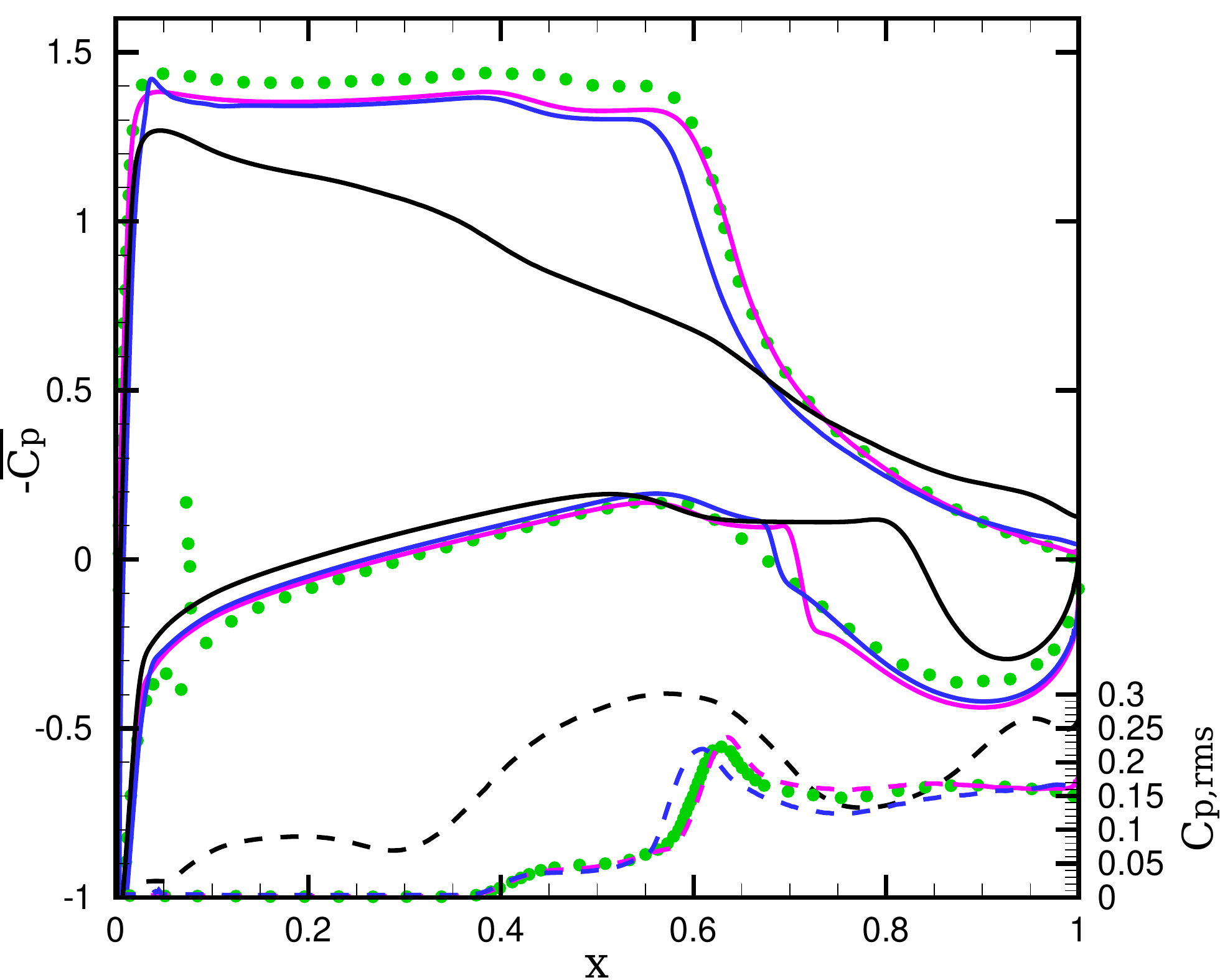}} &  
\imagetop{\includegraphics[width=0.5\columnwidth]{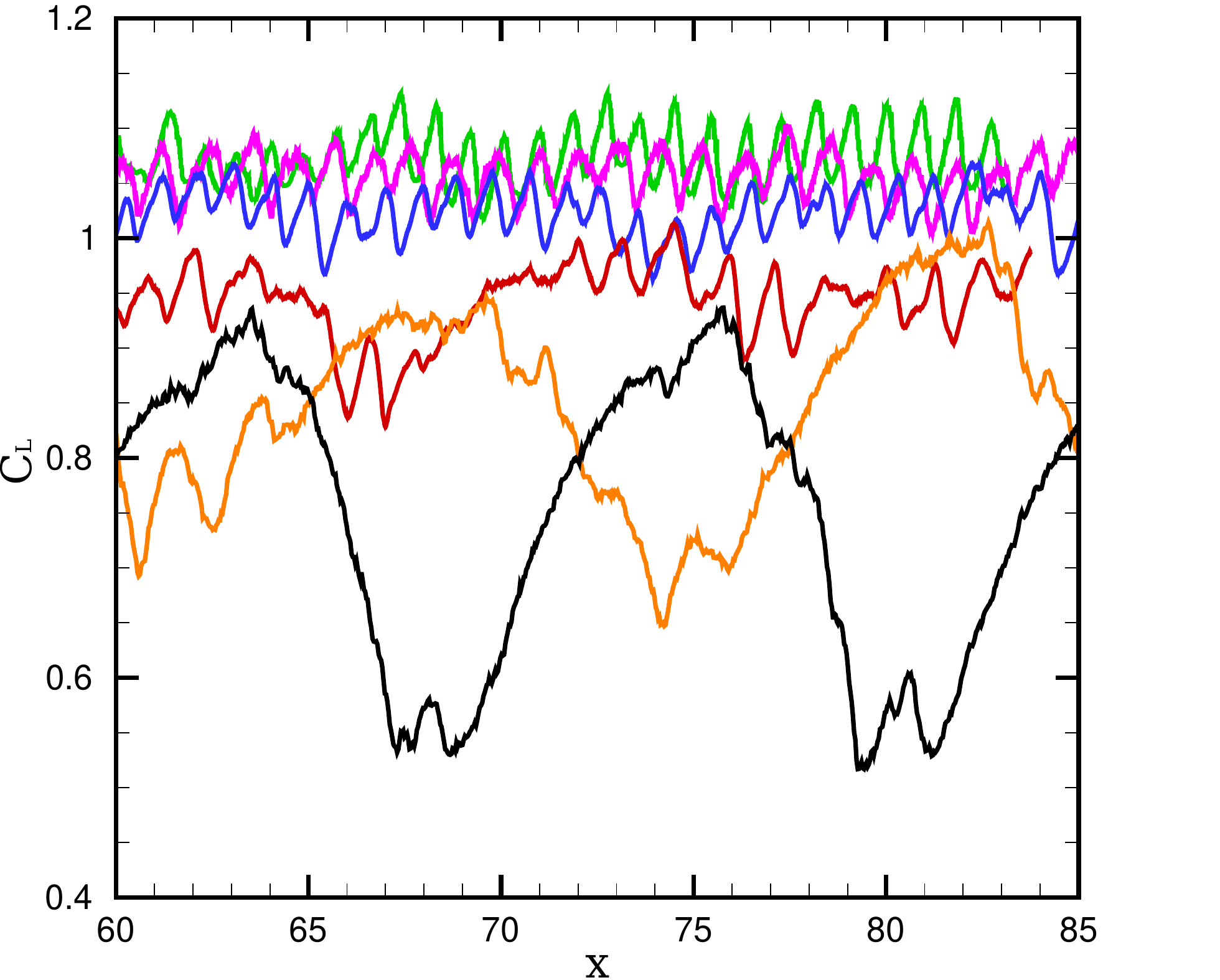}} 
\end{tabular}
  \caption{
  (a) Mean wall-pressure coefficient $\overline{C_p}$ (solid curves) and root-mean-square of fluctuation $C'_{p,rms}$ (dashed curves) as a function of $x$ and
  (b) time histories of $C_L$ are shown for a constant free-stream Mach number $M=0.735$ and various Reynolds numbers, $Re=500,\!000$ (black), $Re=1,\!000,\!000$ (orange), $Re=2,\!000,\!000$ (red) and $Re=3,\!000,\!000$ (blue) using the standard grid. The magenta curves correspond to a simulation at $Re=3,\!000,\!000$ using a refined grid. For comparison, $C_p$ and the $C_L$ history are extracted from \cite{Dandois2018} and denoted by green symbols and curves, respectively.
  }
\label{fig:CL_Reeffect}
\end{figure}

Figure \ref{fig:CL_Reeffect} shows (a) distributions of mean wall-pressure coefficient, $\overline{C_p}$, the root-mean-square of fluctuating component, $C'_{p,rms}$, and (b) histories of lift coefficient $C_L$ for various Reynolds numbers. For now, we only focus on the blue, magenta, and green curves corresponding to $Re=3,\!000,\!000$. The blue and magenta curves denote the present simulation results using standard (90 million grid points) and refined (200 million grid points) grids respectively, while the green curve corresponds to LES results from \cite{Dandois2018} using ONERA's second-order accurate CFD code elsA (400 million grid points, but for a larger domain, see section~\ref{subSec:grid}). Overall, the $\overline{C_p}$ distributions agree well, despite small differences in $\overline{C_p}$ levels over the fore part and shock position. The former may be due to the influence of boundary-layer tripping applied on the pressure side for the case of \cite{Dandois2018}, while the mean shock position appears to be relatively sensitive to the grid resolution. However, the dynamic behaviour of the flow is hardly affected, which is indicated by the good agreement between dashed lines corresponding to $C'_{p,rms}$ as a function of $x$. While the $C'_{p,rms}$ peaks arise from shock-wave oscillations centered around $x\approx0.63$, the plateau further upstream ($0.4<x<0.6$) results from pressure fluctuation within the laminar separation bubble, as will be discussed later in more detail. Also lift fluctuations agree well in figure \ref{fig:CL_Reeffect}(b), where the same colour code is used for different cases. All three cases exhibit clear intermediate-frequency fluctuations (period approximately equal to one time unit), which are slightly modulated by low-frequency undulations (period approximately equal to 12 time units). 

\begin{figure}
\vspace{0.25cm}
\begin{tabular}[t]{@{}l@{}l}
a) & b) \\
\imagetop{\includegraphics[width=0.5\columnwidth]{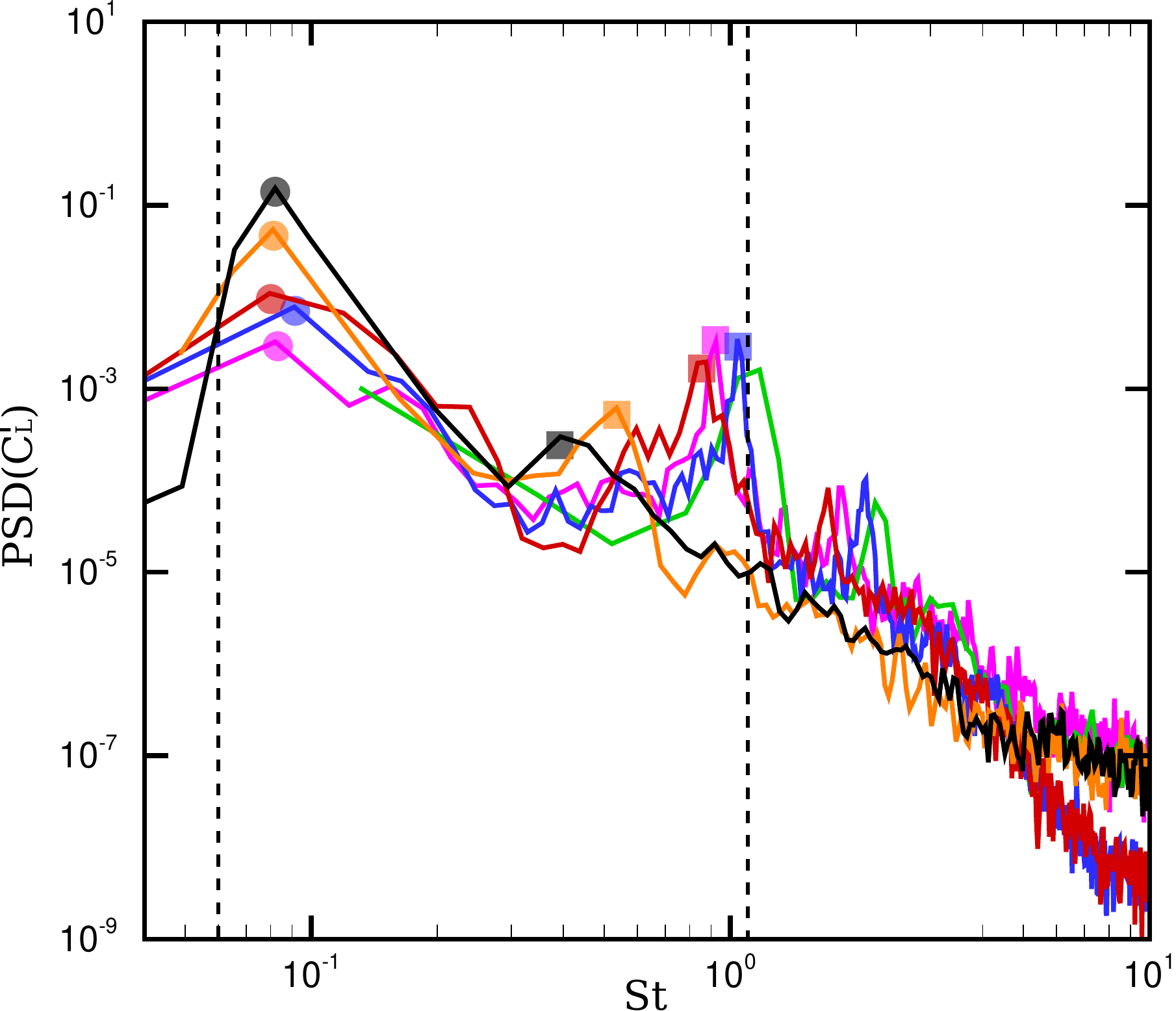}} &  \imagetop{\includegraphics[width=0.5\columnwidth]{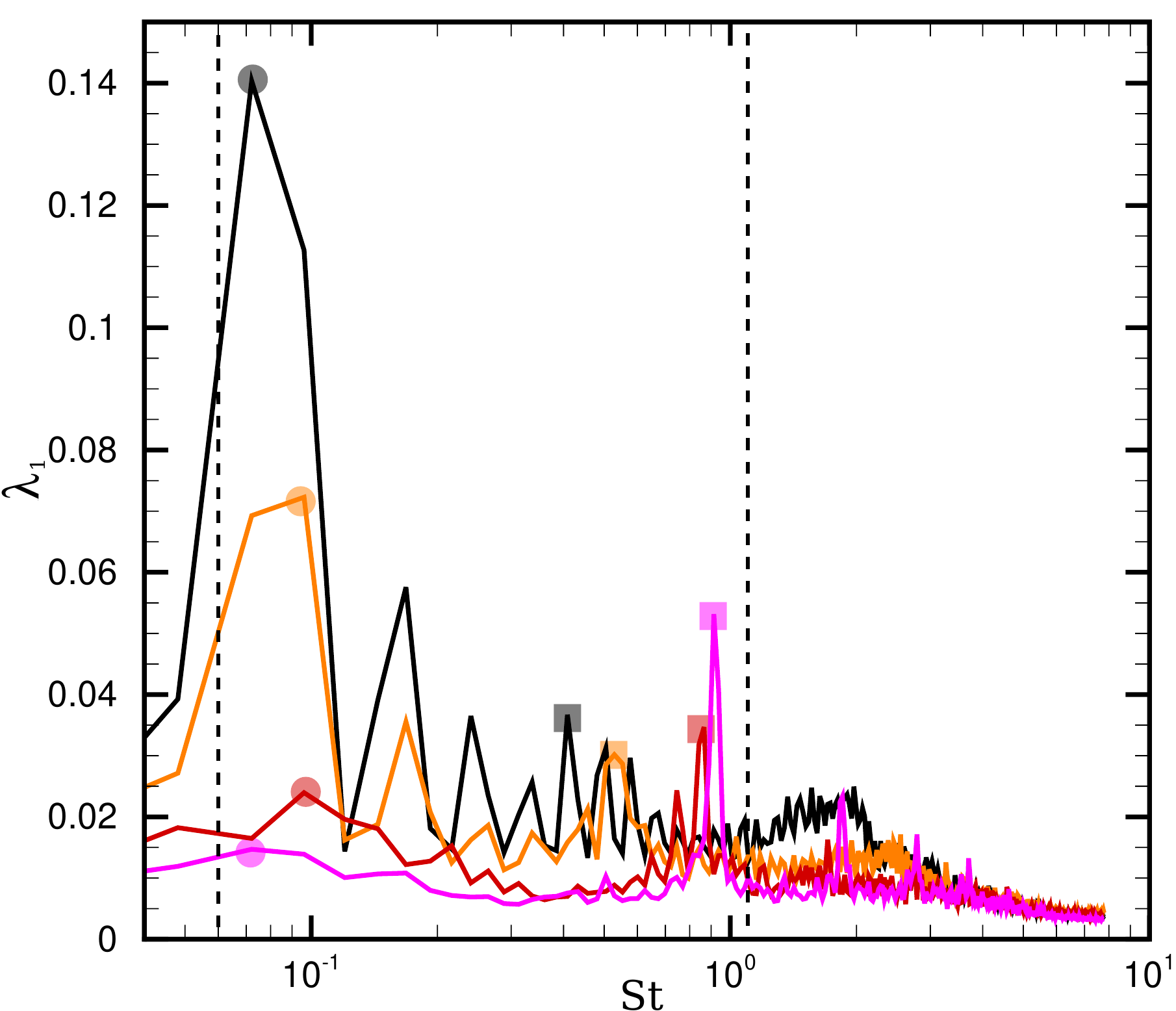}}
\end{tabular}
  \caption{
  (a) Composed Fourier spectra of $C_L$ fluctuations are compared with dominant spectral peaks of experimental results by \cite{Brion2019} indicated by vertical broken lines and simulation data by \cite{Dandois2018}.
  (b) Leading eigenvalues $\lambda_1$ of spectral proper orthogonal decomposition of 2D snapshots are shown for present simulations. Colours correspond to figure \ref{fig:CL_Reeffect}.
  }
\label{fig:Spectra_Reeffect}
\end{figure}
To better assess the frequency content of these lift fluctuations, figure \ref{fig:Spectra_Reeffect}(a) shows Fourier spectra, with spectral peaks observed in experiments of \cite{Brion2019} indicated by vertical black broken lines. Again, we observe a good agreement between blue, magenta and green curves showing intermediate-frequency peaks in spectra in the range of $0.9<St<1.1$ (highlighted by square symbols) with similar power-spectral densities and associated harmonics at $St \approx 2$. The present simulations show a weak peak at low frequencies $St \approx 0.08$ (circles), which is below the cut-off frequency of the spectrum associated with the work of \cite{Dandois2018}. This low-frequency bump is slightly above the Strouhal numbers, $St \approx 0.06$, observed in wind-tunnel tests \cite{Brion2019}. Boundary-layer tripping on the pressure side is not expected to impact the frequency significantly \citep{Moise2023}. Even though \cite{Zauner2020} showed minor sensitivity of low frequencies to the spanwise domain extent, we cannot rule this effect out. The difference may instead be due to wind-tunnel confinement effects such as the presence of side and top walls, which are not captured in the current simulations.  Nevertheless, considering the experimental and numerical challenges when comparing wind-tunnel test and simulation results, the agreement with respect to the low-frequency mode is good.

Although the blue curve in \ref{fig:Spectra_Reeffect}(a) shows some quantitative differences compared to cases using finer grids (the magenta and green curves), we still capture the same physical flow phenomena with very similar scales and amplitudes. This gives us confidence that the standard grid is well suited for simulations carried out at lower Reynolds numbers. Furthermore, we have some justification for the use of narrow domains and periodic boundary conditions in recovering the main flow characteristics observed in wind-tunnel tests.

\paragraph{Reynolds-number effects}

After having verified and validated our observations at high Reynolds numbers of $Re=3,\!000,\!000$ and $M=0.735$, we consider a set of simulations at the same Mach number, but decreasing Reynolds numbers down to $Re=500,\!000$ to study $Re$ scaling effects. All these simulations use the standard grid.

Looking again at figures \ref{fig:CL_Reeffect} and \ref{fig:Spectra_Reeffect}, simulations at $Re=2,\!000,\!000$, $1,\!000,\!000$, and $500,\!000$ are denoted by red, orange, and black curves, respectively.
Looking at figure \ref{fig:CL_Reeffect}(b), we can see that low-frequency oscillations strengthen with decreasing Reynolds numbers, indicating established buffet at low $Re$ and a trend towards buffet offset at high $Re$ for a fixed Mach number and angle of attack. In previous studies of the V2C profile, a similar trend was observed: while the flow was fully stalled at $Re=200,\!000$, buffet was fully developed at $Re=500,\!000$ before it weakened again towards higher $Re$ \citep{Zauner2018a,Moise2022} and eventually dies out \citep{Szubert2016a}.
An opposite trend is observed for intermediate-frequency oscillations, as they become less pronounced during high-lift phases when decreasing $Re$. At $Re=500,\!000$, we eventually observe large-amplitude lift oscillations characteristic of transonic buffet. Averaging of dynamic effects associated with shock-wave motion leads to a rather smooth $\overline{C_p}$ distribution on the suction side shown in figure \ref{fig:CL_Reeffect}(a), where the pressure change associated with the main shock wave is smeared out. Consequently, the $C'_{p,rms}$ peak turns into a broad bump and increased pressure fluctuations are even observed in the fore part of the airfoil. On the pressure side we observe a plateau in $\overline{C_p}$ at $x\approx 0.7$ corresponding to a laminar separation bubble. Looking at spectra shown in figure \ref{fig:Spectra_Reeffect}(a), for decreasing Reynolds numbers we can observe a clear low-frequency peak evolving at a rather constant Strouhal number of $St \approx 0.08$. The intermediate-frequency peak, however, decreases in magnitude and shifts towards lower Strouhal numbers, reaching $St\approx0.4$ at $Re=500,\!000$. Oscillations at similar intermediate frequencies have been also observed for Dassault Aviation's V2C profile at the same Reynolds number, but were less pronounced \cite{Zauner2019c}. 

After having assessed the effect of low- and intermediate-frequency phenomena on aerodynamic wing characteristics, we would now like to study the spatial structure of associated phenomena. Figure \ref{fig:Spectra_Reeffect}(b) shows leading eigenvalues $\lambda_1$ of SPOD of two-dimensional snapshots at different Reynolds numbers. These spectra look qualitatively very similar to those in figure \ref{fig:Spectra_Reeffect}(a). The SPOD eigenvalue spectra are less smooth compared to the Fourier spectra of $C_L$, due to the fixed number of bins used in the SPOD, but the trends in the low and intermediate frequency modes are the same. For the lower $Re$ case, harmonics of the buffet mode coexist in the intermediate-frequency range, which makes the separation of both phenomena very difficult using SPOD. In contrast to the lift spectra, SPOD contains more information about the wake behind the airfoil and we can observe an additional broadband bump at $St \approx 2$, where distinct modes appear at higher Reynolds numbers very close to harmonics of intermediate-frequency phenomena.
These high-frequency phenomena are associated with a von-Karman vortex street and have also been observed and discussed by \cite{Moise2022} for the V2C airfoil.


\begin{figure}
\vspace{0.25cm}
\begin{tabular}[t]{@{}l@{}l}
\imagetop{\includegraphics[width=1.0\columnwidth]{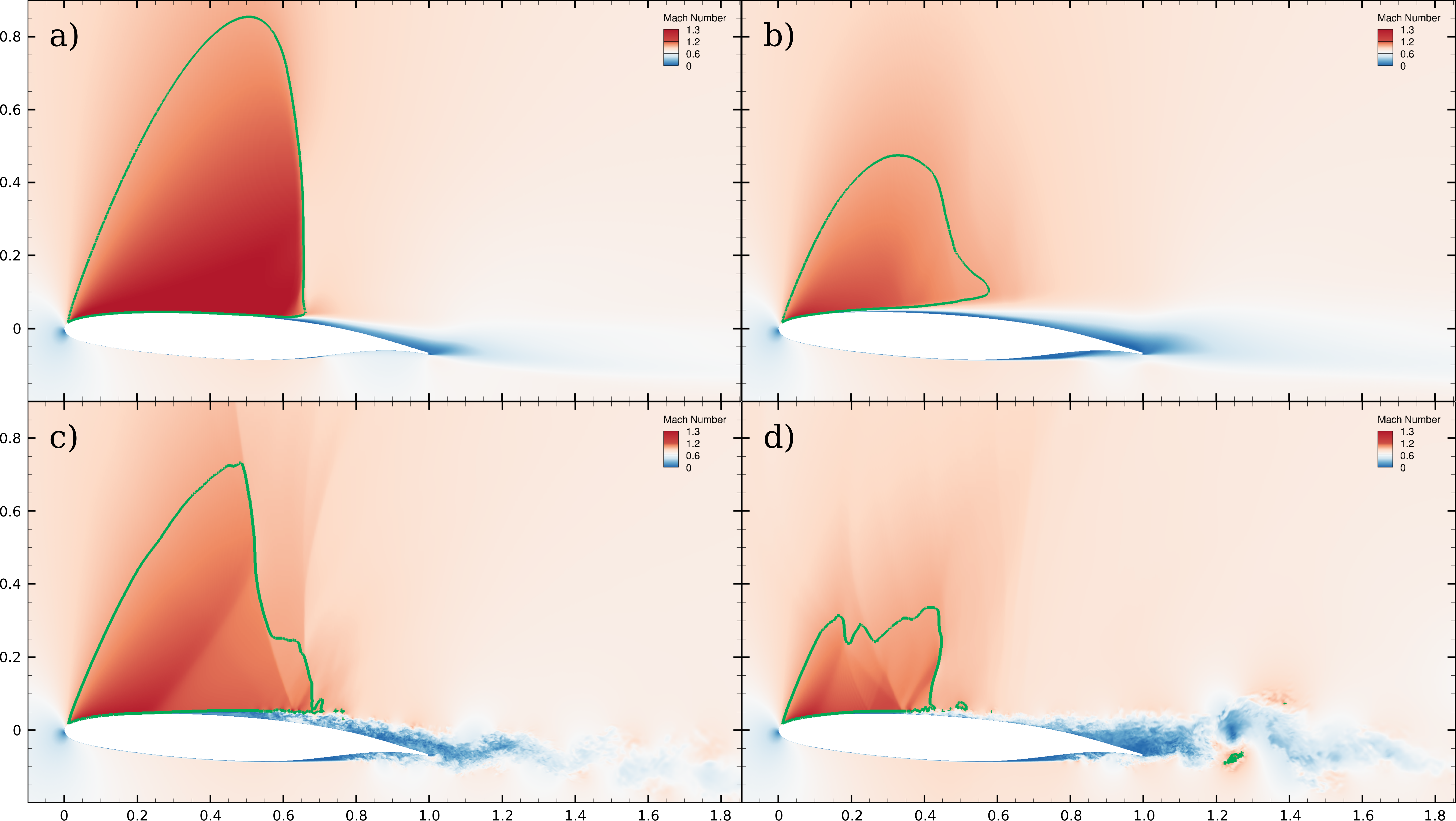}} 
\end{tabular}
  \caption{
  Local mean Mach-number contours shown for (a) $Re=3,\!000,\!000$ and (b) $Re=500,\!000$ at $M=0.735$.
  Instantaneous Mach-number contours shown for $Re=500,\!000$ at $M=0.735$, representative for (c) high-lift and (d) low-lift phases at $t=64.5$ and $69$, respectively.
  }
\label{fig:M_Reeffect}
\end{figure}
Before discussing the shape of dominant coherent structures extracted by SPOD, we first consider the contours of local time- and span-averaged Mach number in figure \ref{fig:M_Reeffect} for representative Reynolds numbers of (a) $Re=3,\!000,\!000$ and (b) $Re=500,\!000$ and instantaneous fields in (c) and (d) at high and low lift conditions respectively for the lower $Re$ case. Based on previous work on V2C wing sections with different domain sizes at similar conditions \citep{Zauner2020,Moise2022} that showed minor variations of coherent features in the spanwise direction, it is sufficient to focus the present work on two-dimensional cross sections (instantaneous at $z=0$ as well as span-averaged), which are obtained by 3D simulations. For the high $Re$ case, we observe a large supersonic region, delineated by the green sonic curve with a distinct shock wave at $x \approx 0.6$. At the shock foot, we can observe a small bump extending the supersonic region slightly in the downstream direction. For the $Re=500,\!000$ case, this bump is more pronounced and extends further away from the surface, which is an indicator for significant flow separation (this can be confirmed, looking at $C_f$ distribution in figure \ref{fig:Cf} in the appendix). Compared to the high $Re$ case, the time-averaged supersonic region appears smaller and smoother due to large-scale unsteadiness. The global flow field at higher $Re$ is not subjected to such significant variations (see $C_L$ histories) and therefore no instantaneous snapshots are shown for brevity. The instantaneous snapshot of the $Re=500,\!000$ case at the high-lift phase figure \ref{fig:M_Reeffect}(c) demonstrates the high-frequency small-scale structures. Behind the shock foot the boundary layer thickens and undergoes transition to turbulence. Then, in the wake region behind the airfoil we see a clear von-Karman vortex street forming. During low lift phases, we observe in figure \ref{fig:M_Reeffect}(d) flow features that are reminiscent of stall.

\begin{figure} 
\centering
\includegraphics[trim={0cm 2cm 0.5cm 2cm},clip,width=.45\textwidth]{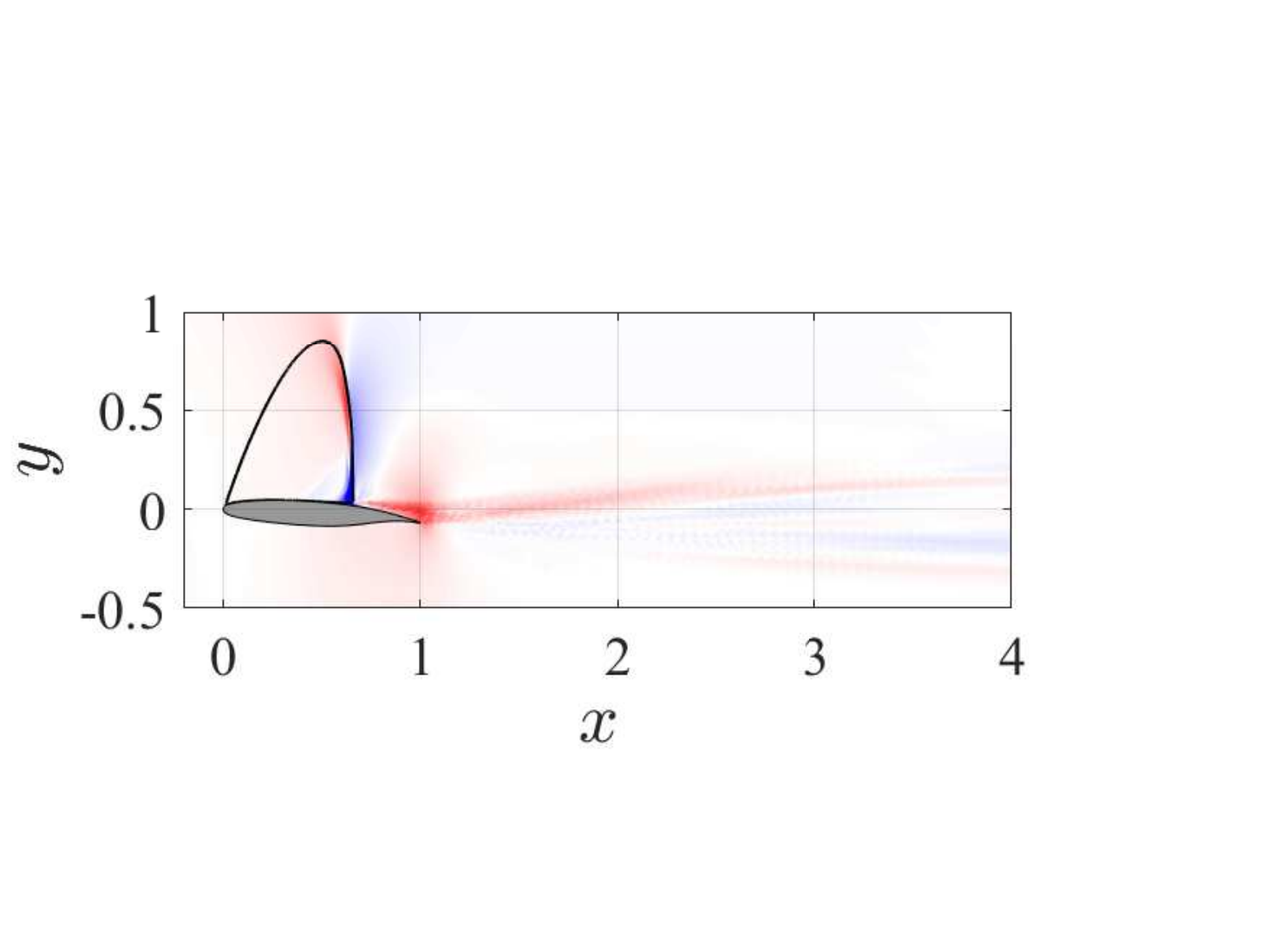}
\includegraphics[trim={0cm 2cm 0.5cm 2cm},clip,width=.45\textwidth]{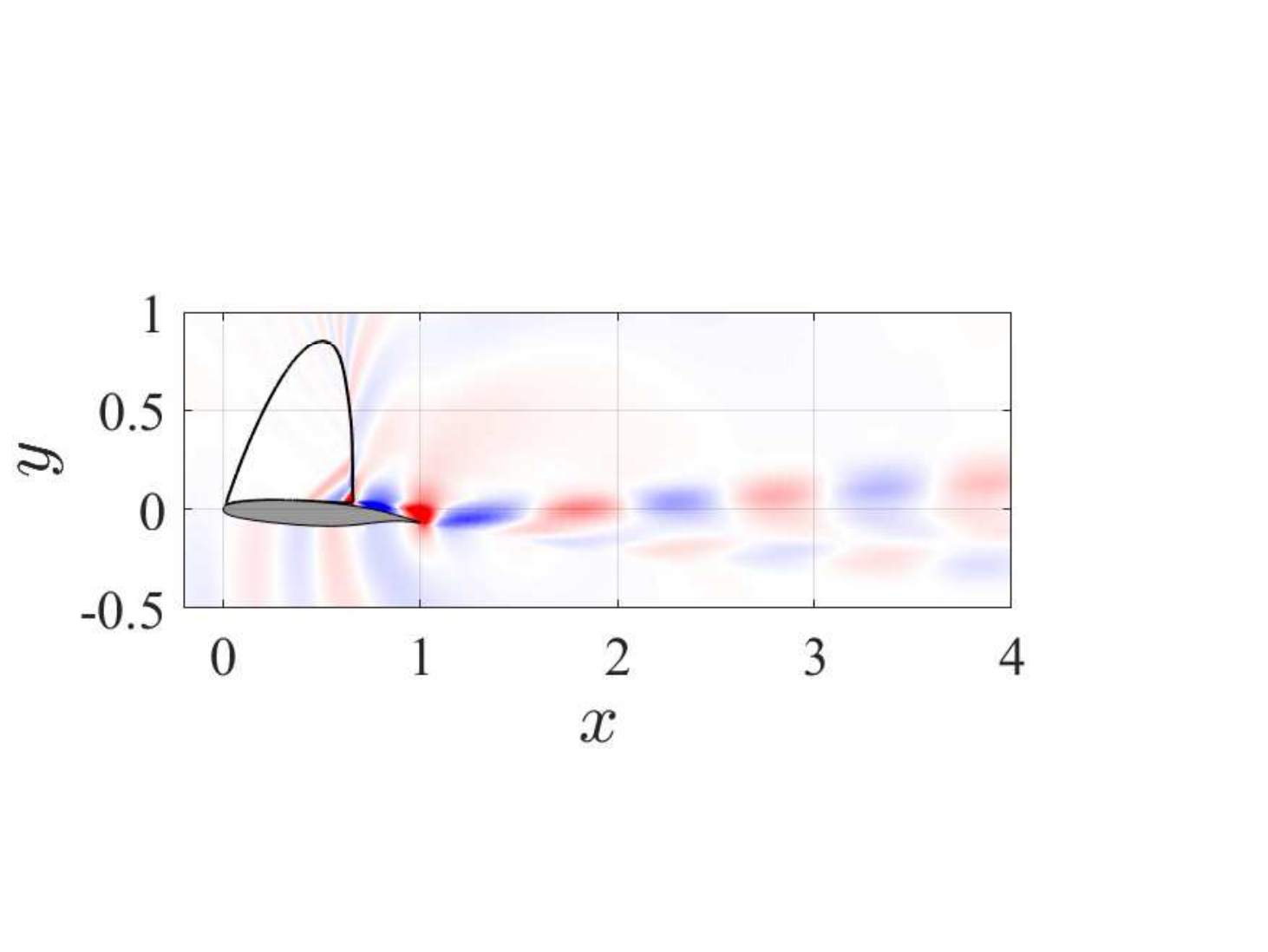}
\includegraphics[trim={0cm 2cm 0.5cm 2cm},clip,width=.45\textwidth]{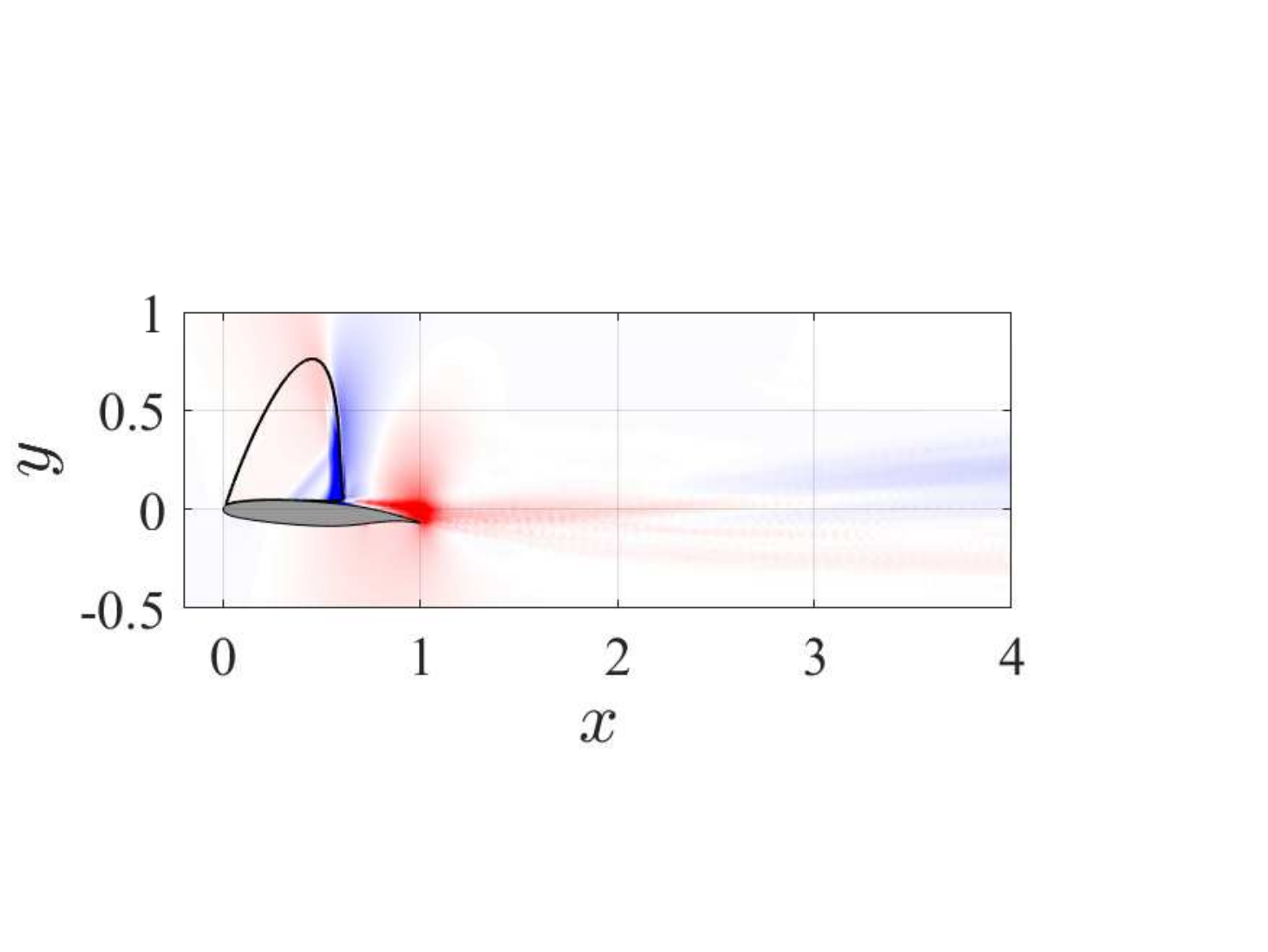}
\includegraphics[trim={0cm 2cm 0.5cm 2cm},clip,width=.45\textwidth]{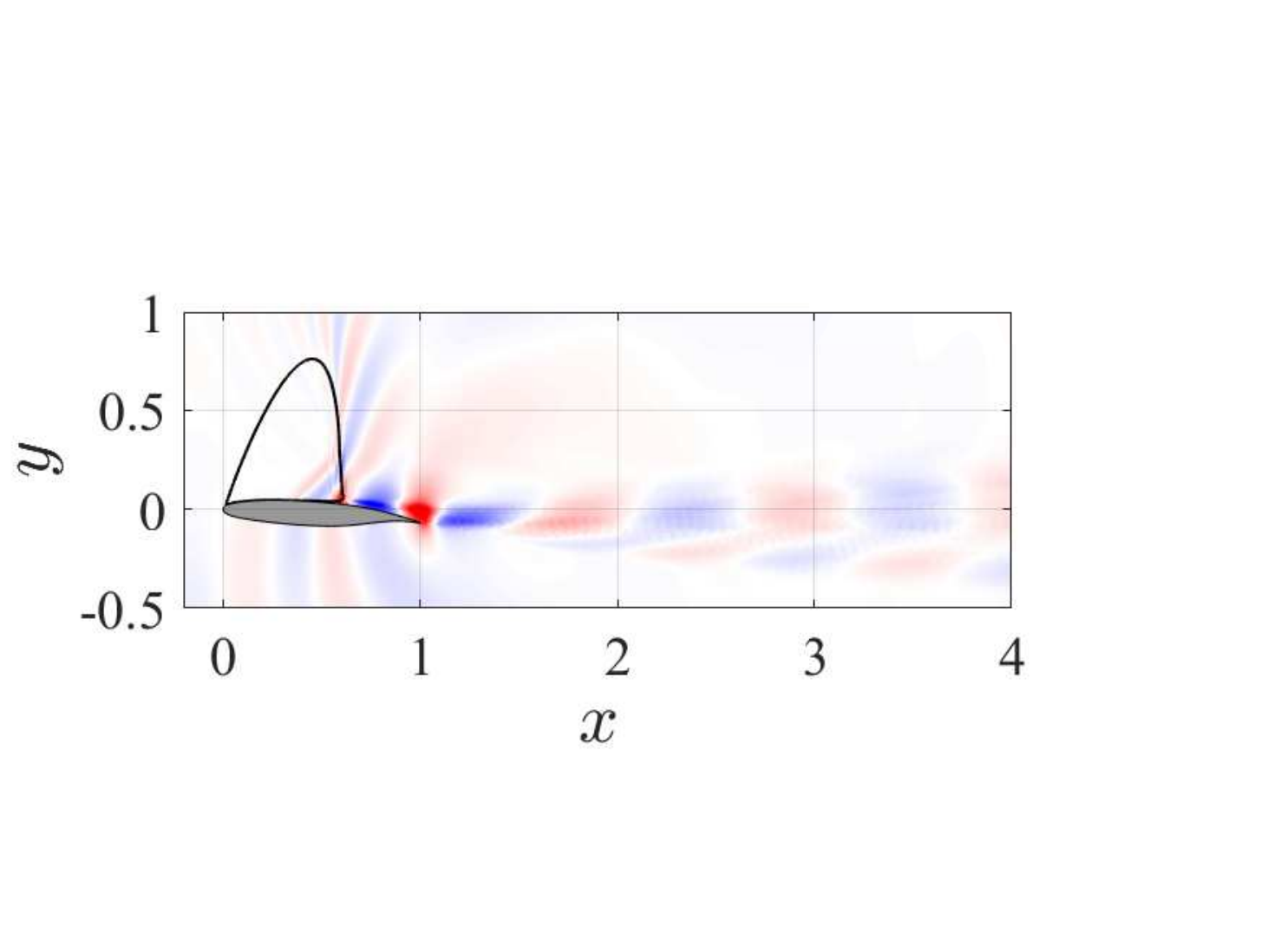}
\includegraphics[trim={0cm 2cm 0.5cm 2cm},clip,width=.45\textwidth]{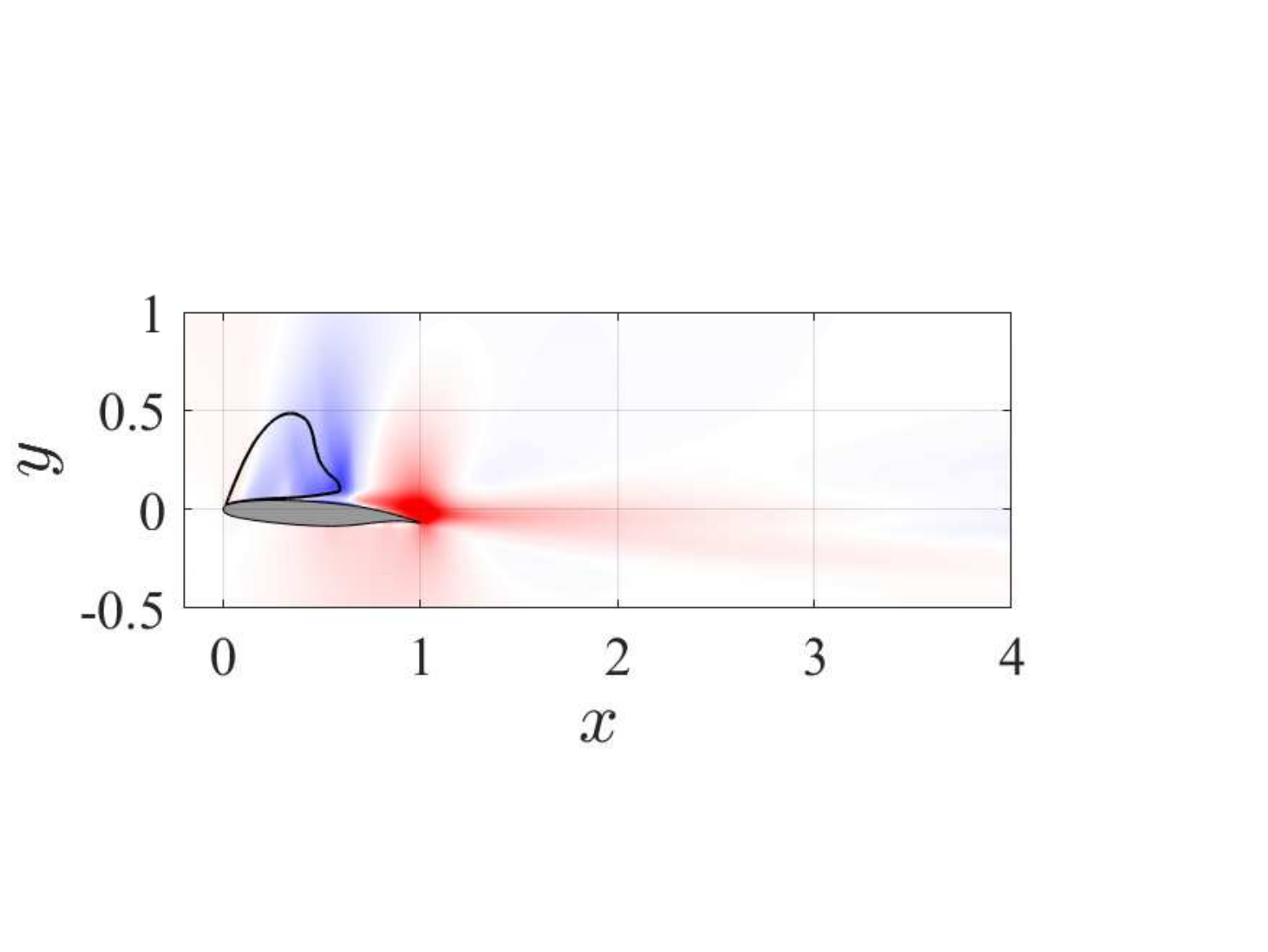}
\includegraphics[trim={0cm 2cm 0.5cm 2cm},clip,width=.45\textwidth]{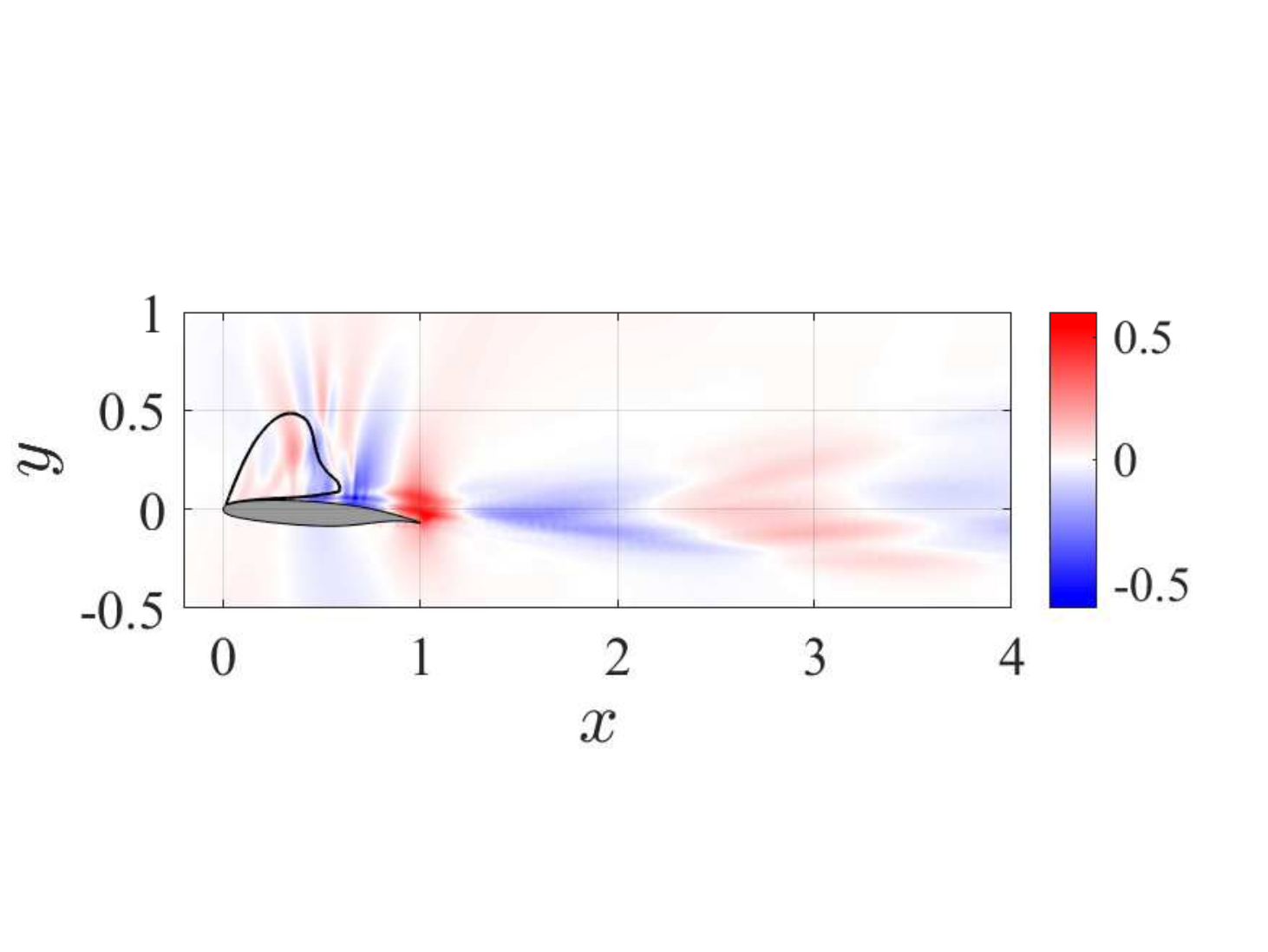}
\caption{Density contours of SPOD modes associated with peaks at low (left) and intermediate (right) frequencies, phased at the trailing edge, are shown for $Re=3,\!000,\!000$ (top), $Re=2,\!000,\!000$ (middle), and $Re=500,\!000$ (bottom).}
    \label{fig:SPODModesReeffect}
\end{figure}
Looking now at SPOD modes showing density-fluctuation contours in figure \ref{fig:SPODModesReeffect}, which correspond to dominant flow oscillations around the mean flow, we can see low- (left column) and intermediate-frequency modes (right column) for representative Reynolds numbers of $Re=3,\!000,\!000$ (top), $Re=2,\!000,\!000$ (middle), and $Re=500,\!000$ (bottom). 
For low-frequency phenomena at high Reynolds numbers, we can confirm rather localised oscillations of the shock wave accompanied by fluctuations of the trailing-edge region, which are $180^{\circ}$ out of phase. Structures in the wake correspond to periodic up- and downward deflection (`flapping'). When reducing the Reynolds number to $Re=2,\!000,\!000$, we observe a $\lambda$-structure occurring within the supersonic region with the leading leg associated with boundary-layer separation. Although less pronounced, this lambda structure also exists at $Re=3,\!000,\!000$. 
The mode shape for the $Re=3,\!000,\!000$ case has a striking resemblance to the globally unstable mode reported in \cite{Sartor2015} (see their figure~12 (a)) for transonic buffet on an OAT15A aerofoil under fully-turbulent conditions. This serves as the justification for referring to the low-frequency modes observed here for free-transition conditions as transonic buffet modes. 

The blue regions around shock and separation waves become thicker for decreasing $Re$, and eventually spread over the entire supersonic region for $Re=500,\!000$.
The spatial organisation of SPOD modes at intermediate frequencies are very similar for the high-$Re$ test cases. For $Re=500,\!000$, however, we observe a larger wavelength since the Strouhal numbers have decreased (recalling the spectra in figure \ref{fig:Spectra_Reeffect}). At higher $Re$, modal features within the supersonic region are limited to oscillations around the separation wave, whereas for $Re = 500,\!000$, the mode structure extends beyond the supersonic region. 

Despite some quantitative variations in wave lengths, Strouhal numbers and extents of regions covered by the mode shape, we observe similar mode structures at various $Re$. Therefore, by studying the moderate $Re$ case in more detail, we expect to be able to project our conclusions at least to $Re$ typical of wind-tunnel experiments.




\section{Mach number at moderate Reynolds number $Re=500,\!000$}
\label{sec:Results}

After having validated our results against reference literature and having confirmed the existence of multiple flow phenomena of interest at moderate as well as high Reynolds numbers, we will now study their Mach-number sensitivity for a fixed $Re=500,\!000$.

\begin{figure}[hbt!]
\centering
\includegraphics[width=1.0\textwidth]{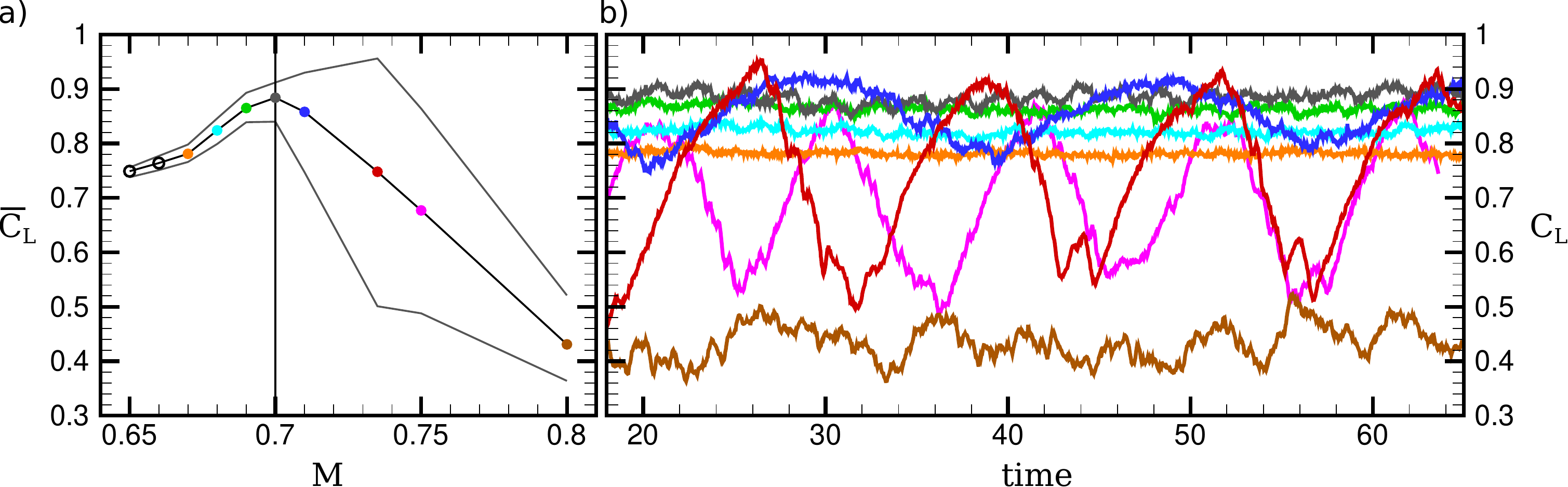}
\caption{(a) The black curve shows time-averaged lift coefficient $\overline{C_L}$ as a function of Mach number, where coloured circles denote simulation data points. Curves below and above show respectively minimum and maximum lift coefficient. Correspondingly coloured histories of $C_L$ are shown in (b). The black vertical line indicates the simulation Mach number, where maximum mean lift is observed and where large-scale oscillations start increasing dramatically.}
    \label{fig:Overview1}
\end{figure}
Figure \ref{fig:Overview1}(a) shows time-averaged lift coefficients as a function of freestream Mach number $M$, denoted by the black curve. Grey curves above and below indicate respectively maximum and minimum instantaneous lift coefficients. Following the colour-code of the Mach number points shown on figure \ref{fig:Overview1}(a), representative $C_L$ histories are shown in figure \ref{fig:Overview1}(b). 
We can distinguish between two main regimes for present test cases, here denoted as \textit{pre-buffet} ($0.67<M<0.7$) and \textit{established buffet} ($M>0.7$). As civil aviation industry is mainly interested in characterising and predicting buffet onset in order to avoid critical flight conditions, we focus our current work on \textit{incipient buffet} observed at $M=0.7$, where the mean lift peaks and low-frequency phenomena set in.

In the pre-buffet regime, the mean lift increases gradually due to compressibility effects (\textit{e.g.}, thinner boundary layers). Initial broad-band fluctuations strengthen and become more regular with increasing $M$, while minimum as well as maximum $C_L$ values increase. Approaching $M=0.70$, fluctuation amplitudes reach $8\%$ of $\overline{C_L}$, while the mean lift remains approximately centered between the extreme values. Looking at figure \ref{fig:Overview1}(b), we can observe slight undulations of the grey curve with a period of $\tau>20$ time units corresponding to Strouhal numbers $St<0.05$, compared to the more pronounced oscillations in the intermediate-frequency range corresponding to $St\approx0.4$ ($\tau \approx 2.5$).
Moving into the developed buffet regime, low-frequency oscillations strengthen rapidly with increasing Mach numbers, progressively dominating the global flow dynamics. While lift maxima further increase with $M$, even though at a much lower rate, lift minima drop dramatically leading to an overall decrease of mean lift. At $M=0.735$, oscillation amplitudes reach more than $68\%$ of $\overline{C_L}$. Further increasing $M$ eventually leads to linearly decreasing maximum $C_L$ values, while also the buffet oscillation amplitude decays. In this study, the maximum Mach number of $M=0.8$ is close to buffet offset and not further increased.

\begin{figure}
\vspace{0.25cm}
\begin{tabular}[t]{@{}l@{}l}
(a) & (b) \\
\imagetop{\includegraphics[width=0.5\columnwidth]{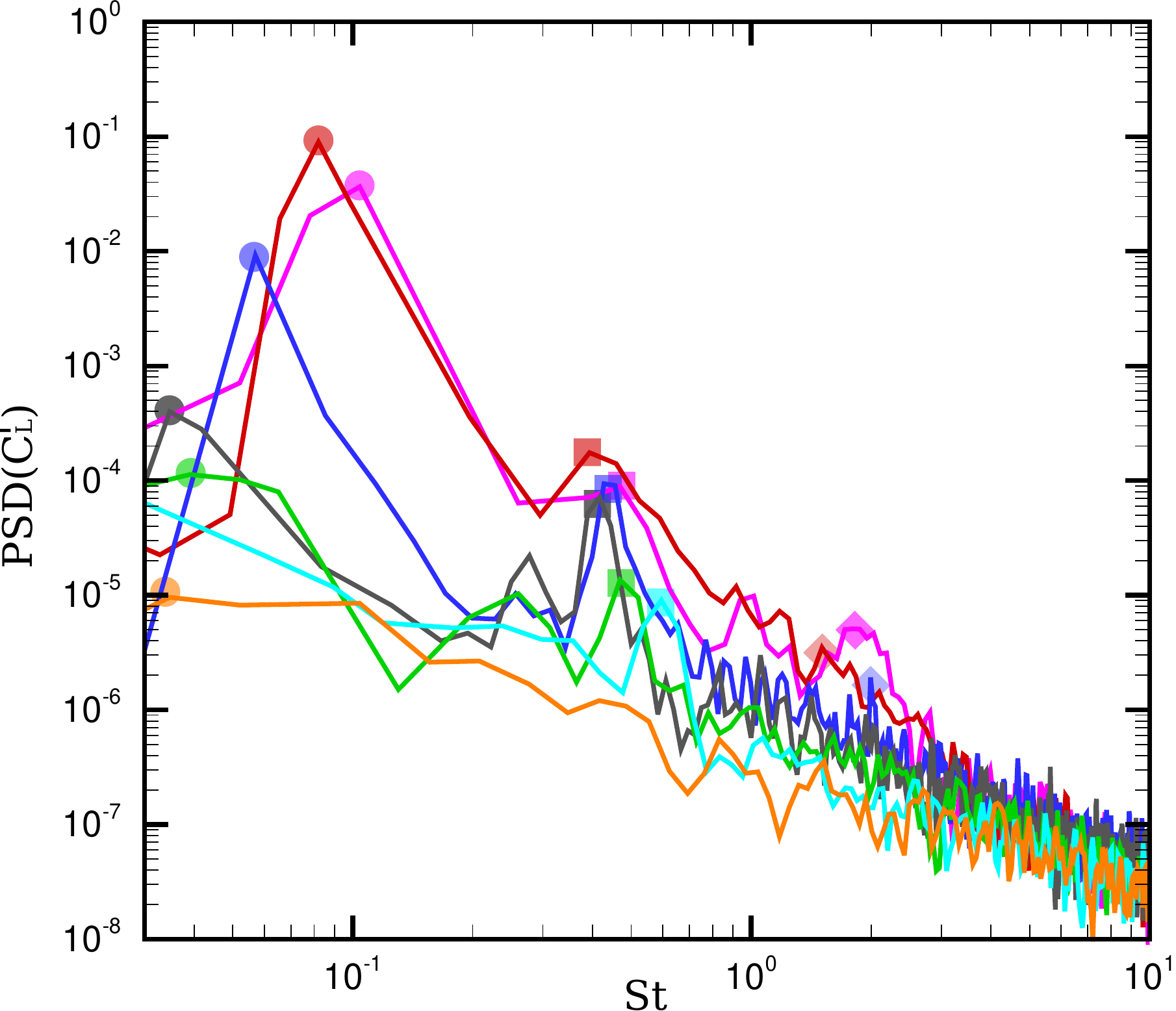}} &  \imagetop{\includegraphics[width=0.5\columnwidth]{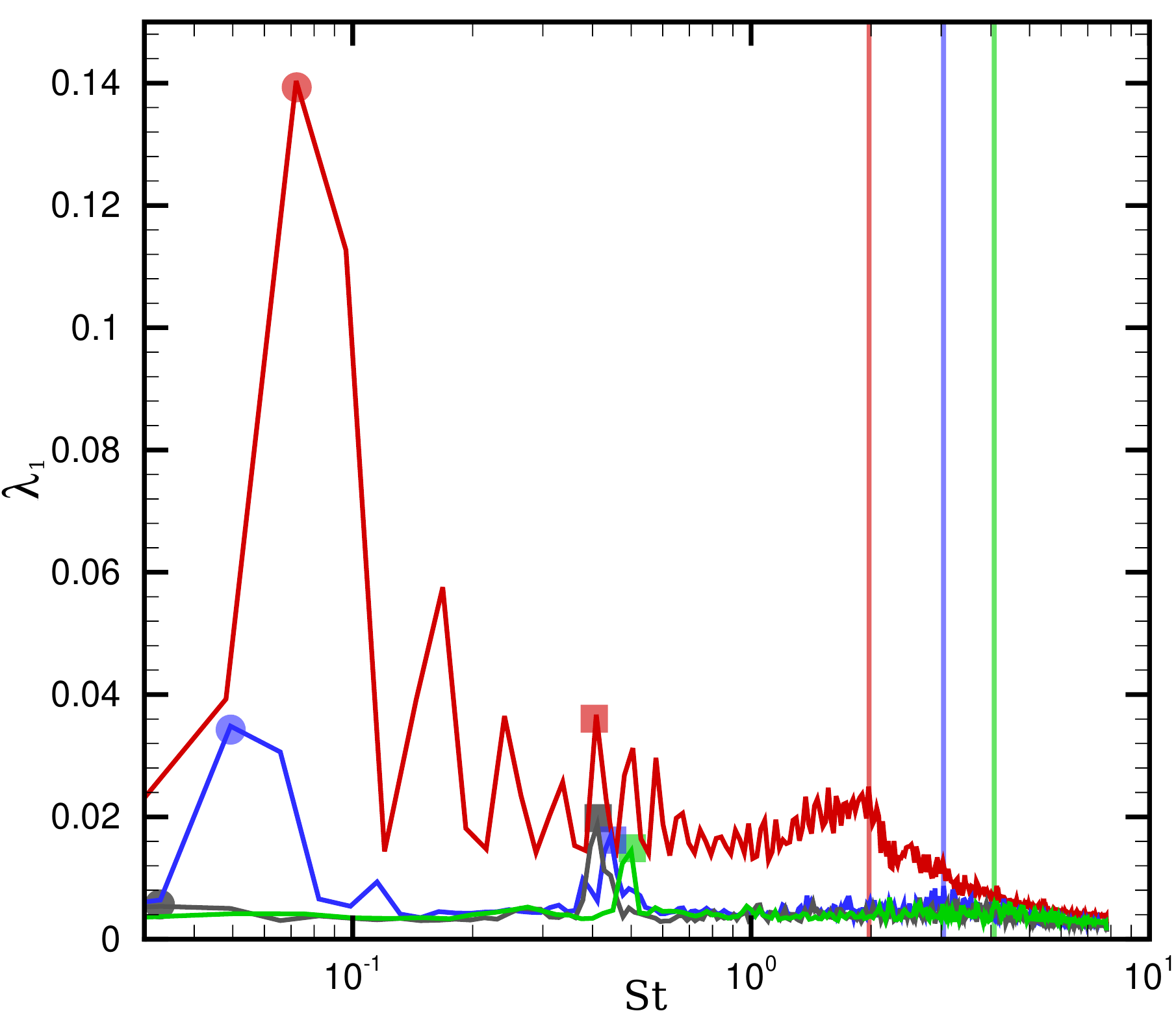}}
\end{tabular}
  \caption{(a) Composed Welch spectra of $C_L$ fluctuations and (b) leading eigenvalues from SPOD of 2D snapshots for $M=0.67$ (orange), $M=0.68$ (light blue), $M=0.69$ (green), $M=0.70$ (grey), $M=0.71$ (blue), $M=0.735$ (red), and $M=0.75$ (magenta).
  }
\label{fig:Spectra}
\end{figure}
Figure \ref{fig:Spectra}(a) shows Fourier spectra of the $C_L$ histories that were shown in figure \ref{fig:Overview1}(b). Even for low Mach numbers ($M<0.7$), we observe an increased low-frequency content in the spectra, but no distinct peaks. In this \textit{pre-buffet} regime, we observe mainly the development of a clear spectral peak at intermediate frequencies around $St\approx0.4$, similar to those reported in the previous section. For $M<0.7$, this intermediate-frequency peak slightly decreases with increasing Mach numbers. At $M=0.7$, where a distinct peak arises at low frequencies $St\approx0.04$, the sharp peak at intermediate frequencies does not significantly change but we observe the presence of an additional peak at $St\approx0.3$, which is associated with the sawtooth-like shape of the $C_L$ history at $M=0.7$. With further increases in the Mach number we observe an increase of Strouhal numbers as well as power-spectral densities associated with the low-frequency phenomenon and it becomes increasingly difficult to separate out the intermediate-frequency. For $M>0.735$, the amplitudes of low-frequency oscillations start decreasing.

We now want to analyse the spatial structures associated with these unsteady flow phenomena. Figure \ref{fig:Spectra}(b) shows corresponding spectra of the leading eigenvalues as a function of Strouhal number obtained by SPOD for representative Mach numbers between $M=0.69$ and $M=0.735$. These spectra look qualitatively similar to those in figure \ref{fig:Spectra}(a). For $M=0.735$, harmonics of the buffet mode coexist in the intermediate-frequency range, which again makes the separation of both phenomena very difficult using SPOD. 
It is interesting to see that during the incipient buffet phase at $M=0.7$, the low-frequency content in global flow oscillations appears to be significantly lower compared to intermediate-frequency fluctuations. As the low-frequency content is much more pronounced in $C_L$ spectra, we can assume that the low-frequency phenomenon has its origin near the airfoil surface.

For $M=0.70$ and $M=0.71$, we have two test cases where the low- and intermediate frequencies are well separated in frequency and the intermediate-frequency oscillations are very similar despite the rapid onset of low-frequency buffet. This is strong evidence that the phenomena arise independently. This would be more difficult to prove at higher $Re$. At $Re=3,\!000,\!000$, it was shown in experimental investigations by \cite{Brion2019} that the amplitude of low-frequency oscillations appears much less sensitive to $M$. Even at $M=0.75$, power-spectral densities of low-frequency oscillations remained more than two orders of magnitude lower compared to intermediate-frequency fluctuations.


\begin{figure}
\vspace{0.25cm}
\begin{tabular}[t]{@{}l@{}l}
a) & b) \\
\imagetop{\includegraphics[width=0.5\columnwidth]{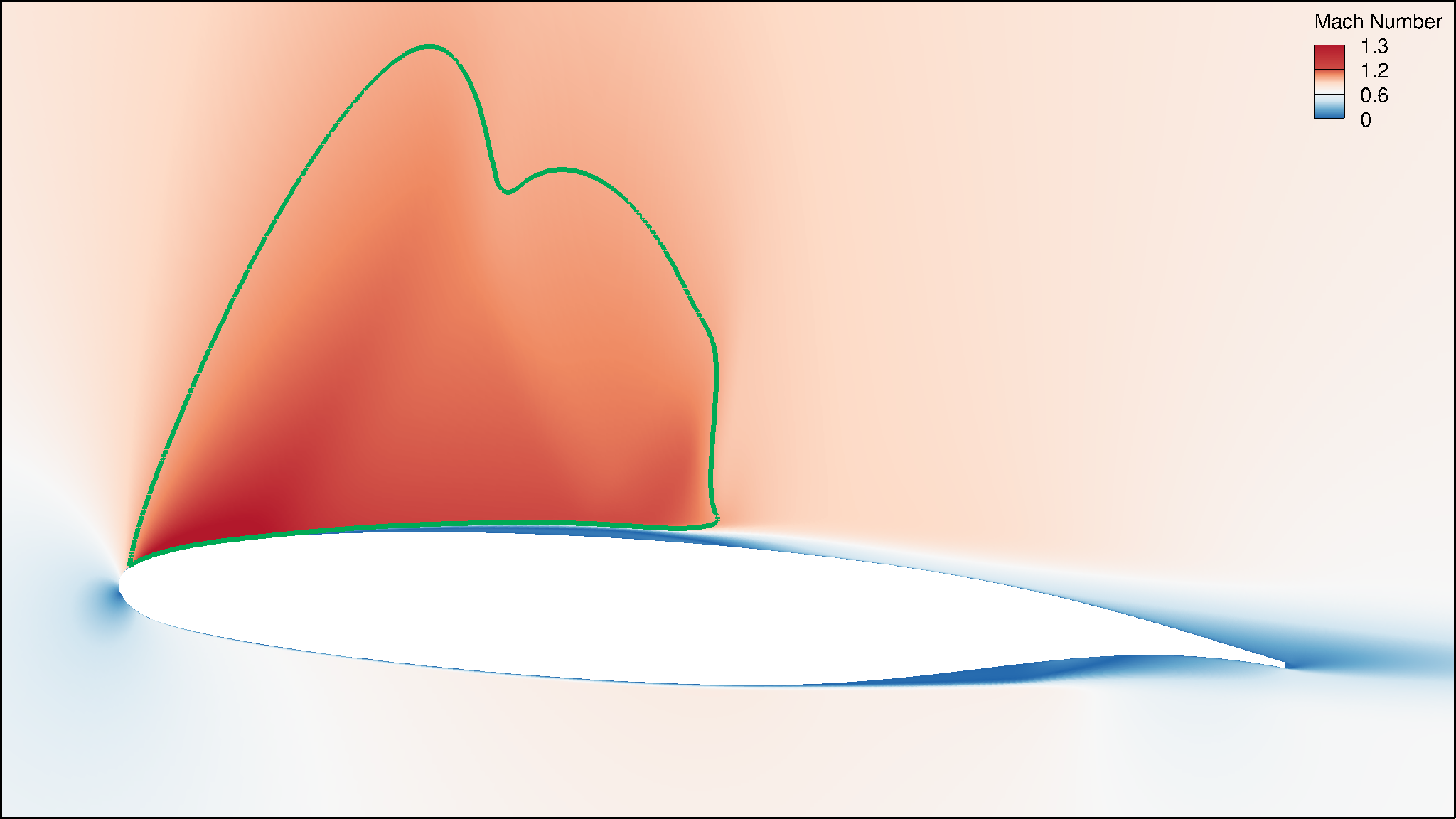}} &  \imagetop{\includegraphics[width=0.5\columnwidth]{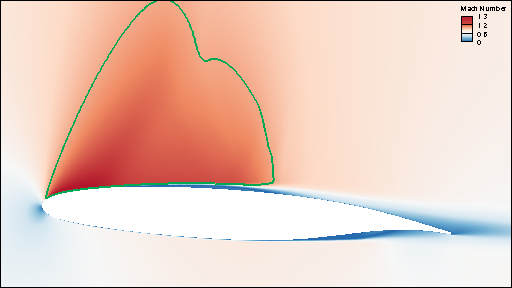}} 
\end{tabular}
  \caption{
  Mach number contours shown for (a) $M=0.70$ and (b) $M=0.71$ at $Re=500,\!000$. The green curves denote sonic lines of $\overline{M}_\mathrm{loc}=1$.  }
\label{fig:M_Meffect}
\end{figure}
Before discussing the modal shapes associated with dominant flow structures at multiple scales for our cases of interest, we introduce the mean-flow characteristics near buffet onset. Figure \ref{fig:M_Meffect} shows time- and span-averaged Mach contours for (a) incipient buffet at $M=0.70$ and (b) developed buffet at $M=0.71$, where the supersonic regions are delineated by green sonic lines ($\overline{M}_\mathrm{loc}=1$). In both cases we distinguish two lobes in the upper boundary of the sonic line. With an increase in the freestream Mach number to $M=0.71$, the supersonic region grows in height and length, while the shock wave moves downstream. For $M \ge 0.71$, the final normal shock bends more upstream with increasing wall distance and the corresponding gradient becomes smoother due to the averaging effect. Shock-induced separation phenomena increase with increasing free-stream Mach numbers so that the re-circulation bubble becomes significantly thicker near the shock foot. Nevertheless, considering that the unsteady characteristics are very different, as shown in $C_L$ histories, it is notable that both contour plots of figure \ref{fig:M_Meffect} are qualitatively very similar.

\begin{figure} 
\centering
\includegraphics[trim={0cm 0cm 0cm 1cm},clip,width=.32\textwidth]{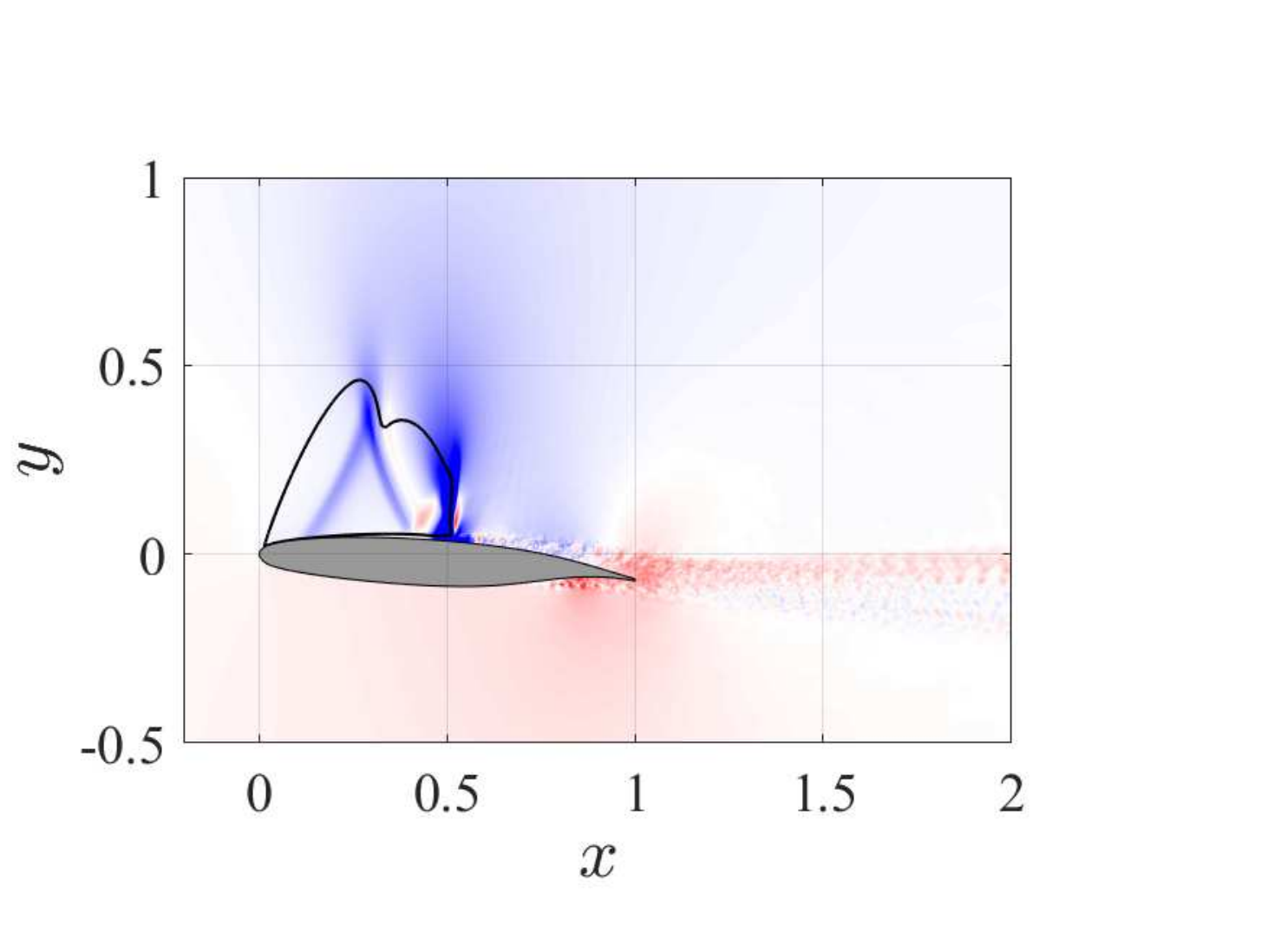}
\includegraphics[trim={0cm 0cm 0cm 1cm},clip,width=.32\textwidth]{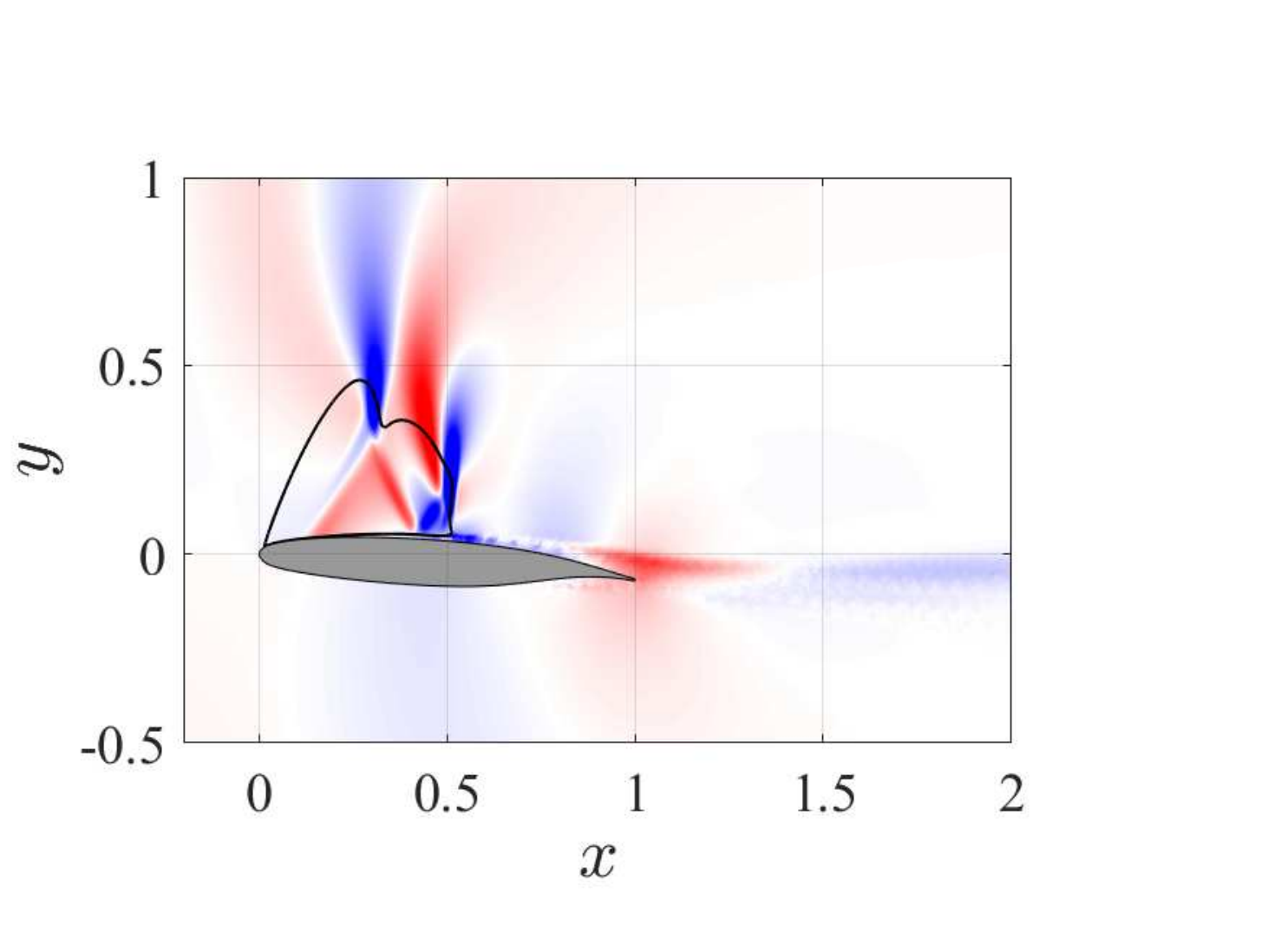}
\includegraphics[trim={0cm 0cm 0cm 1cm},clip,width=.32\textwidth]{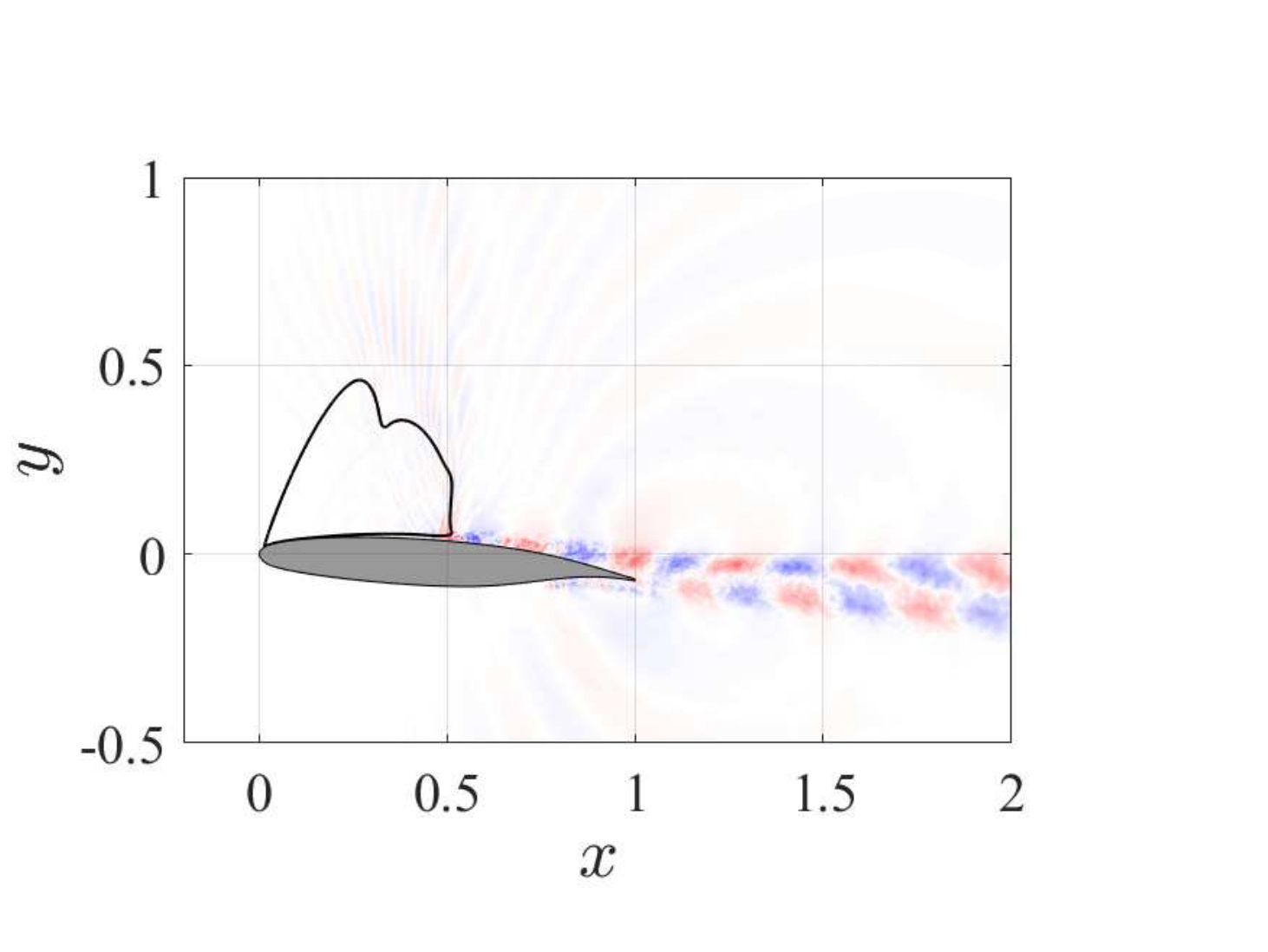}
\includegraphics[trim={0cm 0cm 0cm 1cm},clip,width=.32\textwidth]{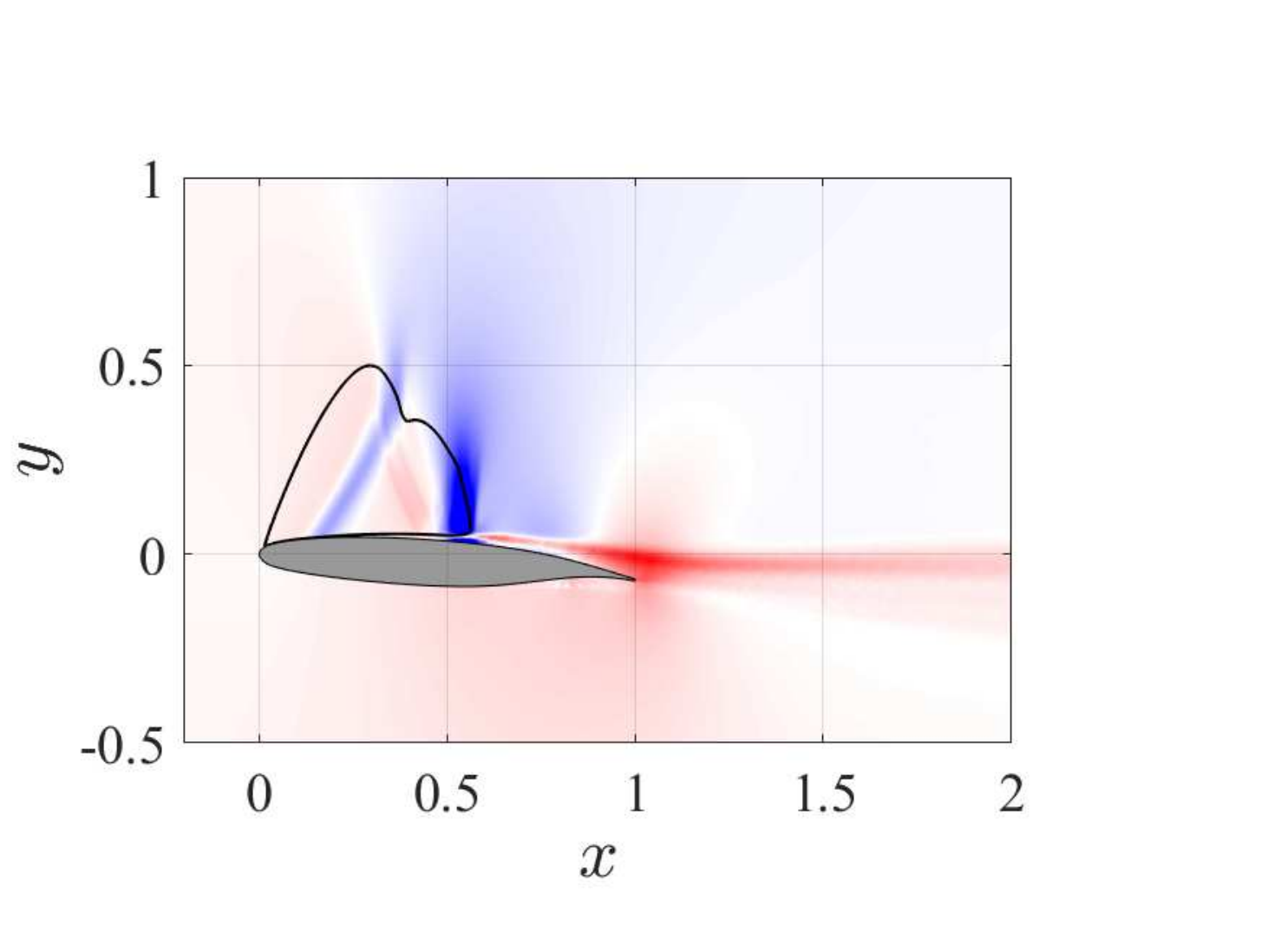}
\includegraphics[trim={0cm 0cm 0cm 1cm},clip,width=.32\textwidth]{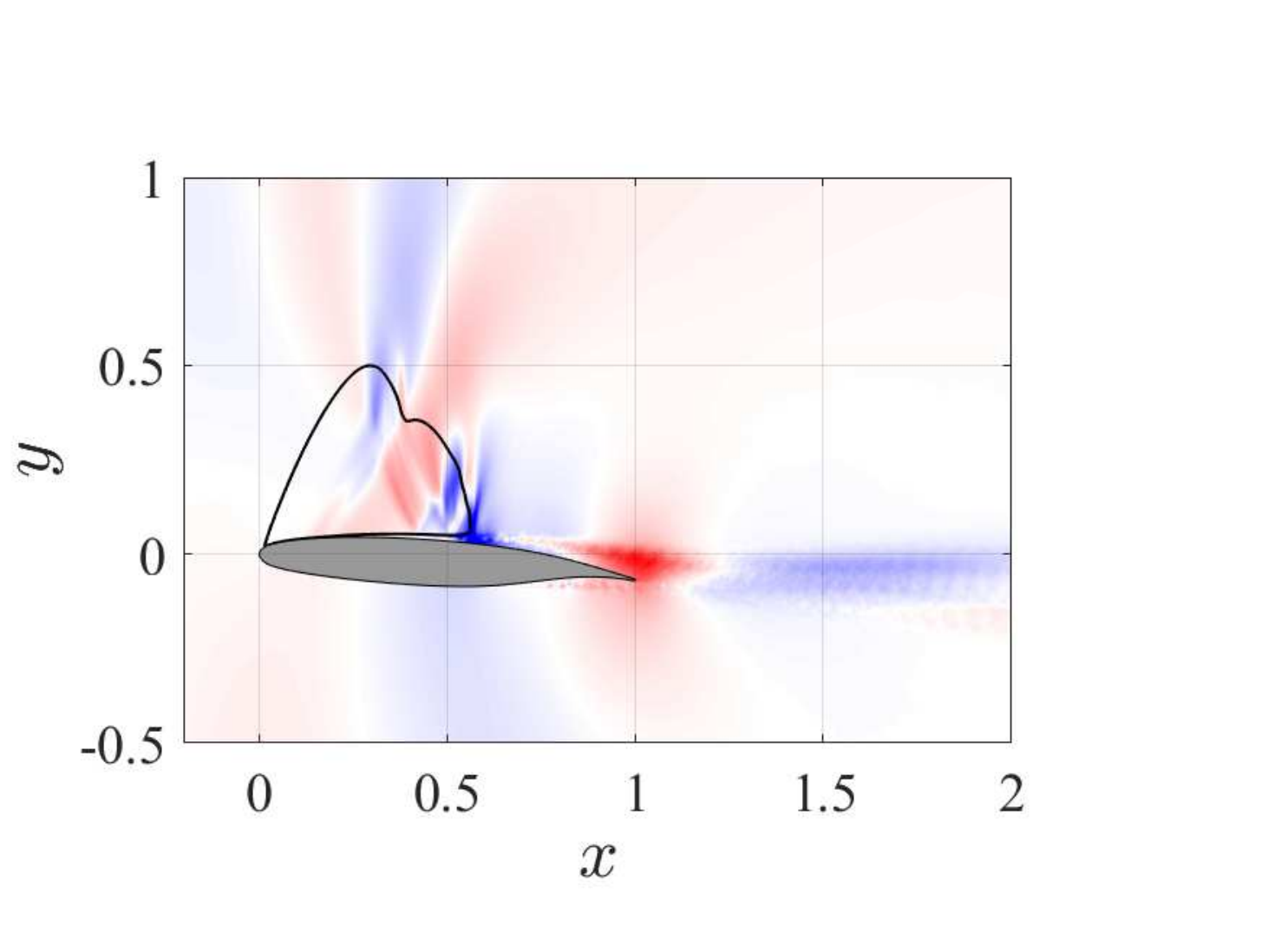}
\includegraphics[trim={0cm 0cm 0cm 1cm},clip,width=.32\textwidth]{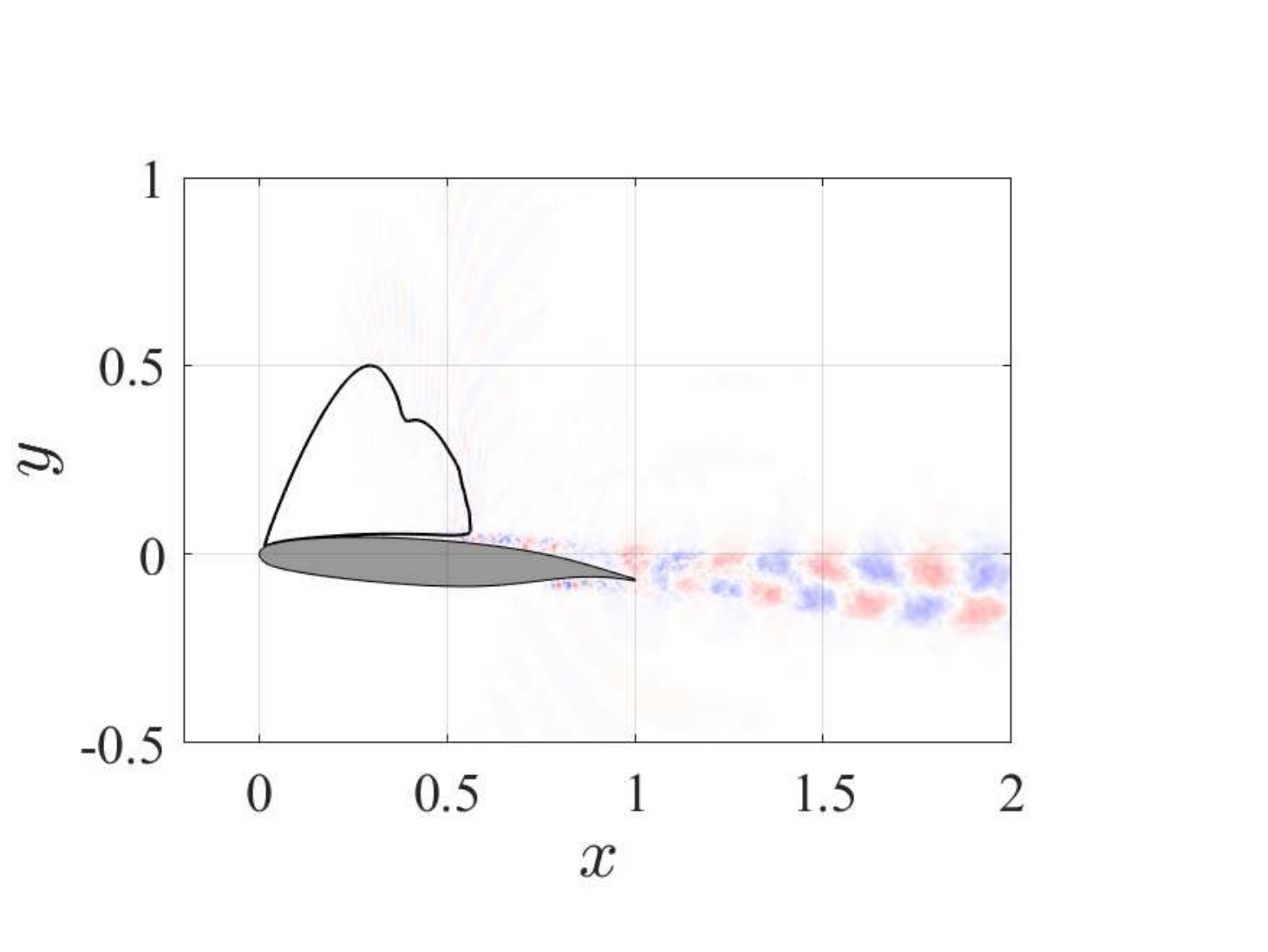}
\includegraphics[trim={0cm 0cm 0cm 1cm},clip,width=.32\textwidth]{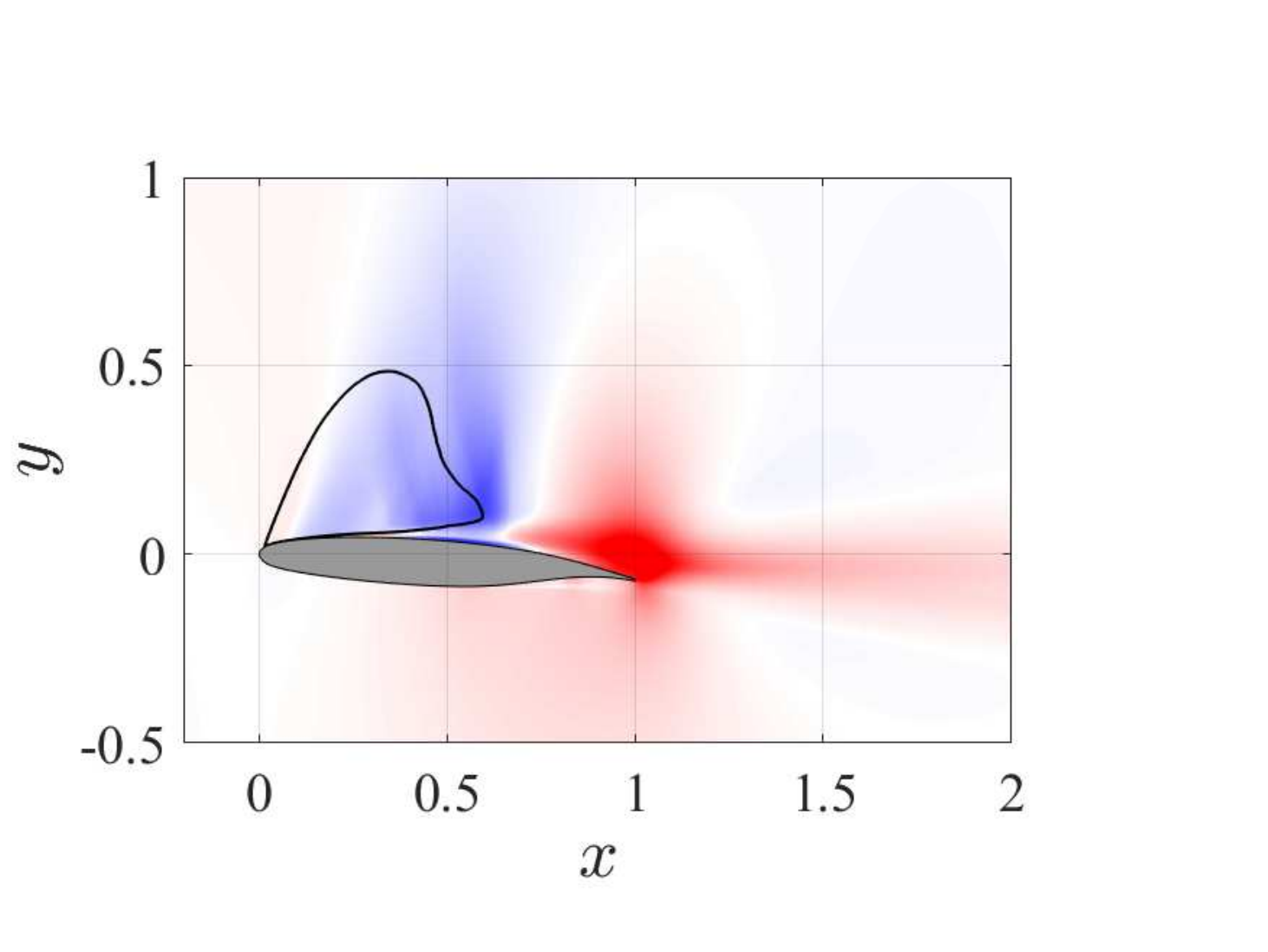}
\includegraphics[trim={0cm 0cm 0cm 2cm},clip,width=.32\textwidth]{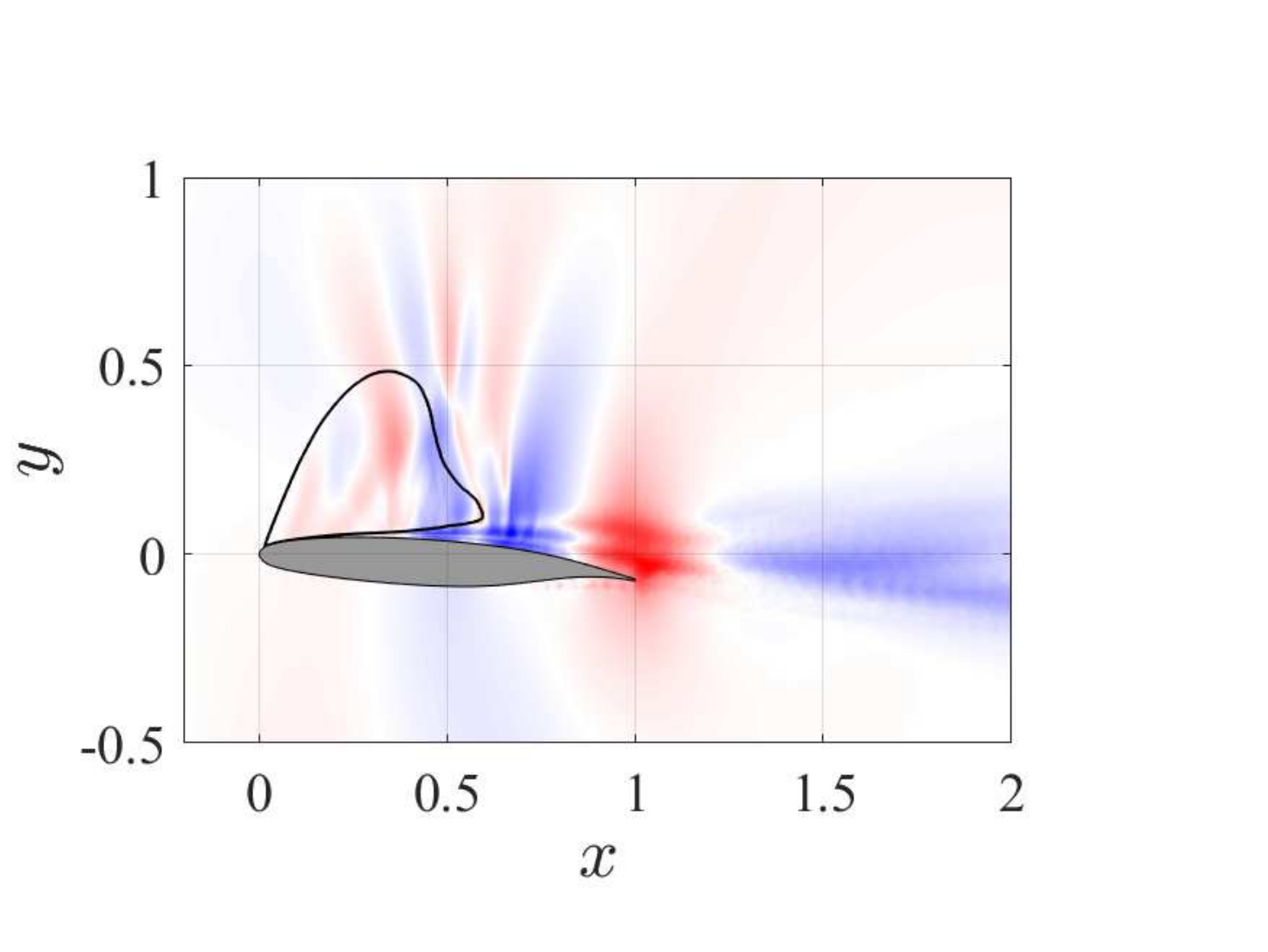}
\includegraphics[trim={0cm 0cm 0cm 1cm},clip,width=.32\textwidth]{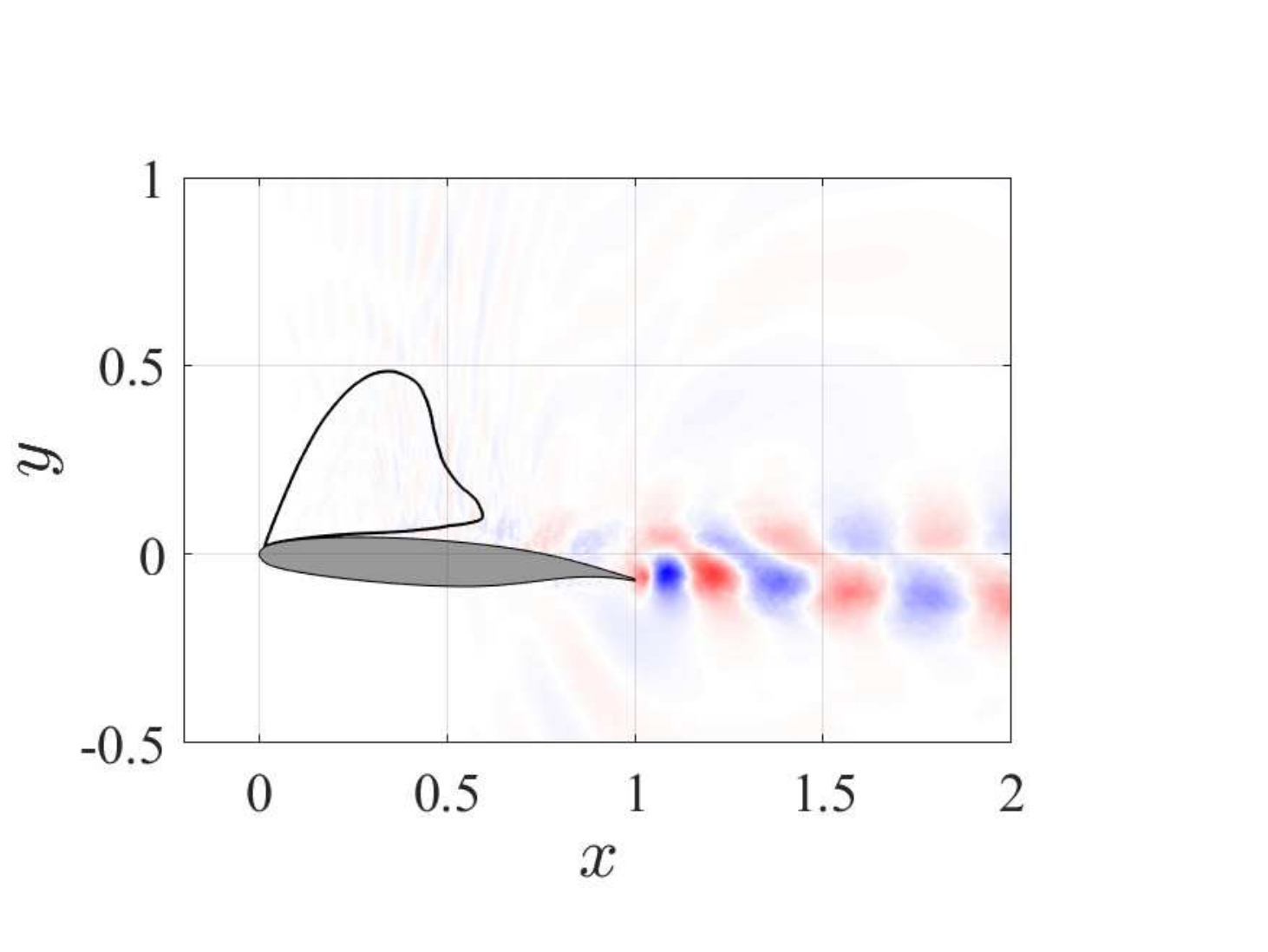}
\includegraphics[trim={0cm 0cm 0cm 1cm},clip,width=.32\textwidth]{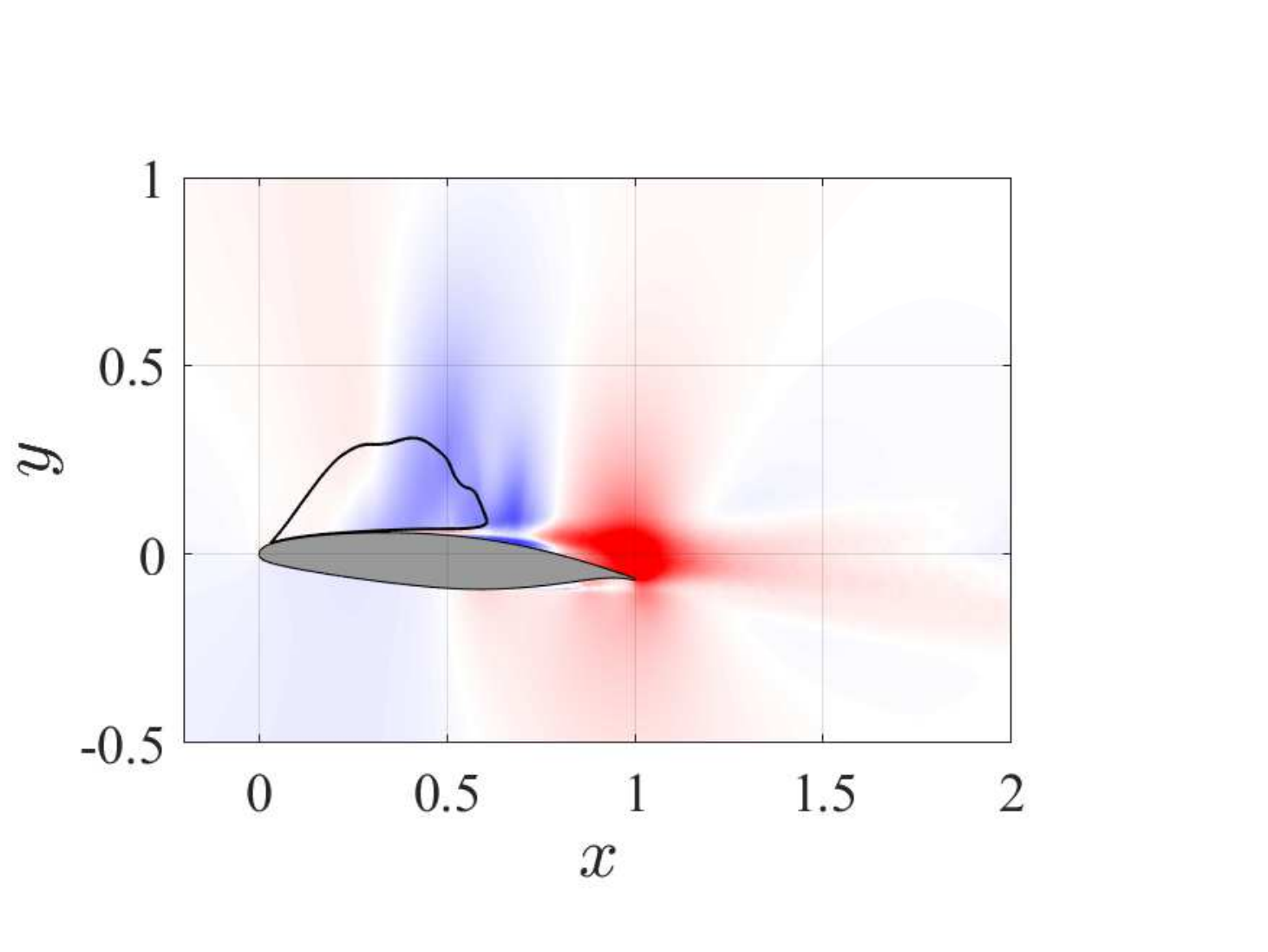}
\includegraphics[trim={0cm 0cm 0cm 2cm},clip,width=.32\textwidth]{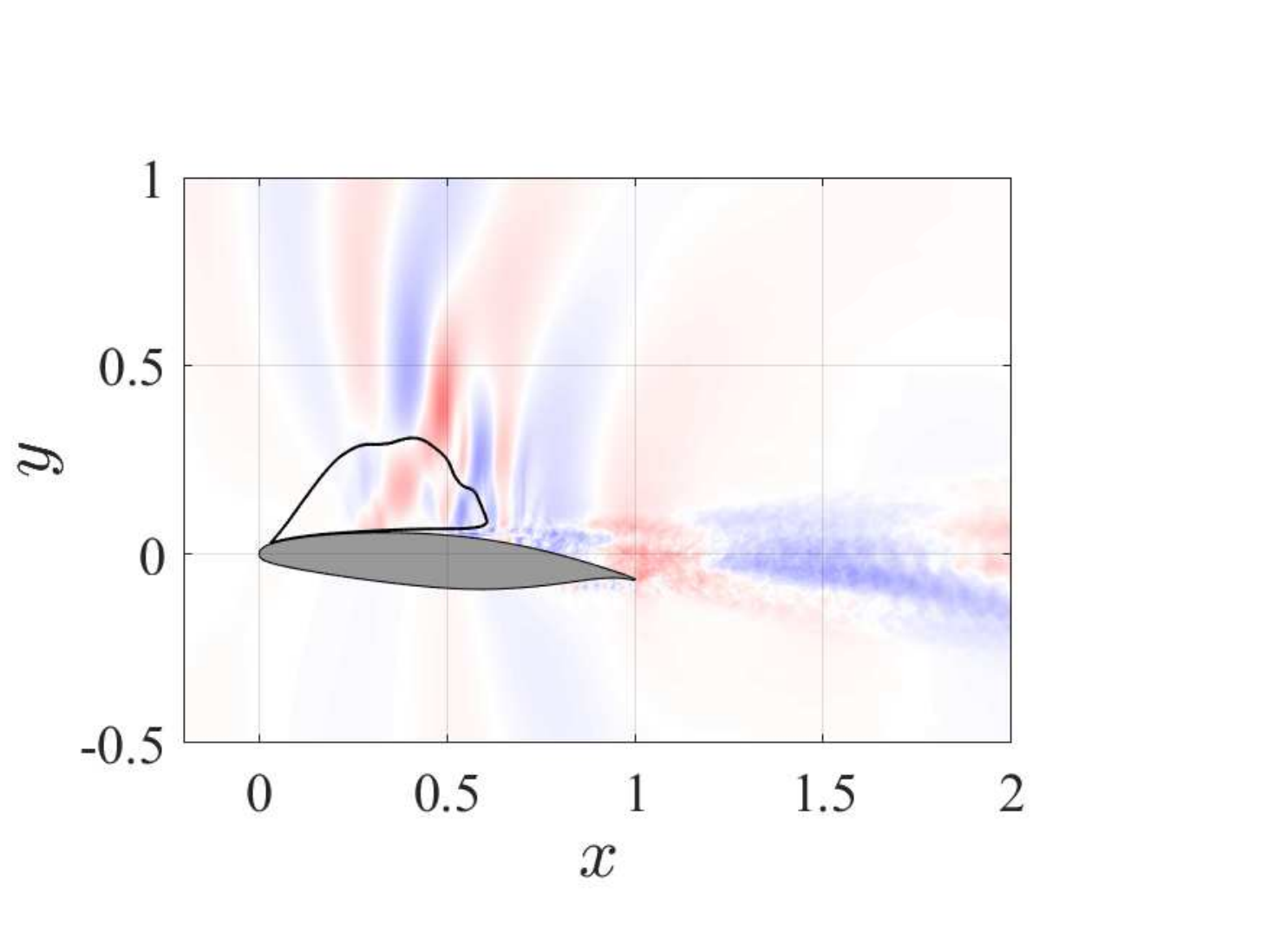}
\includegraphics[trim={0cm 0cm 0cm 1cm},clip,width=.32\textwidth]{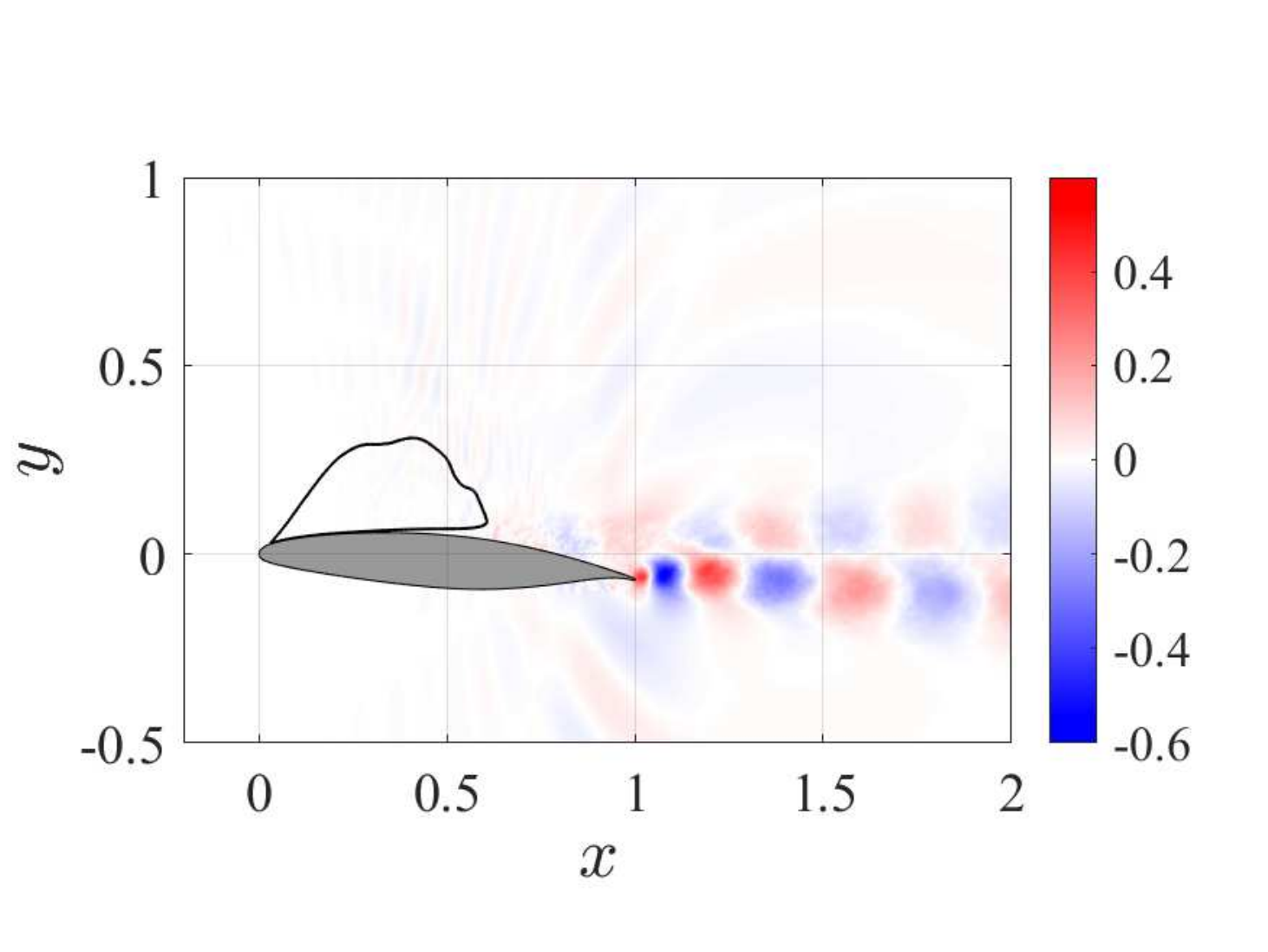}
\caption{SPOD modes associated with peaks at low (left), intermediate (middle) and high (right) frequency, shown using contour plots of the real part of the density field for $M = 0.70$ (top), $M = 0.71$ (second row), and $M = 0.735$ (third row). Corresponding SPOD modes for the V2C profile are shown in the bottom row for the same $Re$ and $\alpha$ for $M=0.735$ \citep{Moise2022}.}
    \label{fig:SPODModesRe500k}
\end{figure}
To assess dominant density fluctuations around the mean flows for incipient and developed buffet conditions, figure \ref{fig:SPODModesRe500k} shows SPOD modes at $M=0.70$, $0.71$, and $0.735$ for the OALT25 profile. In addition, the bottom row of plots shows corresponding SPOD modes of the V2C profile obtained in \cite{Moise2022} at $M=0.735$ and the same $Re$ and $\alpha$. 
Time-averaged sonic curves are indicated by black curves. Left-hand-side, middle, and right-hand-side columns correspond to low-, intermediate-, and high-frequency modes, respectively. These modes are also indicated by squares (low-frequency), circles (intermediate-frequency), and vertical lines (high-frequency) in the spectra of figure \ref{fig:Spectra}(b). Such SPOD modes allow us to assess the regions associated with large fluctuations of the flow field around its mean flow. Movies are available in the supplementary material and online (\href{https://youtu.be/uDQ3QktjveE}{https://youtu.be/uDQ3QktjveE}).

Looking at the low-frequency modes, we can observe at $M=0.70$ and $0.71$ the mode shapes becoming focused around the final shock wave that terminates the supersonic region. Density fluctuations of the $\lambda$-shock structure appear to be in phase with oscillations of the final shock wave and out of phase with the trailing edge. When buffet fully develops, localised regions of high amplitudes become smeared out, while fluctuation amplitudes around the trailing edge on both sides intensify and spread into the wake. 
Even though the buffet mode occurs for the V2C profile at a higher frequency of $St=0.15$ compared to the OALT25 test case ($St=0.082$), the mode shapes look very similar. As \cite{Moise2023} has already established the connection between the V2C mode and the globally unstable buffet mode of \cite{Crouch2007}, we expect the same underlying mechanism for the present low-frequency buffet mode of the OALT25 profile.

While low-frequency oscillations within the supersonic region are mainly in phase for $M=0.70$, we observe that intermediate-frequency oscillations cause significant changes of the shape of shock structures. The main terminating shock and the leading leg of the $\lambda$ structure are not in phase, which is at least one distinguishing feature compared with the transonic buffet mode.
Similar to the DMD mode of the V2C profile at $M=0.7$ reported in \cite{Zauner2019c}, weak peaks can also be observed at a similar frequency of $St=0.6$ in SPOD spectra reported in \cite{Moise2022} for Mach numbers $M=0.7$ and $M=0.735$. Besides the fact that those modes also show no significant sensitivity to $M$, they look qualitatively similar to corresponding SPOD modes of the present OALT25 profile. The differences in terms of amplitude may well be due to geometric characteristics leading to significant differences in mean-flow (indicated by sonic lines) and separation-bubble properties. In contrast to the OALT25 cases shown here, the absence of clear intermediate-frequency oscillations at pre-buffet conditions for the V2C profile indicates that onset properties of this instability can vary for different airfoil geometries independently of those associated with transonic buffet.

High-frequency phenomena are predominantly located within the wake, taking the form of a von-Karman vortex street. The structures can be traced upstream to the separation point and also appear to interact with the trailing edge, causing weak acoustic structures in the SPOD mode. These structures, labelled as \textit{wake modes}, agree well for OALT25 and V2C airfoils and were described in detail in \cite{Moise2022}. They will not be discussed further here, other than to note that they are distinct from the buffet and intermediate-frequency modes and appear at a lower frequency than the structures associated with transition to turbulence ($St>10$). Although it was suggested in \cite{Moise2022} based on results from the V2C profile that these wake modes might be the cause for the oscillations reported in \cite{Dandois2018}, the present direct examination of the OALT25 profile clearly shows that the intermediate-frequency modes (and not the wake modes) are the cause. 

\section{Analysis of intermediate-frequency mode}\label{sec:discussion}


While the low-frequency mode is clearly identified as conventional transonic buffet and the higher frequency modes are explainable as von-Karman wake modes, the mechanism of the intermediate-frequency remains still unclear. It was termed a laminar buffet mode by \cite{Dandois2018} and an analogy was made with the `breathing' mode seen in shock-induced separation bubbles, although the Strouhal numbers based on the separation length ($St_L$) were noted to be different. In the present section, we want to examine this bubble mode from a few different viewpoints.

\subsection{Scaling} 
Firstly we consider the frequency scaling of the mode. Table \ref{tab:Bubble} summarises relevant quantities for four different configurations of the OALT25 (simulation data from \cite{Dandois2018} is added to present results at $Re=3,000,000$) and one V2C test case from \cite{Zauner2019c}. The various dimensionless frequencies are $St=fc/U_\infty$ as obtained from the simulations, $St_L=fL_{sep}/U_\infty$ based on the interaction length $L_{sep}$, defined as the distance from separation to reattachment, and $St_{R}=fL_{sep}\lvert U_R \rvert/U_\infty^2=St_L \lvert U_R\rvert /U_\infty$, where $U_R$ is the maximum velocity in the reverse-flow region (approximated by the minimum $x$-velocity component). We see that $St_L$ is not constant over the different cases and is an order of magnitude higher than the values of this parameter seen in shock-induced separation bubbles. In that case the bubble response is known to be a form of growth and shrinkage known as bubble breathing. Although the precise mechanism is still debated, one line of argument (\textit{e.g.}, \cite{Touber2009}) is that it arises from the response of the separating boundary layer and the associated separation shock to stochastic forcing. Here the separation shock is rather weak in comparison with the shock impinging cases and the Mach number is not far into the supersonic regime. As another difference to cases with shock generators, the shock wave and impingement point for the current case is not fixed and the separation bubble is observed to periodically move back and forth, as it will be shown later in more detail.
Correcting $St_L$ by the factor $\lvert U_R \rvert /U_\infty$ leads to a Strouhal number, where present cases collapse much better, showing $St_{R} \approx 0.022$, even though the onset of low-frequency buffet makes the comparison more difficult for $M>0.7$ at $Re = 500,000$. 
The proposed scaling suggests a role for the reverse-flow vortex in the mechanism for the intermediate mode, but some caution is required, considering the limited number of data points.

\begin{table}
	\caption{Intermediate mode frequency scaling\label{tab:Bubble}}
	\begin{center}
		\begin{tabular}{lccccl|cc}
		    Airfoil & $Re$ & $M$ & $St$ & $L_{sep} / c $ & $U_R/U_{\infty}$ & $St_L$ & $St_{R}$\\
		    \hline
		    OALT25 & $500,\!000$ & 0.68 & 0.60 & 0.274 & -0.141 & 0.164 & 0.023 \\
		    OALT25 & $500,\!000$ & 0.70 & 0.42 & 0.361 & -0.147 & 0.152 & 0.022 \\
		    OALT25 & $500,\!000$ & 0.71 & 0.43 & 0.39 & -0.155 & 0.168 & 0.026 \\
		    OALT25 & $500,\!000$ & 0.735& 0.39 & $0.300^*$ & -0.190$^*$ & 0.117 & 0.022 \\
            OALT25 & $3,\!000,\!000$ & 0.735 & 0.92 & 0.175 & -0.122 & 0.161 & 0.020 \\
            OALT25$^{**}$ & $3,\!000,\!000$ & 0.735 & 1.12 & 0.195 & -0.090 & 0.218 & 0.020 \\
            \hline
		    V2C$^{***}$    & $500,\!000$ & 0.70 & 0.6 & 0.29 & -0.13 & 0.174 & 0.023 \\
		    \hline
		    \multicolumn{8}{l}{$^*$ \ \  estimated during low-lift phase, where bubble mode clearly exists} \\
		    \multicolumn{8}{l}{$^{**}$ \ values taken from \cite{Dandois2018}} \\
            \multicolumn{8}{l}{$^{***}$ values taken from \cite{Moise2022}}
		\end{tabular}
	\end{center}
\end{table}

\subsection{Shock and expansion wave structure}

\begin{figure}[bt!]
\centering
\includegraphics[width=1.0\textwidth]{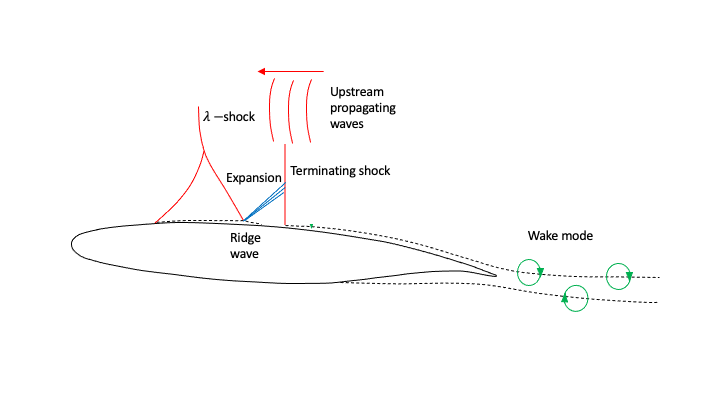}
\caption{Sketch of principal flow features at $M=0.7$, $Re=500,000$.}
    \label{fig:Schematic}
\end{figure}

To examine the intermediate-frequency separation bubble mode in detail we consider the $M=0.7$ case with $Re=500,000$, where this mode is well established. Before discussing the dynamics of the wave structures, we introduce some terminology. Figure \ref{fig:Schematic} extracts the main features that will be useful in the discussion. As mentioned previously, a prominent feature is a $\lambda$-shock structure that extends over the front part of the airfoil. The front leg of the $\lambda$-shock is the compression wave due to boundary-layer separation from the airfoil surface, which can be termed a separation wave. The upper part of the $\lambda$-shock terminates the supersonic region well above the airfoil, resulting in the first lobe structure seen in the average Mach number contours in figure~\ref{fig:M_Meffect}~(a). The rear leg of the 
$\lambda$-shock connects down to the top of the separation bubble on the airfoil surface. This shock reflects from the sonic line at the apex of the separated flow region as an expansion fan (coloured blue in figure \ref{fig:Schematic}), following a structure that is well known from impinging shock wave studies. The turning of the flow due to the expansion wave supports reattachment of the boundary layer. The impact of the shock on the separation is to stop the separation region from growing, so its maximum height is at the impingement location, forming a `ridge' feature. We will use the terminology `ridge wave' as a shorthand for the V-shaped shock-expansion wave pattern formed as the flow turns over the top of the separation bubble. 

Passing the ridge wave, the local freestream is still supersonic and this continues up to the terminating shock wave sketched in figure \ref{fig:Schematic}. The terminating shock determines the rear lobe of the supersonic region in figure~\ref{fig:M_Meffect}~(a). Boundary-layer transition occurs in the vicinity of the reattachment position, which is close to the shock-foot location and, at this Mach number, the boundary layer remains attached up to the trailing edge. The wake mode is sketched as a von-Karman vortex street, developing behind the airfoil. 

One other prominent feature, that leads us into a discussion of the dynamics, is the presence of upstream propagating waves in the subsonic region downstream of the $\lambda$-shock and above and behind the terminating shock. Many acoustic waves are present in the flow, some originating at the trailing edge and some from the transition region. Above the terminating shock, these waves take on a more coherent form and strengthen into upstream-propagating shock waves, with a supersonic flow (relative to their propagation speed) ahead of them.
It is interesting to note that very similar flow features have been observed in experiments of \cite{Boerner2021} over transonic turbine blades in a cascade configuration. In that application, the top bound of the supersonic flow region is restricted by the wake of a neighbouring blade and the upstream-propagation and potential influence of acoustic waves appears significantly restricted. Nevertheless they also observe two distinct phenomena at low and intermediate frequencies, which appear to be related with buffet and separation-bubble instabilities.

\begin{figure}[hbt!]
\centering
\includegraphics[width=0.9\textwidth]{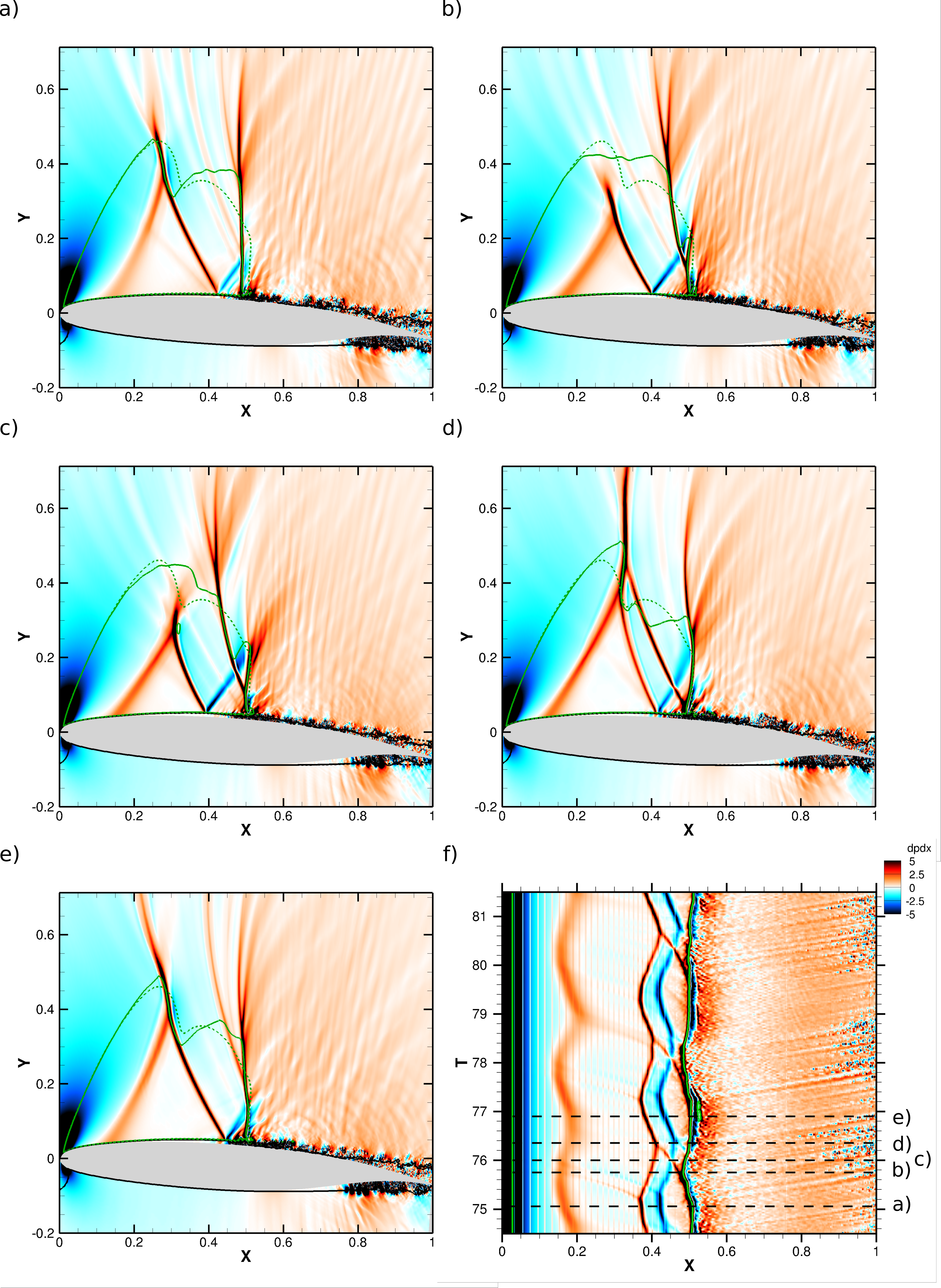}
\caption{Representative 2D snaphots showing contours of $\partial p / \partial x$ for $M=0.7$ at (a) $t=78.5$, (b) $t=79.1$, (c) $t=79.3$, (d) $80.1$, and (e) $80.8$.
Red (blue) contours denote adverse (favourable) pressure gradient. Green solid and black iso-curves denote sonic lines ($M_\mathrm{loc}=1.0$) and $M_\mathrm{loc}=0.5$, respectively. Solid and dashed curves respectively denote instantaneous and time-averaged iso-curves.
(f) Corresponding $x/t$ diagram of data extracted along a curve with a constant wall distance of $\Delta n = 0.05$, where labels indicate time instants of snapshots (a-e). }
    \label{fig:snaps_pre_dpdx}
\end{figure}

Having established a terminology and identified the principal physical features, we can now look at the flow development during a cycle of the intermediate mode. Figure \ref{fig:snaps_pre_dpdx} shows five representative snapshots of the streamwise pressure gradient $\partial p / \partial x$ during one cycle of the mode evolution, together with an $x/t$ diagram in part (f) showing the location of the snapshots with horizontal black lines. 
The $x/t$ diagram contains contours of $\partial p / \partial x$ in a plane located $0.05c$ above the airfoil surface. The method of plotting shows compression regions in red and expansion in blue. A corresponding movie can be found in supplementary material and online (\href{https://youtu.be/_9TwbAgkAU4}{https://youtu.be/_9TwbAgkAU4}). 
It can be seen in figure \ref{fig:snaps_pre_dpdx} (f) that the intermediate mode follows a regular oscillation and frame (e) at $t=80.8$ has a very similar structure as frame (a) at $t=78.5$. The period of $\tau \approx 2.3$ corresponds to the value of $St=0.42$ given in table \ref{tab:Bubble}. The first point to note as we move from frame (a) to (b) is the bifurcation of the terminating shock, the upper part of which moves upstream, merging with one of the upstream propagating waves above the terminating shock. The bifurcation can be seen in (b) at a distance $0.08c$ from the surface. The rear part of the bifurcated shock remains almost stationary, while the front part moves upstream and the bifurcation point moves towards the surface. Eventually, between (d) and (e) the front bifurcated wave merges with the $\lambda$-shock. The top part of the $\lambda$-shock  weakens out as it moves into a region with reducing Mach number and eventually ends up as acoustic radiation into the region upstream of the airfoil. 

A final point to note from figure \ref{fig:snaps_pre_dpdx} (f) is that all the wave features participate in the intermediate mode oscillation. At $x=0.2$ we have oscillations on the separation compression wave. In the region $0.4<x<0.45$ we see the ridge wave structure (shock and expansion waves) moving forward and backwards and at $x=0.5$ we see oscillations in the terminating shock wave. With reference to the ridge wave, the $x$ location of the separation wave lags by roughly $90^\circ$, indicating that the bubble is neither in a pure `breathing' mode, which would require the waves to be out of phase, nor in a simple up- and down-stream motion which would require the waves to be in phase. The movement of the ridge wave is close to being out of phase with the terminating shock location, given by the green sonic line in figure \ref{fig:snaps_pre_dpdx} (f), such that the ridge wave is furthest upstream when the terminating wave is furthest downstream. 

\begin{figure}[hbt!]
\centering
\includegraphics[width=0.95\textwidth]{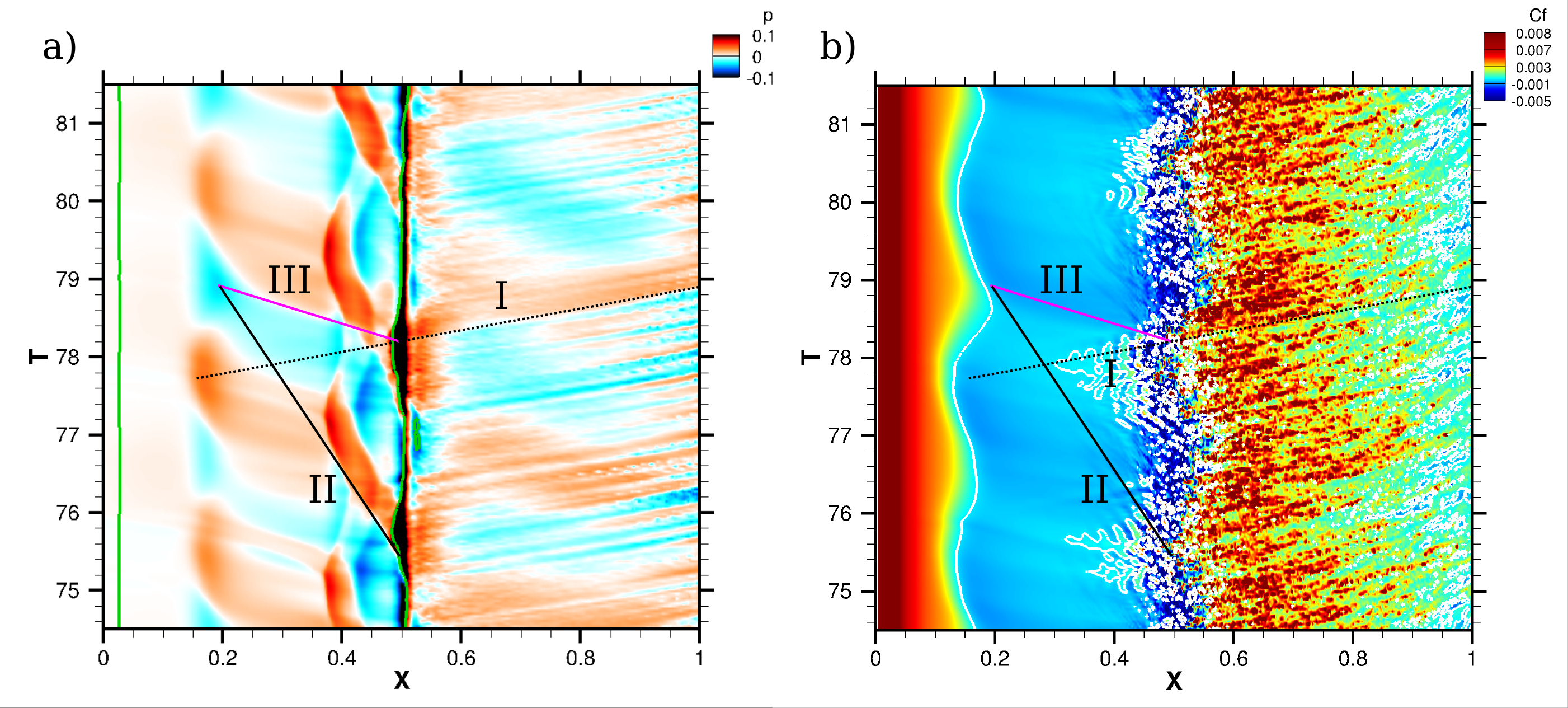}
\caption{Space/time ($x/t$) diagrams showing (a) pressure fluctuations ($p'=p-\Bar{p}$) at a monitor curve with a wall distance of $\Delta n = 0.1$ and (b) contours of skin-friction coefficient. 
Sketched lines emphasise upstream (dashed) and downstream (solid) propagating structures.} 
    \label{fig:xt_pressure}
\end{figure}

\subsection{Separation bubble dynamics}

After having identified the behaviour of distinct flow structures, we now want to further analyse their connections during a cycle of the intermediate frequency oscillation. To assist with this, we will use two more $x/t$ diagrams, shown on figure \ref{fig:xt_pressure}, based on (a) pressure fluctuations at a distance $0.05c$ above the airfoil surface and (b) skin friction contours on the surface. Some propagating features are identified on the figures and will be discussed in the following paragraphs.
Note that, in interpreting figure~\ref{fig:xt_pressure} (a), we need to be careful making connections to shock and expansion waves, as the centre of a shock wave, for example, is located by a rapid change in colour from blue to red, not by the centre of the red region. 


From figure \ref{fig:xt_pressure} (a) we can readily identify the main features. 
Linear instabilities develop in the separation bubble, where velocity profiles contain an inflexion point and form Kelvin-Helmholtz roll-ups with strong spanwise coherence. Such structures have been predicted for similar flows using linear stability analysis \citep{Zauner2017a, Zauner2019c} and typical phase speeds agree well with present convection speeds of $U_c \approx 0.2$, corresponding to the slope of the dotted line labelled $I$ in figure \ref{fig:xt_pressure}.
Even after break-down to turbulence, these vortex structures maintain spanwise coherence, leaving traces over the aft-section of the airfoil at $St>10$. Further details are provided by \cite{Zauner2020}, where Q-criteria visualisations of these phenomena for a different airfoil are shown in their figure 11.
The pressure oscillation caused by movement of the separation compression wave is seen at $x=0.2$. 
We also note the bifurcation of the terminating shock, with the leading wave moving upstream along the lines labelled $II$, eventually merging with the ridge wave at $x=0.4$. Near the separation wave, we again observe features in the contours that appear to align with the extension of line $II$. A third feature that appears is the line labelled $III$, which can be associated with traces of acoustic waves circumventing the supersonic region at a speed corresponding to the slope of the magenta line of approximately $U_a \approx U_{\infty}-U_{\infty}/M=0.43$ ($U_{\infty}$ denotes the freestream velocity). These waves pass through the supersonic regions at significantly higher frequencies compared to those corresponding to the bubble mode. However, the acoustic field may well be moderated by unsteadiness at lower frequencies, leading to changes in contour patterns of figure \ref{fig:xt_pressure} corresponding to acoustic speeds.
The velocity associated with line $II$ ($U_{II} \approx 0.087$) is about 20 times slower than typical acoustic waves, but of the same order as the maximum reverse-flow velocity ($U_R=-0.147$). 
The mechanism leading to the slope of features aligned with $II$ is not clear, but one could speculate about some near-wall structures being responsible.


More insights are possible from the skin friction plot in figure~\ref{fig:xt_pressure} (b), where sketched lines are identical with (a). Here the separation point is shown by the white line at $x \approx 0.15$. The reattachment is obscured by the unsteadiness, but can be traced as the colour change from deep blue to red at $x \approx 0.525$. 
During the upstream motion of the separation wave (\textit{e.g.} $76.6<t<77.8$) the $C_f$ is dropping, which favours the development of the linear instabilities mentioned before. As a consequence, the reattachment region starts spreading upstream. This observation aligns with local linear stability results of \cite{Zauner2019c}, where laminar boundary layers become unstable further upstream during low-lift phases (i.e. when the separation wave is close to its most upstream position). 
The region just upstream of reattachment, where transition is starting, forms a roughly triangular-shaped region, bounded by line $I$ and a slope corresponding to velocities similar to maximum reverse-flow velocities of $U_{r,max} \approx -0.147$ which may indicate an upstream convection of disturbances helping to sustain the process of transition to turbulence.

The above discussion has concerned only the $M=0.7$ $Re=500,000$ case. From analysis of SPOD mode shapes, some changes in this mode can be observed as $Re$ increases, the most obvious being that the separation bubble reduces in size and the upper portion and trailing shock of the $\lambda$-shock merge with the termination shock, as can be seen in the SPOD modes on figure \ref{fig:SPODModesReeffect}. This suggests that the shock bifurcation is not a crucial part of the mechanism, which instead involves a connection between fluctuations inside the separated boundary layer and upstream-propagating waves travelling up the main shock and around the top of the supersonic flow region. According to the Strouhal number arguments above, the period is set by the separation length and the strength of the reverse flow vortex. It would be interesting to investigate whether such modes can be detected in shock-induced separation bubbles. Furthermore, global stability and/or resolvent analysis are needed to gain better understanding of this intermediate-frequency separation-bubble mode.

\section{Conclusion}


An extensive parameter study has been performed for ONERA's OALT25 profile at a constant angle of attack of $\alpha=4^{\circ}$, covering $500,\!000 < Re < 3,\!000,\!000$ at $M = 0.735$ and $0.67 < M < 0.80$ at $Re = 500,\!000$.
Present results at $Re = 3,\!000,\!000$ and $M=0.735$ agree well with simulations and experiments of \cite{Dandois2018} and \cite{Brion2019}, respectively. 
With longer sampling time in the current work, a secondary low-frequency peak could be observed, in addition to the intermediate-frequency mode that was labeled as `laminar buffet' in \cite{Dandois2018}. 

The low-frequency mode matches, in terms of frequency and mode shape, the conventional global instability associated with transonic buffet \citep{Crouch2007}. This mode has previously been observed for cases subjected to tripped or fully turbulent boundary layers and often referred to as `turbulent buffet', even though we confirm here its occurrence for laminar upstream boundary layers as well.
The frequency of this low-frequency mode shows minor $Re$ sensitivity, but its amplitude increases with decreasing $Re$. When reducing $M$, on the other hand, the spectral peak shifts to reduced $St$ with decaying amplitude.

The intermediate-frequency mode appears to be focused on the laminar separation bubble, but does not show the same frequency or $180^{\circ}$ phase shift between separation and reattachment, which is characteristic of the commonly observed low-frequency `breathing' mode of shock-induced separation bubbles. 
Furthermore, we observe the formation and convection of linear instabilities within the separation bubble and upstream-propagating structures, which are significantly slower than acoustic waves in the freestream. Their role with respect to the bubble mode remains unclear.
Frequencies and amplitudes of established bubble modes show minor $M$ sensitivity, but weaken and shift to slightly higher frequencies for $M<0.7$. When reducing $Re$, the spectral peak shifts to reduced $St$ with decaying amplitude.

A new scaling $St_{R}=f L_{sep} U_R/U_{\infty}^2 \approx 0.02$ for the bubble mode is proposed, which emphasises the role of interaction length $L_{sep}$ and strength of the reverse-flow vortex $U_R$. Good agreement of this scaling for different $Re$ suggests a relation between the slowly upstream-propagating structures and the strength of the reverse-flow vortex, but requires further examination by global stability or resolvent analysis.

In addition to low- and intermediate-frequency modes, for cases at moderate Reynolds numbers we observe distinct wake modes at higher frequencies, which are similar to those described by \cite{Moise2022} for the V2C airfoil. At increased Reynolds numbers, however, this wake mode seems to interact with the first harmonic of the bubble mode, which causes strong vortex shedding from the separation bubble. 

Comparing SPOD modes of OALT25 and V2C profiles, we can identify similar modal features for both airfoil geometries. While their shapes and sensitivity to $M$ agree well, onset conditions as well as frequencies and amplitudes seem to be strongly influenced by geometric aspects, leading to significant differences in mean-flow characteristics.

\backmatter

\bmhead{Supplementary information}

\bmhead{Acknowledgments}
The authors would like to thank V. Brion and J. Dandois for insightful discussions and ONERA for providing the OALT25 airfoil geometry. PM was supported by an EPSRC grant entitled ``Extending the buffet envelope: step change in data quantity and quality of analysis'' (Grant ID: EP/R037027/1). The simulations were performed on the Iridis5 cluster at the University of Southampton and on the UK national supercomputer facility ARCHER2, using computer time provided via the UK Turbulence Consortium grant EP/R029326/1. 

\bmhead{Data availability}
Pertinent data will be made openly available on the University of Southampton repository.

\begin{appendices}

\section{Wall-pressure and skin-friction coefficient at moderate Reynolds numbers.}

\begin{figure}[hbt!]
\centering
\includegraphics[width=.495\textwidth]{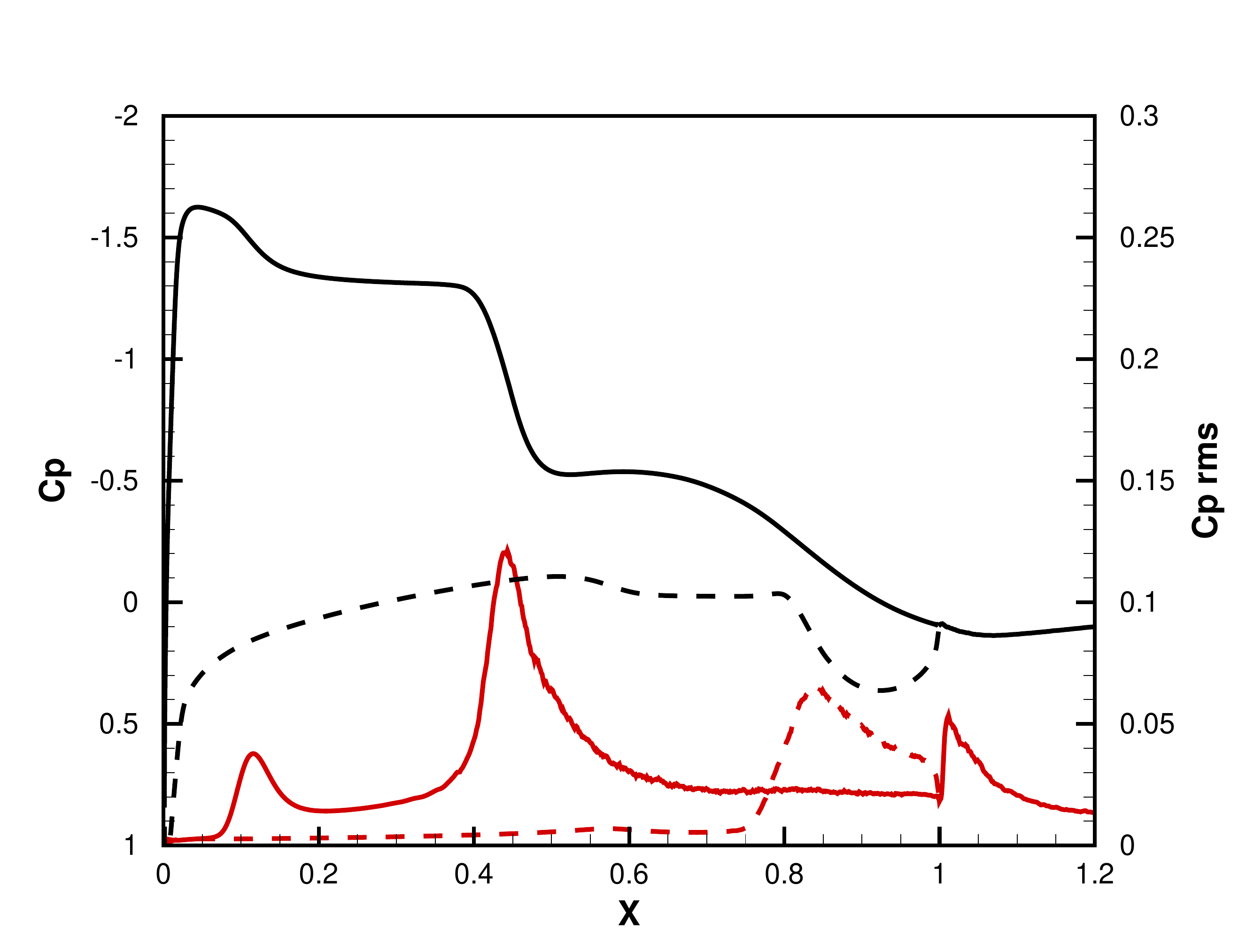}
\includegraphics[width=.495\textwidth]{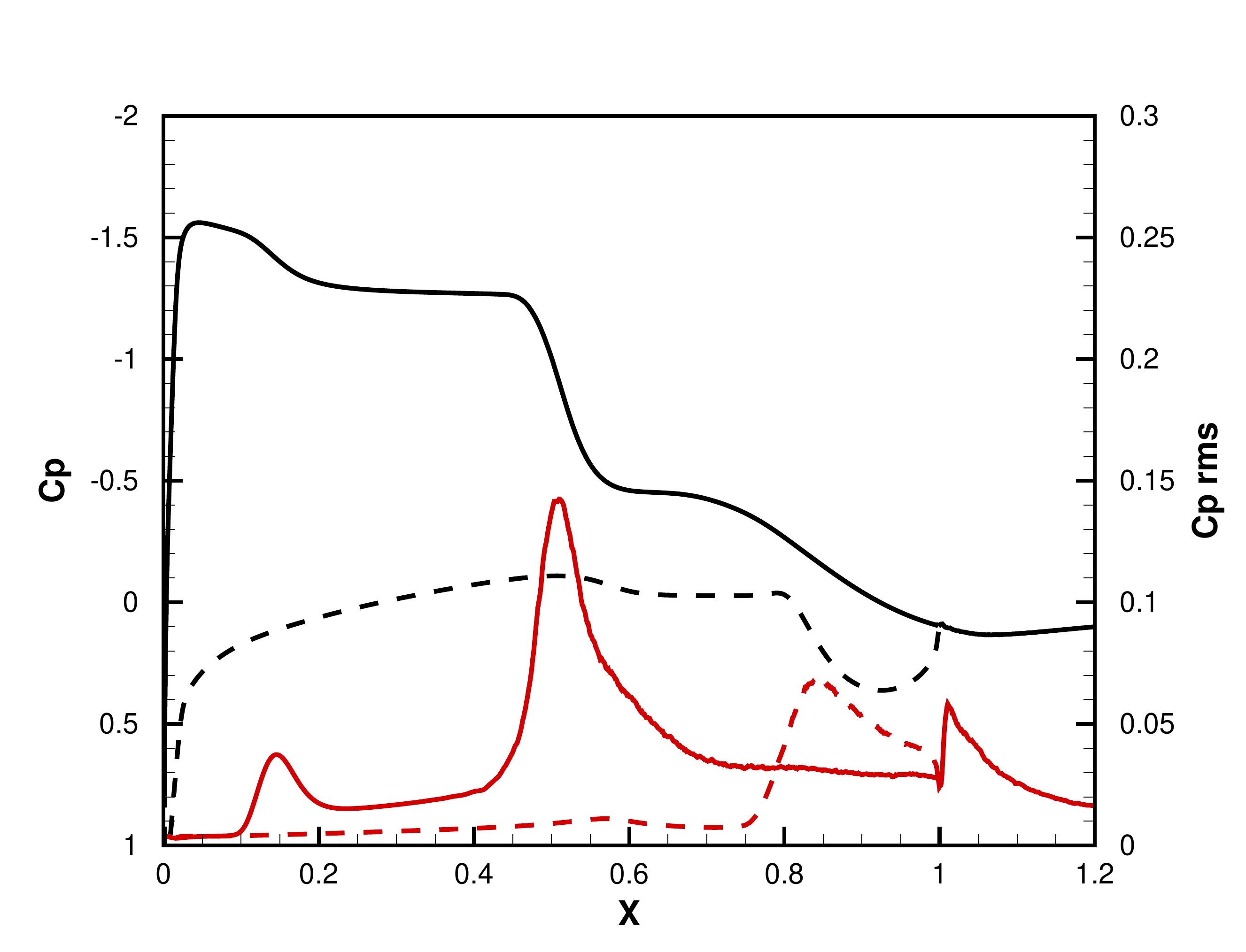}
\includegraphics[width=.495\textwidth]{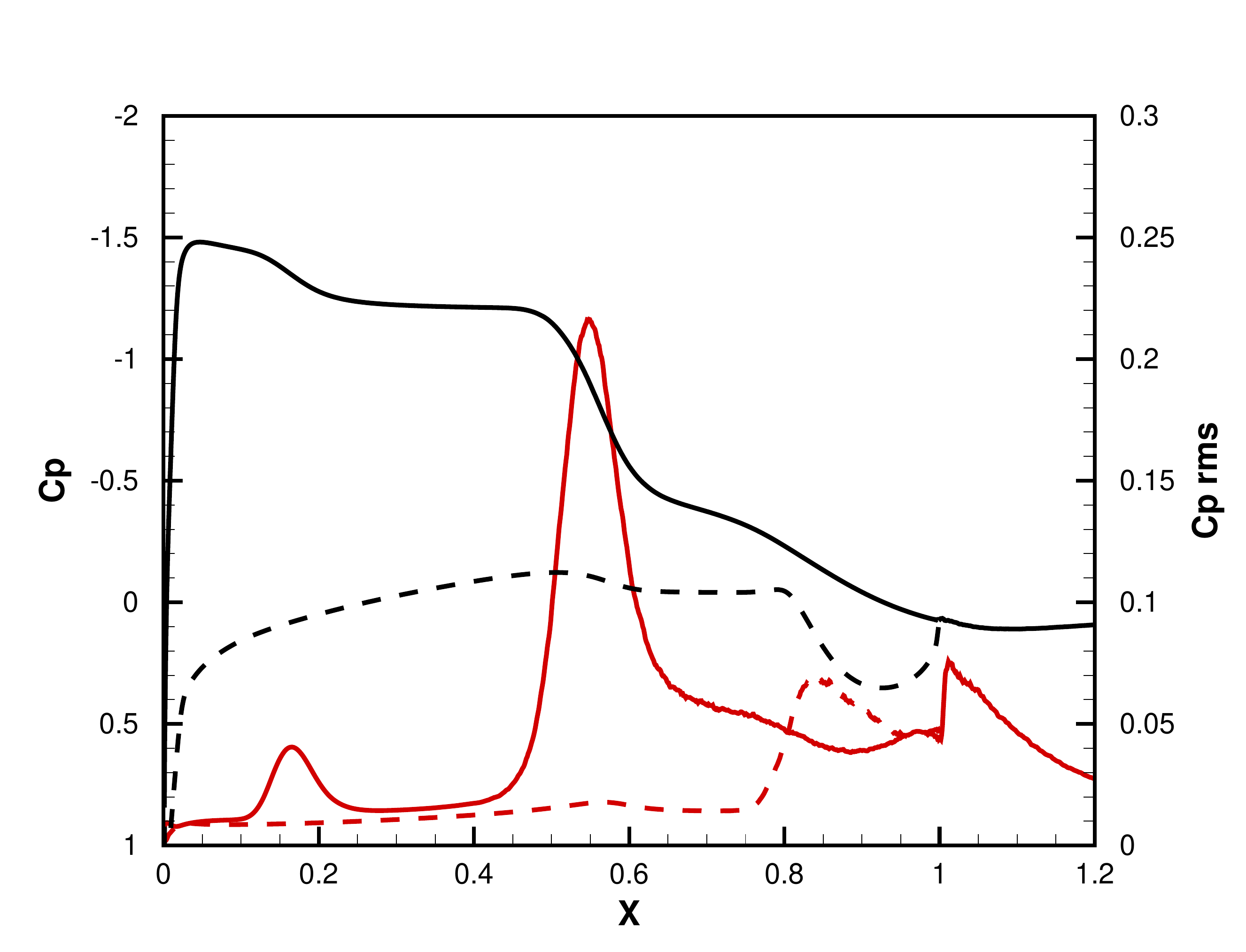}
\includegraphics[width=.495\textwidth]{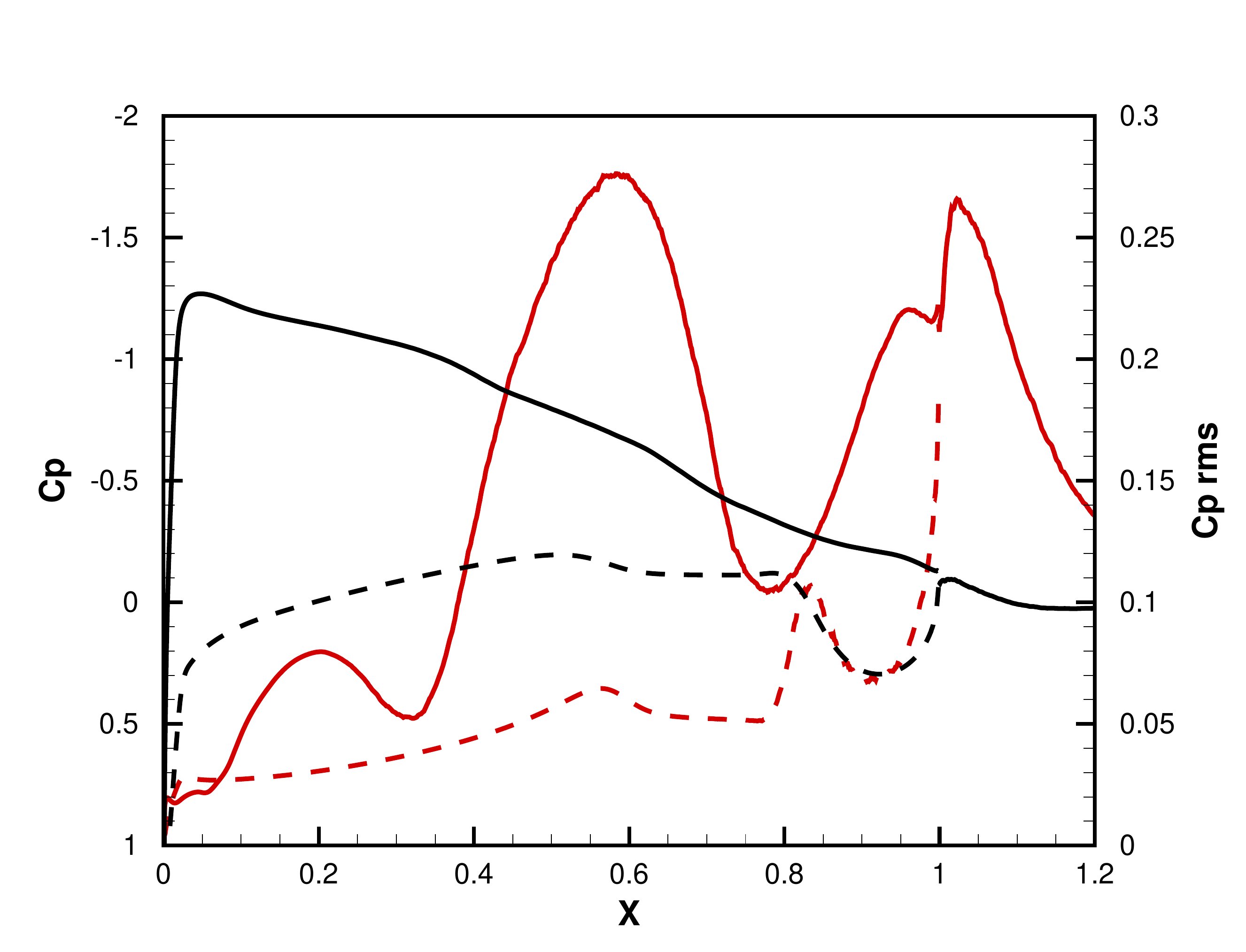}
\caption{Black curves show mean wall-pressure coefficient $C_{p,w}$ as a function of chord position and $C_p$ along a short portion of the wake center line for free-stream Mach numbers of (a) $M=0.69$, (b) $M=0.70$, (c) $M=0.71$, and (d) $M=0.735$. Red curves show corresponding root-mean-square values of the pressure coefficient denoted as $C_{p,rms}$. While solid curves correspond to the upper side of the airfoil, dashed lines correspond the lower side.} 
    \label{fig:Cp}
\end{figure}
Wall-pressure coefficient (black curves) together with corresponding root-mean-square ($rms$) fluctuations (red curves) are shown in figure \ref{fig:Cp} as functions of $x$. Solid and dashed curves denote suction and pressure sides, respectively. The $C_p$ curves are extended into the wake along the grid line originating from the center of the blunt trailing edge. Figure \ref{fig:Cp}(a) shows the wall-pressure distribution at $M=0.69$, before buffet develops. We observe a local minimum on the suction side at $x\approx0.05$ followed by a increase caused by a laminar separation bubble (as shown later in corresponding $C_f$ plots). The main shock is located around $x = 0.4-0.5$ and terminates the supersonic region as well as the (mainly) laminar boundary layer. 
While the shock wave and its corresponding pressure jump is strongly dependent on local Mach numbers of the supersonic flow upstream, downstream pressure recovery also depends on the airfoil geometry. In order to recover freestream conditions downstream of the airfoil, the flow along a streamline needs to accelerate again. This (weak) post-expansion leads to a small bump in the $C_p$ curve downstream of the shock wave centered around $x \approx 0.7$.
Wall-pressure fluctuations on the upper side denoted by the red solid curve show three distinct peaks where the flow separates ($x\approx0.15$), at the shock foot ($x\approx0.4$), and just downstream of the trailing edge. Fluctuations are strongest near the shock foot, where the boundary layer reattaches. Between these peaks, pressure fluctuations decay to relatively low amplitudes. 

While changes in $Cp$ for $M\le0.70$ are moderate, Mach-number effects become more pronounced with the onset of developed buffet.
For $M=0.735$ (figure \ref{fig:Cp}(d)), the $C_p$ curve over the suction side looks qualitatively very different compared to the cases before. Due to large-scale shock motion, we cannot identify the shock position as clearly as before, but the center of the (broad) $rms$ peak at $x \approx 0.6$ suggests no significant changes of the mean shock position compared to the $M=0.70$ case. We observe a global increase of pressure fluctuations around the airfoil for high $M$, particularly at the trailing edge and in the wake. It is interesting to see a small local maximum in the fluctuation intensities slightly upstream of the trailing edge at $x\approx0.95$. 



\begin{figure}[hbt!]
\centering
\includegraphics[width=.495\textwidth]{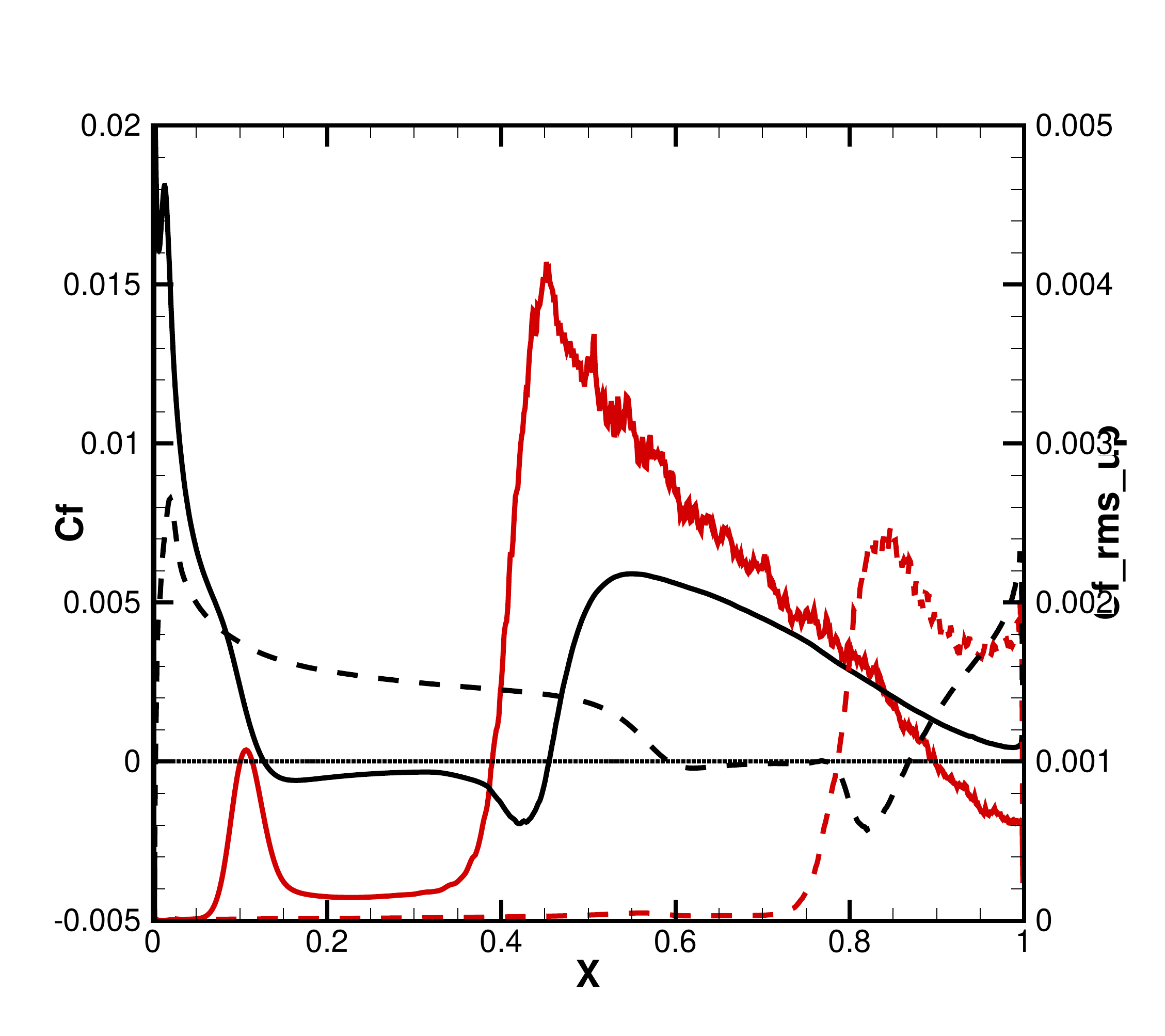}
\includegraphics[width=.495\textwidth]{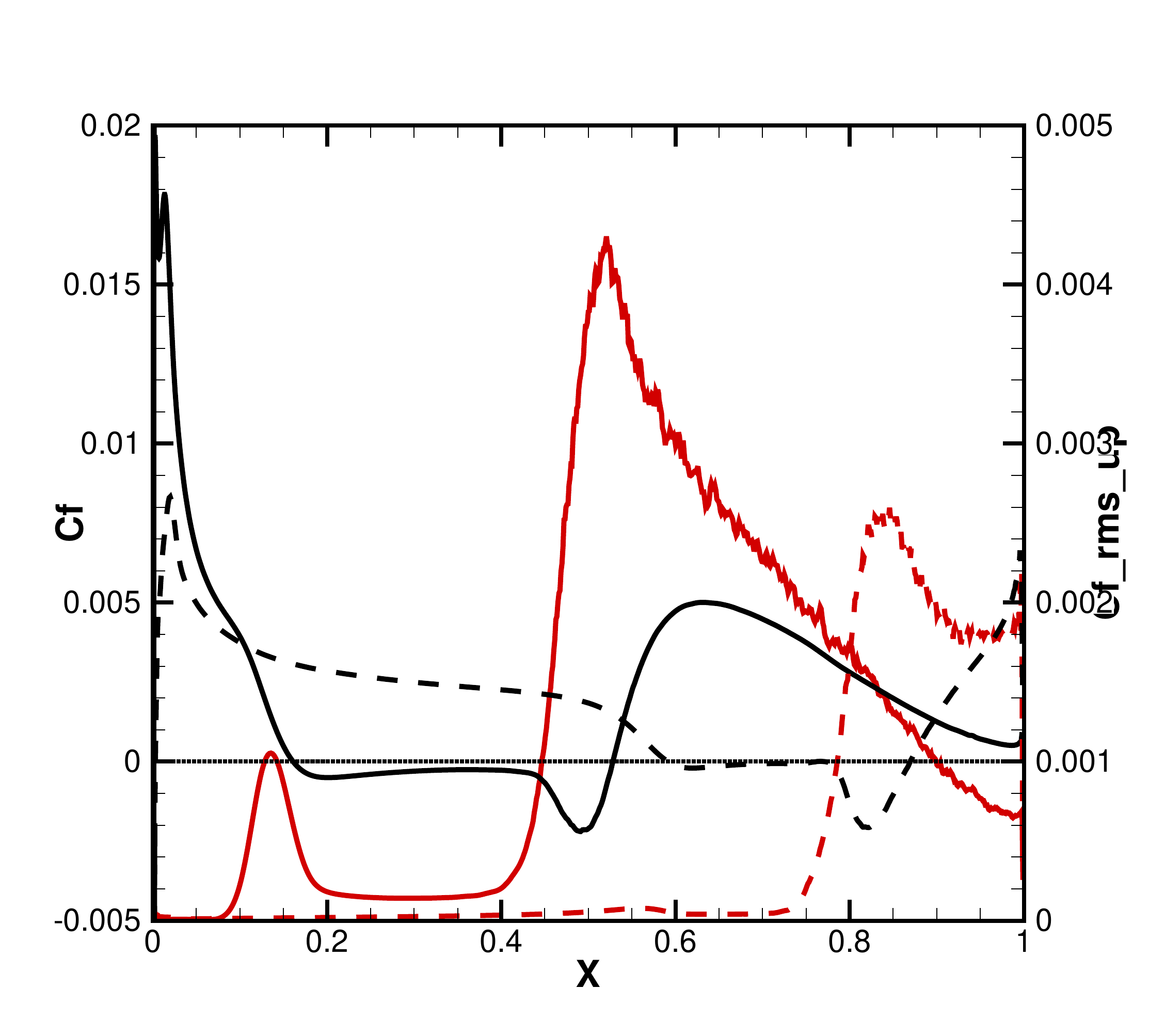}
\includegraphics[width=.495\textwidth]{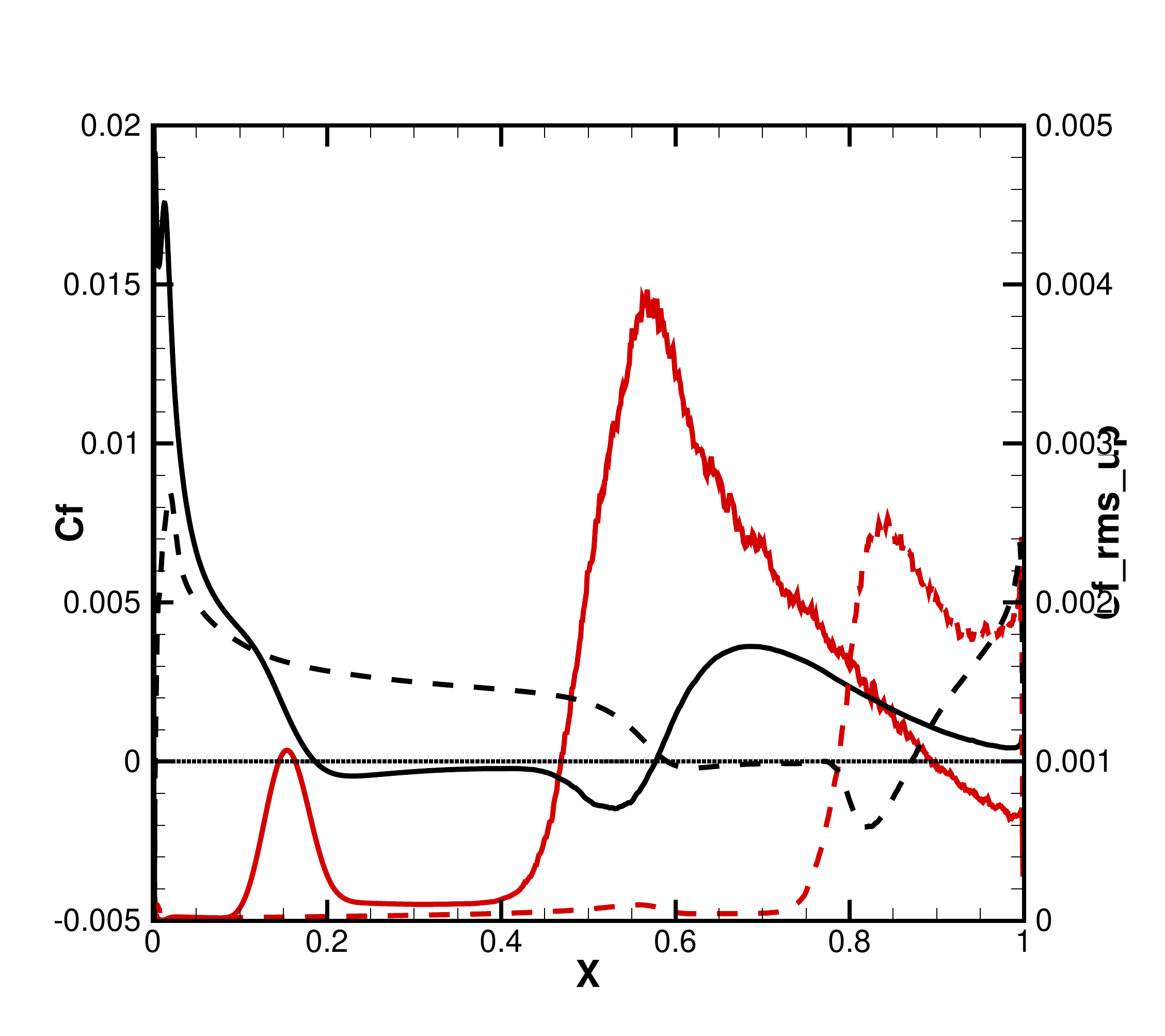}
\includegraphics[width=.495\textwidth]{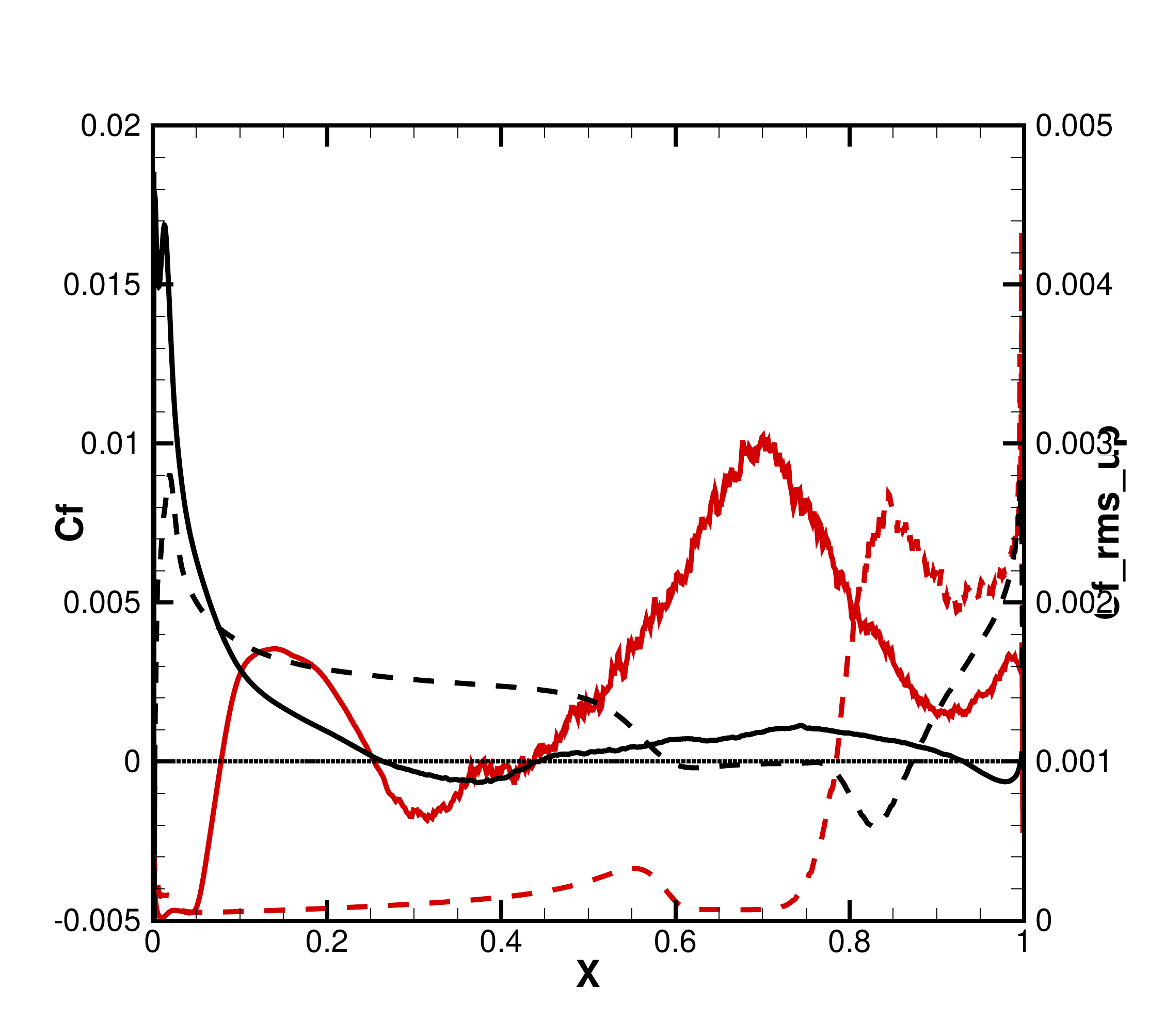}
\caption{Black curves show mean skin-friction coefficient $C_f$ as a function of chord for free-stream Mach numbers of (a) $M=0.69$ (left top), (b) $M=0.70$, (c) $M=0.71$, and (d) $M=0.735$. Red curves show corresponding root-mean-square values of the skin-pressure coefficient $C_{f,rms}$. While solid curves correspond to the upper side of the airfoil, dashed lines correspond the lower side.}
    \label{fig:Cf}
\end{figure}

In a similar manner as before for $C_p$, figure \ref{fig:Cf} shows the time-averaged skin-friction coefficient (black curves) as well as root-mean-squared $C_f$ fluctuations (red curves). Solid and dashed lines correspond respectively to suction and pressure sides. 
For $M \le 0.71$, the separation bubble ($C_f<0$) grows while shifting downstream. A local maximum in $C_f$ after reattachment ($C_f>0$) decreases with increasing $M$. For $M=0.735$, we don't observe such typical characteristics of a laminar separation bubble due to significant unsteady effects (intermittent flow separation). 
However, it should be emphasised that averaged values need to be treated with care after transonic buffet sets in. 

Time-averaged $C_f$ on the pressure side look very similar for cases before and after buffet onset, showing a separation bubble well upstream of the trailing edge. For almost all simulated cases at $M<0.735$, the fluctuation levels on the pressure side are similar and for $x<0.8$ (where the flow is mainly laminar) almost negligible.

On the suction side, we can observe local peaks in $C_f$ fluctuation-intensities at similar positions as for $C_p$, corresponding to separation and reattachment region (near the shock foot). It is remarkable that the trend for the global maximum of fluctuation intensities is now reversed though, as the peak decreases between $M=0.70$ and $M=0.735$ at the reattachment point (coinciding with the shock foot) with increasing Mach number and buffet intensity. The local maximum of fluctuation intensity at the separation point, however, first decreases with increasing $M$ before it increases again for $M>0.7$.
A third local maximum arises at the TE for $M=0.735$, while the remaining peaks become wider due to the large-scale unsteadiness mentioned before.

\end{appendices}

\bibliography{SEILES_paper_V3} 

\end{document}